\documentclass[a4paper,10pt]{article}
\pdfoutput=1 

\usepackage{jheppub}

\usepackage[T1]{fontenc}
\usepackage{comment}

\usepackage{amsmath,amssymb,color,graphicx,subcaption}

\newcommand{\Vsc}[1]{V_{\rm sc}(\{\tilde{p}\},#1)}

\newcommand{\cNP}[1]{\langle h_{#1}\rangle}
\newcommand{\cNPtilde}[1]{\langle \mathfrak{h}_{#1}\rangle}
\newcommand{\cNPct}[1]{\langle h^{\mathrm{c.t.}}_{#1}\rangle}
\newcommand{\cNPctimp}[1]{\langle h^{\mathrm{c.t.}}_{#1}\rangle_{\rm imp.}}
\newcommand{\cNPDMS}[1]{\langle h_{#1}\rangle_{\rm DMS}}

\author[a]{Andrea Banfi,}
\author[b]{Basem Kamal El-Menoufi,}
\author[a]{Ryan Wood}

\affiliation[a]{Department of Physics and Astronomy, University of Sussex,\\Sussex House, Brighton, BN1 9RH, U.K.}
\affiliation[b]{Lancaster-Manchester-Sheffield Consortium for Fundamental Physics, Department of Physics and Astronomy, University of Manchester, \\ Manchester M13 9PL, U.K.}

\emailAdd{a.banfi@sussex.ac.uk}
\emailAdd{basem.el-menoufi@manchester.ac.uk}
\emailAdd{rw380@sussex.ac.uk}

\title{\bf Interplay between perturbative and non-perturbative effects with the ARES method}

\abstract{We present a new semi-numerical method to compute leading
  hadronisation corrections to two-jet event shapes in $e^+e^-$
  annihilation. The formalism we present utilises the dispersive
  approach, where the magnitude of power corrections is controlled by
  suitable moments of an effective strong coupling, but it can be
  adapted to other methods. We focus on observables where the
  interplay between perturbative and non-perturbative effects is
  crucial in determining the power corrections.  A naive treatment of
  power corrections for some of these observables gives rise to an
  unphysical behaviour in the corresponding distributions for moderate observable values, thus considerably
  limiting the available range to fit the non-perturbative moments.
  We present a universal treatment to handle such observables, based
  on a suitable subtraction procedure, and compare our results to the
  analytic result in the case of total broadening. Finally, for the
  first time we present predictions for the thrust major, which cannot
  be handled with analytic methods.}

\begin{document} 

\maketitle
\flushbottom

\section{Introduction}
\label{sec:intro}

Final-state observables such as event-shape distributions and jet
rates are powerful probes of strong-interaction dynamics, as they span
a wide range of energy scales in a continuous fashion. Due to their
direct sensitivity to the strong coupling and the lack of additional
uncertainties due to parton distribution functions, in $e^+e^-$
annihilation they provide some of the most precise determinations of
$\alpha_s$ (see e.g.~\cite{Workman:2022ynf}). This is due to the
continuous improvements in their description in perturbative QCD. In
particular, distributions in event-shape variables and jet resolution
parameters, as well as their moments, can be computed at fixed order
in QCD perturbation theory. For observables that vanish in the two-jet
limit, fixed-order calculations are available at
next-to-next-to-leading order (NNLO)
accuracy~\cite{Gehrmann-DeRidder:2008qsl,Weinzierl:2008iv}, i.e.\ up
to order $\alpha_s^3$. However, fixed-order calculations are not adequate to describe
distributions in the two-jet region, where large logarithms of an
observable's value $v$ appear at all orders in perturbation theory. The
state-of-the-art accuracy for such resummations is
next-to-next-to-logarithmic
(NNLL)~\cite{Becher:2012qc,Banfi:2014sua,Banfi:2016zlc,Banfi:2018mcq},
which accounts for terms up to $\alpha_s^n \ln^{n-1}(1/v)$ in the
\emph{logarithm} of the observable's rate, defined as the fraction of events for
which an observable's value is less than a threshold $v$. For the
thrust, the heavy-jet mass and $C$-parameter, this accuracy can be
pushed to N$^3$LL
accuracy~\cite{Becher:2008cf,Chien:2010kc,Abbate:2010xh,Hoang:2015hka}.
Most $e^+e^-$ data lie at the LEP1 centre-of-mass energy
$Q=91.2\,$GeV, corresponding to the mass of the $Z$-boson, $M_Z$. At
this energy, all $\alpha_s$ determinations have to take into account
the presence of non-perturbative corrections due to
hadronisation. These are suppressed by inverse powers of the
centre-of-mass energy $Q$, and are hence known as ``power
corrections''. Leading power corrections to final-state observables
are linear in $1/Q$~\cite{Manohar:1994kq,Webber:1994cp}. Their effect
is to change both observable means and distributions by an amount of
order 10\% at LEP1 energies~\cite{Dokshitzer:1995zt,Dokshitzer:1997ew}.

Hadronisation is a phenomenon whose description lies beyond what can
be achieved with QCD perturbation theory. Therefore, in the absence of
a non-perturbative description of strong interactions, some modelling
is required.  One possibility is to use Monte Carlo parton-shower
event generators. These are multi-purpose tools that simulate fully
exclusive collider events down to the hadron level. In this approach,
one computes the ratio between an observable distribution or moment
at hadron and at parton level, and applies this correction to the
corresponding perturbative prediction. Such a procedure has been used
for instance in the determination of $\alpha_s$ via a comprehensive
fit of event-shape distributions and the two-jet rate, whose distributions have been computed at NNLO, supplemented with a resummation of
large logarithms in the two-jet region at the next-to-leading
logarithmic (NLL) accuracy~\cite{Dissertori:2009ik}. The uncertainty
in the fit due to hadronisation corrections was estimated by changing
the event generator and/or the hadronisation model, and was three times smaller than that on the
perturbative predictions. This was not the case for a most recent
determination using the two-jet rate, where the resummation of large
logarithms was performed at the NNLL
accuracy~\cite{Verbytskyi:2019zhh}. There the situation is reversed,
with hadronisation uncertainties being twice as big as perturbative
uncertainties. A reliable Monte Carlo determination of hadronisation
corrections requires that parton level distributions are in reasonable
agreement with the corresponding perturbative predictions. While this
seems acceptable for less accurate NLL resummations, the precision of
NNLL resummations adds extra tension and, quoting the Particle Data
Group (PDG), ``the parton level of a Monte Carlo simulation is not
defined in a manner equivalent to that of a fixed-order
calculation''~\cite{Workman:2022ynf}. Studies of event-shape moments
showed also a tension between Monte Carlo and analytic determination
of hadronisation corrections~\cite{Gehrmann:2010uax}.

Another possibility is to employ analytic hadronisation models. These
rely on the observation that perturbative series in QFT show a factorial
divergence, so they cannot be pushed to arbitrarily high orders. Such
divergence in QCD is called an ``infrared renormalon'' and affects all
distributions in event shapes and jet resolution
parameters~\cite{Nason:1995np,Beneke:1998ui,Caola:2021kzt}. An estimate of the size
of the error one makes by truncating the series before the factorial
divergence takes over leads to a contribution that is non-analytic in
$\alpha_s$, suppressed at most by $1/Q$ (see e.g.~\cite{Beneke:1998ui}). Further assumptions as to the physical origin of such a contribution led to
analytic models where hadronisation corrections are expressed in terms
of matrix elements of ultra-soft gluons. These matrix elements cannot be computed from
first principles but exhibit some degree of universality and hence, in principle, 
some of their features can be extracted from
data~\cite{Manohar:1994kq,Akhoury:1995sp,Dokshitzer:1995zt,Dokshitzer:1997ew,Dokshitzer:1997iz,Dokshitzer:1998pt}. The
most ambitious models include many powers of $1/Q$, written in terms
of soft matrix elements, and expressed in the form of a
non-perturbative shape function. A simpler but
more popular model considers $1/Q$ hadronisation corrections
only. These are ``computed'' using the matrix element for the emission
of a single ultra-soft gluon where the QCD coupling was replaced by a
soft effective coupling~\cite{Dokshitzer:1995qm}. The main result is
that $1/Q$ corrections are proportional to the average
$\alpha_0(\mu_I)$ of this effective coupling up to a low scale
$\mu_I$~\cite{Dokshitzer:1995zt,Dokshitzer:1997ew}. The corresponding
corrections are added to perturbative event-shape distributions and
mean values, and simultaneous fits of $\alpha_s(M_Z)$ and
$\alpha_0(\mu_I)$ are performed. Such fits show a mild degree of
universality for $\alpha_0(\mu_I)$, which for most observables lay
between 0.3 and 0.7, whilst giving a value for $\alpha_s(M_Z)$
compatible with the world average, although with large
uncertainties~\cite{L3:1995eyy,ALEPH:1996sio,DELPHI:1996oqw,OPAL:1996fae,OPAL:1997asf,L3:1997bxr,L3:1997dkv,MovillaFernandez:1997fr,Wicke:1998nkq,ALEPH:2003obs,DELPHI:2003yqh}. These
determinations use NLO fixed-order predictions, matched to NLL
resummations for distributions in the two-jet region, although there
exist recent examples where NNLL resummations are
considered~\cite{Gehrmann:2012sc}. A similar approach uses
soft-collinear effective theory (SCET) to compute the N$^3$LL
resummation for the $C$-parameter and the thrust distributions. SCET
also predicts the occurrence of a $1/Q$-suppressed correction, written
in terms of a non-perturbative parameter $\Omega$ in the
effective
theory~\cite{Abbate:2010xh,Becher:2013iya,Hoang:2015hka}. This leads
to simultaneous fits of $\alpha_s(M_Z)$ and the non-perturbative
parameter $\Omega$~\cite{Abbate:2010xh,Hoang:2015hka}. Quoting the PDG
again, such fits ``quote surprisingly small overall experimental,
hadronisation, and theoretical uncertainties of only 2, 5, and 9
per-mille, respectively, which calls for an independent
confirmation''~\cite{Workman:2022ynf}.

The above simultaneous fits of $\alpha_s$ and a non-perturbative (NP)
parameter have an intrinsic source of uncertainty related to the fact
that the same NP corrections computed in the two-jet region, where
resummation effects are important, are assumed to hold for larger
observable's values, i.e.\ in the three-jet region, where the fits are
actually performed. In fact, for most event shapes, leading power
corrections in the two-jet region correspond to a rigid shift of the
corresponding distributions~\cite{Dokshitzer:1997ew}. The jet
broadenings provide a notable exception~\cite{Dokshitzer:1998qp}, in
that the NP shift shows a mild dependence on the observable's value.
It has been argued~\cite{Gehrmann:2012sc}, and then shown
explicitly~\cite{Caola:2021kzt,Caola:2022vea}, that the emission of hard
perturbative gluons cause NP corrections to be proportional to a
coefficient that strongly depends on the observable's value and is
not the same as that computed in the two-jet limit, but only tends to
that when the observable's value is vanishingly small. With this
approach it was possible to determine $1/Q$ power corrections in the
three-jet region for the thrust and the
$C$-parameter~\cite{Caola:2022vea}. The outcome of this analysis is
that, in the three-jet region, leading hadronisation corrections still
depend on the same parameter appearing in the two-jet region, but with
an observable-dependent coefficient that can be computed. This led to a simultaneous fit of $\alpha_s(M_Z)$ and $\alpha_0(\mu_I)$ in the three-jet region for the thrust, $C$-parameter and three-jet resolution~\cite{Nason:2023asn}. This result
is an example of the fact that hadronisation corrections can never be
thought of in isolation, but always as additional contributions to given
perturbative configurations. Such interplay between perturbative and
non-perturbative effects is not only at work in the three-jet, but
also in the two-jet region. In fact, there are observables such as the
total broadening or the thrust major in which one needs to take
into account the recoil of a hard quark or anti-quark due to multiple
perturbative soft and collinear
emissions~\cite{Dokshitzer:1998qp}. The approach of~\cite{Caola:2022vea}, which considers only the emission of a
single hard gluon, is not able to account for these effects.

The aim of this paper is to present a general, semi-numerical method
to determine $1/Q$ hadronisation corrections in the two-jet region for
a large class of event-shape variables, including the interplay with
perturbative QCD radiation. The method follows the strategy of ARES,
the Automated Resummer for Event Shapes~\cite{Banfi:2014sua}, where
NNLL resummation is performed by means of a Monte Carlo simulation of
multiple soft and collinear emissions, accompanied by at most one
``special'' emission triggering NNLL corrections. In the present case, the special emission is an
ultra-soft gluon. This is the first step to having predictions for
event-shape distributions, valid in the two- and the three-jet regions, which
will be needed for a new and more accurate global simultaneous fit of
$\alpha_s$ and $\alpha_0$.

The paper is organised as follows. In section~\ref{sec:npcor} we
describe the method and its scope. In section~\ref{sec:observables} we
use the method to compute the NP corrections to known event-shape
distributions and to the thrust major. The latter, which gives the
main novel result of this paper, does not admit an analytic treatment,
and can only be handled with the semi-numerical approach presented
here. For the total broadening and the thrust major, the predictions
presented in section~\ref{sec:observables} cannot be naively matched
to the three-jet region due to an unphysical divergence arising in the
three-jet region. Such divergence can be ``cured'' in a general way,
which is the topic of section~\ref{sec:subtraction}. After a brief
discussion in section~\ref{sec:means} on how to treat the mean values
of event shapes, in section~\ref{sec:pheno} we perform simultaneous
fits of $\alpha_s(M_Z)$ and $\alpha_0(\mu_I)$ using NLL resummations
matched to NLO and our new determination of $1/Q$ power
corrections. The concluding section~\ref{sec:the-end} contains a
summary of our main findings and further research directions.

\section{Leading hadronisation corrections in the two-jet region}
\label{sec:npcor}

In this section we introduce a general method to compute leading power
corrections to a large class of event shapes in the two-jet
region. Before explaining the method in its full generality, it is
instructive to look at the known examples of the thrust and heavy-jet
mass, which display some of the theoretical issues involved.

\subsection{Two examples: the thrust and heavy-jet mass}
\label{sec:tau-rho}
Starting with kinematics, the thrust axis splits an event in two hemispheres $\mathcal{H}_1$ and $\mathcal{H}_2$. Let $\rho_1$ be the invariant mass squared of hemisphere $\mathcal{H}_1$ (normalised to $Q^2$, the centre-of-mass energy squared of the $e^+e^-$ collision) and similarly $\rho_2$ the invariant mass squared of hemisphere $\mathcal{H}_2$, in formulae
\begin{equation}
	\rho_\ell \equiv 
	\frac{1}{Q^2}\left(\sum_{i \in \mathcal{H}_\ell} q_i\right)^2\,, \qquad \ell=1,2 \ \ ,
\end{equation}
where each $q_i$ denotes the momentum of a final-state hadron. From $\rho_1$ and $\rho_2$, we can construct the heavy-jet mass $\rho_H=\max(\rho_1,\rho_2)$. Also, close to the two-jet limit, where the thrust $T$ is close to one, we have $1-T\simeq \rho_1+\rho_2$.

The bulk of events contributing to the distributions in the thrust and heavy-jet mass comes from soft and/or collinear partons, still with momenta much larger than $\Lambda_{\rm QCD}$, so that their emission probabilities can be safely computed in perturbative QCD. In this regime, one can approximately write the fraction of events such that $1-T<\tau$ or $\rho_H<\rho$ in terms of the distribution $J_q(k^2)$, the probability that a quark-initiated jet has an invariant mass $k^2$ \cite{Catani:1990rr}, as follows
\begin{equation} \label{eq:tau-rho-dist}
	\begin{split}
		\Sigma(\tau) &= 
		\int_0^\infty \!\! dk_1^2 \,J_q(k_1^2) \int_0^\infty \!\! dk_2^2 \,J_q(k_2^2) \,\Theta\Big(\tau Q^2 -k_1^2-k_2^2\Big)=\int_0^{\tau Q^2} \!\! \!\! dk_1^2 \,J_q(k_1^2) \int_0^{\tau Q^2-k_1^2} \!\! \!\! \!\! dk_2^2 \,J_q(k_2^2) \ \ , \\
		\Sigma(\rho) &= 
		\int_0^\infty \!\! dk_1^2 \,J_q(k_1^2) \int_0^\infty \!\! dk_2^2 \,J_q(k_2^2) \,\Theta\Big(\rho Q^2 -\max(k_1^2,k_2^2)\Big)=\left[\int_0^{\rho Q^2}\!\! \!\! dk^2 \,J_q(k^2)\right]^2 \ \ .
	\end{split}
\end{equation}
The above distributions are better expressed in terms of the Laplace transform of $J_q(k^2)$, defined as
\begin{equation} \label{eq:Jnu}
	\tilde J_q(\nu) = 
	\int_0^\infty \!\! dk^2 \, e^{-\nu k^2} \, J_q(k^2) \ \ .
\end{equation}
This gives
\begin{equation} \label{eq:tau-rho-Laplace}
	\Sigma(\tau) = 
	\int\frac{d\nu}{2\pi i \nu} \, e^{\nu \tau Q^2} \left[\tilde J_q(\nu)\right]^2\,, \qquad 
	\Sigma(\rho) = 
	\left(\int\frac{d\nu}{2\pi i \nu} \, e^{\nu \rho Q^2} \, \tilde J_q(\nu)\right)^2 \ \ ,
\end{equation}
where the contour of the $\nu$-integration runs parallel to the imaginary axis, to the right of all the singularities of the integrand. 

Following the approach of \cite{Dokshitzer:1997ew}, looking at the explicit expression of $\tilde J_q(\nu)$ reported there, we observe that it contains a contribution arising from soft radiation, as follows
\begin{align}\label{eq:Jtilde-soft}
	\ln \tilde J_q(\nu)\simeq 
	2 C_F \int_0^Q\frac{dq}{q} \, \frac{\alpha_s(q)}{\pi} \int_{q^2/Q^2}^{q/Q} \frac{du}{u} \left[e^{-\nu u Q^2}-1\right] \ \ .
\end{align}
The coupling $\alpha_s(q)$ in
eq.~\eqref{eq:Jtilde-soft} is to be interpreted as a
non-perturbative effective coupling. The main result of
\cite{Dokshitzer:1997ew} is that one can single out in the $q$-integral
 an ultra-soft contribution with $q<\mu_I$, where
 $\Lambda_{\rm QCD}\lesssim\mu_I< \tau Q$. This ultra-soft
 contribution produces a number of corrections suppressed by powers of
 $1/Q$, which arise from expanding $e^{-\nu u Q^2}$ in
 eq.~\eqref{eq:Jtilde-soft}. Following~\cite{Dokshitzer:1997ew}, the first order in the expansion gives the leading $1/Q$ non-perturbative correction
 \begin{equation}\label{eq:dJtilde-soft-expanded}
 	\delta\ln \tilde J_q(\nu)\simeq 
 	2 C_F \int_0^{\mu_I} \frac{dq}{q} \, \frac{\alpha^{\rm NP}_s(q)}{\pi} \int_{0}^{q/Q}\frac{du}{u} \left[-\nu u Q^2\right] 
 	= -\nu Q^2 \Delta \, ,
\end{equation}
where 
\begin{equation} \label{eq:Delta-def}
	\Delta \equiv 
	\frac{2 C_F}{\pi} \frac{\mu_I}{Q}\int_0^{\mu_I} \! \frac{dq}{\mu_I}\, \alpha^{\rm NP}_s(q) \, ,
\end{equation}
where $\alpha^{\rm NP}_s(q)$ is the non-perturbative component of the physical coupling $\alpha_s(q)$. The representation of
eq.~\eqref{eq:dJtilde-soft-expanded} assumes that ultra-soft
contributions to $\tilde J_q(\nu)$ have the same form as perturbative
contributions, just with a different coupling. One can take a more
general approach and model the whole $\delta\ln \tilde J_q(\nu)$ as a
non-perturbative shape function. In this case, the first term in the
expansion in $\nu$ will still have the form as in
eq.~\eqref{eq:dJtilde-soft-expanded}, but $\Delta$ will not be
interpreted as the moment of an effective coupling, but rather as a
matrix element of Wilson lines~\cite{Korchemsky:1994is}. Let
us now write $\ln \tilde J_q(\nu)$ as the sum of a fully perturbative
contribution, $\ln \tilde J^{\rm PT}_q(\nu)$,\footnote{The actual form of
	$\ln \tilde J^{\rm PT}_q(\nu)$, which depends on the perturbative
	order considered, is not relevant for our discussion. The only
	important aspect is that the coupling appearing in the integrals
	defining $\ln \tilde J^{\rm PT}_q(\nu)$ is perturbative. This implies that $\ln \tilde J^{\rm PT}_q(\nu)$ produces a renormalon divergence, proportional to $1/Q$, which by construction cancels against $\delta\ln \tilde J_q(\nu)$.}  and the
leading NP correction $-\nu Q^2 \Delta$. For the thrust distribution,
this gives
\begin{equation} \label{eq:tau-Laplace-shift}
	\Sigma(\tau) = 
	\int\frac{d\nu}{2\pi i \nu} \, e^{\nu (\tau-2\Delta) Q^2}\left[\tilde J^{\rm PT}_q(\nu)\right]^2 = \Sigma_{\rm PT}(\tau-2 \Delta) \ \ ,
\end{equation}
where
\begin{equation} \label{eq:tau-PT}
	\Sigma_{\rm PT}(\tau) \equiv 
	\int\frac{d\nu}{2\pi i \nu} \, e^{\nu \tau Q^2}\left[\tilde J^{\rm PT}_q(\nu)\right]^2 \ \ .
\end{equation}
Keeping the leading power of $\nu$ in $\delta\ln \tilde J_q(\nu)$
gives rise to a shift in the perturbative thrust distribution. If we
analyse more closely the assumptions under which the shift
approximation works, we see that it holds mathematically because we
have been able to expand the exponential $e^{-\nu u Q^2}$ in
eq.~\eqref{eq:Jtilde-soft}. But what is the physical meaning of such
expansion? It is known that the values of $\nu$
giving the largest contribution in eq.~(\ref{eq:tau-PT}) are of the order $1/(\tau Q^2)$. Also,
the dimensionless variable $u$ is just the contribution to the thrust
due to the ultra-soft gluon. This is of the order $\mu_I$ and hence
much smaller than $\tau Q^2$. In this limit, we have
$\Delta\sim \mu_I/Q \ll \tau \sim 1/(\nu Q^2)$, hence we can expand
$\Sigma(\tau)$ as follows
\begin{equation} \label{eq:tau-Laplace-shift-expanded}
	\begin{split}
		\Sigma(\tau)&\simeq 
		\int\frac{d\nu}{2\pi i \nu} \, e^{\nu \tau Q^2}\left(1-2 \nu Q^2 \Delta\right)\left[\tilde J^{\rm PT}_q(\nu)\right]^2= \Sigma_{\rm PT}(\tau)-2\Delta \frac{d\Sigma_{\rm PT}(\tau)}{d\tau}\simeq \Sigma_{\rm PT}(\tau-2 \Delta) \ \ .
	\end{split}
\end{equation}
The approximation in the last line of the above equation is not needed
for event shapes such as the thrust, heavy-mass, $C$-parameter
\cite{Dokshitzer:1998pt} and $D$-parameter \cite{Banfi:2001pb}, whose
shift is just a constant value, independent of the Laplace variable
$\nu$. The distributions in those event shapes can be expressed in
terms of a Laplace transform, where the NP shift appears manifestly in
an exponentiated form as in eq.~\eqref{eq:tau-Laplace-shift}.  There
are cases, such as the jet broadenings, see
e.g.~\cite{Dokshitzer:1998qp,Banfi:2001sp}, where the shift itself
depends on the Laplace variable $\nu$. In that case, under the
assumptions that $\mu_I/Q$ is much less than the observable's value,
one only obtains a shift in the approximate way indicated by
eq.~\eqref{eq:tau-Laplace-shift-expanded}.

At this point, an important remark is in order. While in the case of
the thrust the non-perturbative shift is completely uncorrelated with
perturbative contributions, this is not the case for the heavy-jet
mass. In fact, for its integrated distribution, we obtain
\begin{equation} \label{eq:rho-Laplace-shift}
	\Sigma(\rho) = 
	\left(\int\frac{d\nu}{2\pi i \nu} \, e^{\nu (\rho-\Delta) Q^2} \, \tilde J^{\rm PT}_q(\nu)\right)^2 =\Sigma_{\rm PT}(\rho-\Delta) \ \ ,
\end{equation}
with
\begin{equation} \label{eq:rho-PT}
	\Sigma_{\rm PT}(\rho) \equiv 
	\left(\int\frac{d\nu}{2\pi i \nu} \, e^{\nu \rho Q^2} \, \tilde J^{\rm PT}_q(\nu)\right)^2 \ \ .
\end{equation}
Why is the shift of the heavy-jet mass half of that of the thrust?
This is because the heavy-jet mass collects contributions from all
hadrons in the heavier hemisphere. If an ultra-soft gluon is emitted
in the lighter hemisphere, as long as $\rho Q > \mu_I$, it will not be
able to contribute to the heavy-jet mass. This is the simplest example
of interplay between perturbative and non-perturbative contributions,
which is the main topic of this article.

Let us comment now on the physical meaning of the variables $q$ and $u$ in eq.~\eqref{eq:Jtilde-soft}. The variable $q$ is the transverse momentum $k_t$ of the soft emission with respect to the emitting parton (quark or antiquark). For the thrust and heavy-jet mass, this is the same as the transverse momentum with respect to the thrust axis. The variable $u Q^2$ is the contribution of the ultra-soft gluon to the invariant mass of the jet. It can be used to define the rapidity $\eta$ with respect to the emitting quark or antiquark as
\begin{equation} \label{eq:eta-kt-u}
	\eta \equiv \ln\frac{k_t}{uQ} \ \ .
\end{equation} 
Using the variables $k_t$ and $\eta$, we can express $\Delta$ as follows
\begin{equation} \label{eq:Delta-def}
	\Delta \equiv 
	\frac{2 C_F}{\pi} \frac{\mu_I}{Q}\int_0^{\mu_I} \! \frac{dk_t}{\mu_I}\, \alpha^{\rm NP}_s(k_t)\int_0^\infty \!\! d\eta \, e^{-\eta} \ \ .
\end{equation}
Since the $\eta$ integral is exponentially damped in rapidity, the
rapidity integral can be extended up to infinity instead of the
appropriate kinematic limit. The exponential damping implies also that the
only ultra-soft emissions that give an appreciable contribution to
$\Delta$ are those at $\eta\sim 0$. Note also that, in the assumption that
the distribution of ultra-soft gluons is the same as that of
perturbative soft gluons except for a different coupling, the rapidity
integral (alternatively the $u$ integral in
eq.~\eqref{eq:Jtilde-soft}) can be performed.

In the following, we assume that leading non-perturbative corrections to event-shape distributions have the same origin as for the thrust and the heavy-jet mass, to summarise:
\begin{itemize}
\item they are due to ultra-soft emissions with $k_t\ll v^p Q$, where $v$ is the value of the considered event shape and $p$ is some positive power, typically one. This is in contrast to perturbative emissions where $k_t \sim v^p Q$.
\item they give rise to a shift of the corresponding perturbative distributions; the shift can be computed by considering a single ultra-soft emission (with accompanying virtual corrections). 
\item the dynamics of these ultra-soft emissions is not known; in the case that non-perturbative corrections are modelled as a shape function, an ultra-soft emission $k$ is produced with an unknown matrix element squared $\mathcal{M}^2_{\rm NP}(k)$; with the additional assumption that the distribution of ultra-soft emissions follows that of soft perturbative gluons, they are emitted uniformly in rapidity and azimuth, with an unknown dependence on transverse momentum, which can be embodied in a soft effective coupling.
\end{itemize} 

As a closing remark, we note that different assumptions on the
distribution of ultra-soft emissions result in various degrees of
observable dependence for the shift. For instance, if we assume that
non-perturbative corrections to the thrust and heavy-jet mass arise
from a shape-function modifying the invariant mass distribution of a
single jet, the shift in the distributions of these observables will
depend on the same non-perturbative quantity $\Delta$. As we will see
later, assuming that ultra-soft emissions are produced
uniformly in rapidity makes it possible to relate the shifts of many more observables.

\subsection{Kinematics of ultra-soft emissions}
\label{sec:kinematics}

In many cases, event shapes are defined using the thrust axis as a
reference axis to project momenta upon. However, the thrust axis is
generally not a direction that corresponds to a singularity of the QCD
matrix elements, which are better parameterised in terms of the
directions of the actual final-state particles. It is therefore useful
to relate different parameterisations of soft-gluon momenta.  To be
precise, we consider 2-jet events in $e^+ e^-$ annihilation. The
thrust axis defines two light-like vectors $(p_1,p_2)$, in terms of
which the final-state $q\bar{q}$ pair, $(\tilde{p}_1,\tilde{p}_2)$,
perturbative emissions, $k_i$, and the ultra-soft gluon, $k$, have the
following Sudakov decomposition
\begin{align}
	\nonumber
	k &= z^{(1)} p_1 + z^{(2)} p_2 + k_{\perp} \;\;, \quad 
	k_i = z_i^{(1)} p_1 + z_i^{(2)} p_2 + k_{\perp i} \,\ \ , \\
	\tilde{p}_1 &= z^{(1)}_p p_1 + z^{(2)}_p p_2 + p_{\perp 1} \,, \quad 
	\tilde{p}_2 = \bar{z}^{(1)}_p p_1 + \bar{z}^{(2)}_p p_2 + p_{\perp 2} \ \ ,
\end{align}
where $k_{\perp}$ etc.\ are space-like vectors with purely transverse components, i.e.\ $k_{\perp} = (0,\vec{k}_t,0)$, $p_{\perp 1} = (0,\vec{\tilde{p}}_{t,1},0)$ and so on. The rapidity of the emission with respect to the thrust axis, $\eta$, is given by
\begin{align}
\eta \equiv \frac{1}{2}\ln\frac{z^{(1)}}{z^{(2)}}
  \ \ . 
\end{align}
Let us consider the case in which emission $k$ is collinear to
$\tilde{p}_1$.  The collinear singularity is encoded in the following
transverse momentum
\begin{align}\label{eq:kappadef}
	\vec{\kappa}^{(1)} \equiv \vec{k}_t - \frac{z^{(1)}}{z_p^{(1)}} \, \vec{\tilde{p}}_{t,1} \ \ .
\end{align}
Here $\kappa^{(1)} \equiv |\vec{\kappa}^{(1)}|$ defines the transverse momentum with respect to $\tilde{p}_1$.
From the properties of the thrust axis, we have that the total transverse momentum vanishes in each hemisphere and thus we have
\begin{align}
	\vec{\tilde{p}}_{t,1}  = -\vec{k}_t + \vec{p}_{t,1} \ \ ,
\end{align}
where $\vec{p}_{t,1}$ denotes the recoil in the hemisphere due to all perturbative emissions. More precisely, for each leg $\ell$ we can define
\begin{equation}
  \label{eq:ptell}
  \quad \vec{p}_{t,\ell}  \equiv - \sum_{i \in \mathcal{H}_\ell} \vec{k}_{ti} \ \ .
\end{equation}
Using the fact that these are soft, we can approximate $z_p^{(1)}\simeq 1$ in eq.~\eqref{eq:kappadef} up to corrections suppressed by powers of $v$. Neglecting contributions quadratic in $\kappa^{(1)}$, we obtain
\begin{equation}
  \label{eq:kappa1-approx}
  \vec{\kappa}^{(1)} \simeq \vec{k}_t-z^{(1)} \vec{p}_{t,1}\ \ .
\end{equation}
Given $\kappa^{(1)}$, we can define the rapidity $\eta^{(1)}$ with
respect to the $\tilde{p}_1$ direction as follows
 \begin{align}
   \eta^{(1)} \equiv \ln\frac{z^{(1)}Q}{\kappa^{(1)}} \ \ .
\end{align}
Similarly, we define suitable kinematic variables for the case in which $k$ is collinear to $\tilde p_2$
\begin{equation}
  \label{eq:kappa2-eta2}
  \vec{\kappa}^{(2)} \equiv \vec{k}_t - \frac{z^{(2)}}{\bar z_p^{(2)}} \, \vec{\tilde{p}}_{t,2}\simeq \vec{k}_t-z^{(2)} \vec{p}_{t,2} \ \ ,\quad \eta^{(2)} \equiv \ln\frac{z^{(2)}Q}{\kappa^{(2)}} \ \ .
\end{equation}
The boundaries for $\eta^{(\ell)}$ ($\ell=1,2$) are a bit
involved. Nevertheless, identifying the region collinear to
$\tilde p_\ell$ with hemisphere $\mathcal{H}_\ell$, we have that the
limits on the Sudakov variable $k_t/Q < z^{(\ell)} < 1$ can be recast
in terms of $\eta^{(\ell)}$ and $\kappa^{(\ell)}$ as follows
\begin{align}
  \label{eq:etalimits}
	\ln\left( \left(1- \frac{p^2_{t,\ell}}{Q^2 } \sin^2\phi \right)^{1/2} +\frac{p_{t,\ell}}{Q } \cos\phi  \right) < \eta^{(\ell)} < \ln \frac{Q}{\kappa^{(\ell)}} \ \ ,
\end{align}
where $\phi$ is the azimuthal angle with respect to the direction of $\vec{p}_{t,\ell}$. In the presence of soft emissions only, $p_{t,\ell} \ll Q$ and thus  the lower limit of $\eta^{(\ell)}$ vanishes.

\subsection{General treatment of leading non-perturbative corrections}
\label{sec:NP-general}

We now consider a generic recursive infrared and collinear safe (rIRC)
safe~\cite{Banfi:2004yd} observable, denoted by $V(\{\tilde p\},k_1,\dots,k_n)$, in $e^+e^-$
annihilation. Here $\{\tilde p\}=\{\tilde p_1, \tilde p_2\}$ are the
momenta of a hard quark-antiquark pair and $k_1,\dots,k_n$ subsequent
emissions. Without any additional emissions
$\{\tilde p_1,\tilde p_2\}$ are the momenta of a back-to-back
quark-antiquark pair. In this case $V(\{\tilde p\})=0$, whereas in
general $V(\{\tilde p\},k_1,\dots,k_n)\ge 0$. 
We consider the region in which
$V(\{\tilde p\},k_1,\dots,k_n)=v\ll 1$, i.e.\ we are close to the Born
limit, but $v$ is not too small, so that the value of the observable
is determined by perturbative QCD emissions. To enforce this condition, we
restrict ourselves to the region $v\gg \Lambda_{\rm QCD}/Q$, with
$Q\sim\sqrt s$ the typical hard scale of the process. In this region, the
observable cumulant $\Sigma(v)$, the fraction of events such that
$V(\{\tilde p\},k_1,\dots,k_n)<v$, can be in largest part computed in
perturbative QCD, with small non-perturbative corrections, as
follows:
\begin{equation} \label{eq:Sigmatot}
	\Sigma(v) = 
	\Sigma_{\rm PT}(v)+\delta \Sigma_{\rm NP}(v) \ \ ,
\end{equation}
where
\begin{equation} 
	\Sigma_{\rm PT}(v) = 
	\int dZ[\{k_i\}] \, \Theta\big(v-V(\{\tilde p\},\{k_i\})\big) \ \ .
\end{equation}
Here $dZ[\{k_i\}]$ is an integration measure, associated with multiple
emissions and the corresponding virtual corrections, with the
normalisation
\begin{equation} \label{eq:dZ-normalisation}
	\int dZ[\{k_i\}] =1 \ \ .
\end{equation}
In particular, if we consider only soft and collinear emissions widely separated in angle, the relevant configurations giving NLL accuracy~\cite{Banfi:2004yd}, we have
\begin{equation} \label{eq:dZ}
	dZ[\{k_i\}] \simeq 
	dZ_{\rm sc}[\{k_i\}] \equiv e^{-\int[dk] M^2(k)} \sum_{n=0}^\infty \frac{1}{n!}  \prod_{i=1}^n[dk_i] M^2(k_i) \ \ ,
\end{equation}
where $[dk]$ is the Lorentz-invariant measure for a gluon of four-momentum $k=(\omega,\vec k)$ 
\begin{equation} \label{eq:dk}
	[dk] = \frac{d^3\vec k}{(2\pi)^3 \, 2\omega} \ \ .
\end{equation}
 In terms of these
variables, the matrix element squared for a soft emission off the $q\bar{q}$ dipole, $M^2(k)$, is given by 
\begin{equation} \label{eq:M2}
	M^2(k) = 
	\frac{16\pi \, C_F \, \alpha^{\rm CMW}_s(\kappa)}{\kappa^2} \ \ ,
\end{equation}
where $\alpha_s^{\rm CMW}$ is the strong coupling in the physical CMW
scheme \cite{Catani:1990rr}. Here $\kappa$ is the invariant transverse
momentum of emission $k$ with respect to the hard $q\bar{q}$ pair, defined as
\begin{equation}
  \label{eq:kappa}
  \kappa^2 \equiv 2\frac{(\tilde p_1 k)(\tilde p_2 k)}{(\tilde p_1 \tilde p_2)}\ \ .
\end{equation}
For emissions collinear to leg $\ell$, $\kappa\simeq \kappa^{(\ell)}$ introduced in section~\ref{sec:kinematics}.  Last, for such configurations, for any
function $G(\{\tilde p\},k_1,\dots,k_n)$, we define
\begin{equation} \label{eq:dZsc-G}
	\int dZ_{\rm sc}[\{k_i\}] \, G(\{\tilde p\},\{k_i\})\equiv 
	e^{-\int[dk] M^2(k)} \sum_{n=0}^\infty \frac{1}{n!}  \prod_{i=1}^n[dk_i] \, M^2(k_i) \, G(\{\tilde p\},k_1,\dots,k_n) \ \ .
\end{equation}
As for the cases of the thrust and heavy-jet mass, we model leading NP corrections  as the
contribution of an ultra-soft emission $k$, whose transverse momentum is much smaller than the typical transverse momentum of soft emissions contributing to $\Sigma_{\rm PT}(v)$. The latter is of the order $v^p Q$, with $p$ some positive power. The ultra-soft emission is produced with an unknown
matrix element squared $\mathcal{M}^2_{\rm NP}(k)$ and gives the following correction to $\Sigma(v)$
\begin{equation}
  \label{eq:SigmaNP}
	\delta\Sigma_{\rm NP}(v) = 
	\int [dk] \, \mathcal{M}^2_{\rm NP}(k) \int dZ[\{k_i\}] \, \Big[\Theta\big(v-V(\{\tilde p\},k,\{k_i\})\big)-\Theta\big(v-V(\{\tilde p\},\{k_i\})\big)\Big] \ \ ,
\end{equation}
where virtual corrections are implemented via unitarity, i.e.\ by
imposing that this contribution is zero if there are no constraints on
any emissions.\footnote{This approximation corresponds to the fact that power
  corrections to the total cross section for $e^+e^-$ annihilation
  into hadrons occur with higher powers of $1/Q$  \cite{Dokshitzer:1995qm}.}

The difference between the observable with an additional ultra-soft
emission and without it is in general much smaller than $v$ in the
region we are interested in. Therefore, we can approximate the
difference between the two step functions in eq.~\eqref{eq:SigmaNP} as
follows
\begin{equation} \label{eq:diffstep}
	\Theta\big(v-V(\{\tilde p\},k,\{k_i\})\big) - \Theta\big(v-V(\{\tilde p\},\{k_i\})\big) \simeq 
	-\delta V_{\rm NP}(\{\tilde p\},k,\{k_i\}) \, \delta\big(v-V(\{\tilde p\},\{k_i\})\big) \ \ ,
\end{equation}
where we have introduced the change in the observable due to a NP
emission $k$, as follows
\begin{equation} \label{eq:dV-NP} 
	\delta V_{\rm NP}(\{\tilde p\},k,\{k_i\}) \equiv 
	V(\{\tilde p\},k,\{k_i\})-V(\{\tilde p\},\{k_i\}) \ \ .
\end{equation}
Substituting the approximation in eq.~(\ref{eq:diffstep}) into eq.~(\ref{eq:SigmaNP}), we obtain
\begin{equation} \label{eq:dSigma-approx}
	\begin{split}
		\delta\Sigma_{\rm NP}(v) & \simeq 
		-\int [dk] \mathcal{M}^2_{\rm NP}(k) \int dZ[\{k_i\}] \, \delta V_{\rm NP}(\{\tilde p\},k,\{k_i\}) \, \delta\big(v-V(\{\tilde p\},\{k_i\})\big) \\ 
		&=- \langle\delta V_{\rm NP}\rangle \int dZ[\{k_i\}] \, \delta\big(v-V(\{\tilde p\},\{k_i\})\big) \ \ ,
  \end{split}
\end{equation}
where we have introduced the average of $\delta V_{\rm NP}$ over all
perturbative configurations, defined as 
\begin{equation}
	\langle\delta V_{\rm NP}\rangle \equiv 
	\frac{\int [dk] \mathcal{M}^2_{\rm NP}(k) \int dZ[\{k_i\}] \, \delta V_{\rm NP}(\{\tilde p\},k,\{k_i\}) \, \delta\big(v-V(\{\tilde p\},\{k_i\})\big) } {\int dZ[\{k_i\}] \, \delta\big(v-V(\{\tilde p\},\{k_i\})\big)} \ \ .
\end{equation}
Observing also that
\begin{equation} \label{eq:dSigmaPTdv}
	\int dZ[\{k_i\}] \, \delta\big(v-V(\{\tilde p\},\{k_i\})\big) = 
	\frac{d\Sigma_{\rm PT}}{dv} \ \ ,
\end{equation}
we can write
\begin{equation} 
	\Sigma(v) \simeq 
	\Sigma_{\rm PT}(v)-\langle\delta V_{\rm NP}\rangle \frac{d\Sigma_{\rm PT}}{dv}\simeq \Sigma_{\rm PT}(v- \langle\delta V_{\rm NP}\rangle) \ \ .
\end{equation}
This means that, whenever $v \gg \langle\delta V_{\rm NP}\rangle$, NP corrections
amount to a shift of the perturbative distribution by an amount given
by the average $\langle\delta V_{\rm NP}\rangle$.

In principle, $\langle\delta V_{\rm NP}\rangle$ is different for each
observable and might depend non-trivially on unknown NP
dynamics. However, here we make the assumption that the production rate of ultra-soft emissions is driven by PT
dynamics, i.e.\ it is uniform in rapidity, defined with respect to final-state partons, and azimuth, as follows
\begin{equation} \label{eq:M2NP}
	[dk] \mathcal{M}^2_{\rm NP}(k) = 
	\sum_\ell \frac{d\kappa}{\kappa}  M^2_{\rm NP}(\kappa) \, d\eta^{(\ell)} \, \frac{d\phi}{2\pi} \ \ ,
\end{equation}
where $\eta^{(1)}$ ($\eta^{(2)}$) have been defined in
Sect.~\ref{sec:kinematics}.  In general, the ultra-soft emission $k$
is accompanied by soft and/or collinear emissions. In this paper, we
assume these accompanying emissions are soft, collinear and widely
separated in angle. This is appropriate to achieve NLL accuracy for
$\Sigma_{\rm PT}(v)$. We also restrict ourselves to event-shape
variables (i.e.\ we exclude jet-resolution parameters) and
furthermore only to those event shapes for which the NP contribution to the
observable is linear in $\kappa$, as follows 
\begin{equation} \label{eq:deltaV}
	\delta V_{\rm NP}(\{\tilde p\},k,\{k_i\}) = 
	\frac{\kappa}{Q} \, h_V\big(\eta^{(\ell)},\phi,\{ \tilde{p}\},\{k_i\}\big) \ \ .
\end{equation}
This implies 
\begin{equation} \label{eq:cV}
	\langle\delta V_{\rm NP}\rangle = 
	\frac{\langle \kappa\rangle_{\rm NP}}{Q} \, \langle h_V\rangle \ \ ,
\end{equation}
where 
\begin{equation} 
	\langle \kappa\rangle_{\rm NP} = 
	\int d\kappa \, M^2_{\rm NP}(\kappa) \ \ ,
\end{equation}
and
\begin{equation} \label{eq:cV-computed}
	\cNP{V} \equiv 
	\frac{\sum_\ell \int d\eta^{(\ell)} \frac{d\phi}{2\pi} \int dZ_{\rm sc}[\{k_i\}] \, h_V(\eta^{(\ell)},\phi,\{\tilde{p}\},\{k_i\})\, \delta\Big(1-\frac{\Vsc{\{k_i\}}}{v}\Big)}{\int dZ_{\rm sc}[\{k_i\}] \, \delta\Big(1-\frac{\Vsc{\{k_i\}}}{v}\Big)} \ \ .
\end{equation}
Furthermore, we restrict ourselves to observables where the rapidity
integral in eq.~\eqref{eq:cV-computed} is convergent when the upper
bound is pushed to infinity. For such observables, setting the
rapidity boundary to the actual kinematic limit $\ln (Q/\kappa)$ would
give a contribution that has a further suppression in $\kappa$, i.e.\
a sub-leading power correction.

In eq.~\eqref{eq:cV-computed}, $\Vsc{k_1,\dots,k_n}$ is the value that $V(\{\tilde p\},k_1,\dots,k_n)$ assumes when all emissions are soft and collinear. Formally, in the presence of only soft and collinear emissions $k_1,\cdots,k_n$, 
\begin{equation} \label{eq:Vsc}
	\frac{ \Vsc{k_1,\dots,k_n}}{v} \equiv 
	\lim_{v\to 0} \frac{V(\{\tilde p\},k_1,\dots,k_n)}{v} \ \ .
\end{equation}
An important remark is in order here. Since $V(\{\tilde p\},k_1,\dots,k_n)$ is a rIRC safe observable, all soft and collinear emissions $k_1,\dots,k_n$ have momenta in the region $V(\{\tilde p\},k_i) \sim v$. This is in contrast with the contribution of the NP emission, $\delta V_{\rm NP} \sim \kappa \ll vQ$. 

In the dispersive approach of~\cite{Dokshitzer:1995zt} the relation between $\langle \kappa\rangle_{\rm NP}$ and the phenomenological parameter $\alpha_0$ is given by 
\begin{align}
		\langle \kappa\rangle_{\rm NP} = \frac{4C_F}{\pi^2} \mu_I \left(\alpha_0(\mu_I)- \alpha_s - 2 \beta_0\, \alpha_s^2\left(1+\ln\frac{Q}{\mu_I} + \frac{K}{4\pi\beta_0}\right)\right) \ \ ,
\end{align} 
where $\beta_0  = (11 C_A - 4 T_R n_f)/12 \pi$.

Before moving forward, we comment on the relevance of the result in eq.~\eqref{eq:cV}. It means that, for all event shapes for which $\delta V_{\rm NP}$ has the property in eq.~\eqref{eq:deltaV}, the NP shift to the corresponding distributions is given by the product of a genuine NP quantity $\langle \kappa\rangle_{\rm NP}$,  the average transverse momentum of ultra-soft emissions, and a \emph{calculable} coefficient $\cNP{V}$. 
The aim of this paper is precisely to devise a semi-numerical procedure to
compute $\cNP{V}$ for a generic rIRC safe event shape whose leading NP
corrections are given as the shift in eq.~\eqref{eq:cV}. 
The general procedure is the same as used to compute NLL and NNLL corrections to event-shape distributions. First, we rewrite the measure  $dZ_{\rm sc}[\{k_i\}] $ as
\begin{equation} \label{eq:dZsc-split}
	dZ_{\rm sc}[\{k_i\}] = 
	e^{-R(v)} e^{-\int^v [dk] M^2(k)} \sum_{n=0}^\infty \frac{1}{n!} \prod_{i=1}^n[dk_i] M^2(k_i) \ \ ,
\end{equation}
where the ``radiator'' $R(v)$ is given by
\begin{equation} \label{eq:radiator}
	R(v) = 
	\int_v [dk] M^2(k) \equiv 
	\int [dk] M^2(k) \, \Theta(\Vsc{k}-v) \ \ .
\end{equation}
For {\em perturbative} emissions which are soft and collinear,
it is convenient to express any soft momentum $k$ in terms of the leg
$\ell$ it is collinear to, the variable $\zeta = \Vsc{k}/v$ and its
azimuth $\phi$ with respect to a suitably chosen plane.  For event
shapes only, which is what we consider here, we can freely integrate
over the rapidity fraction of emission $k$ with respect to the total
available rapidity for fixed $\zeta$ and $\phi$
\cite{Banfi:2004yd}. This gives
\begin{equation} \label{eq:M2-simp}
	[dk] M^2(k) = 
	\sum_\ell \frac{d\zeta}{\zeta} \, R'_{\ell} (\zeta v) \, \frac{d\phi}{2\pi} \ \ ,
\end{equation}
where $R'_{\ell} (\zeta v)$ is the Jacobian arising from the integration over the rapidity fraction. Therefore, the soft-collinear phase space measure reads
\begin{equation} \label{eq:dZsc-final}
	dZ_{\rm sc}[\{k_i\}] = 
	e^{-R(v)} \, d\mathcal{Z}[\{R'_{\ell_i},k_i\}] \ \ ,
\end{equation}
where we have introduced a modified soft-collinear measure normalised in such a way that
\begin{equation}
	\int d\mathcal{Z}[\{R'_{\ell_i},k_i\}] \, \Theta\Big(1- \max_i\{\zeta_i\}\Big) = 1 \ \ .
\end{equation}
Up to NLL accuracy $R'_{\ell} (\zeta v) \simeq R'_{\ell} (v)$, which gives
\begin{equation} \label{eq:dZ-rescaled}
	d\mathcal{Z}[\{R'_{\ell_i},k_i\} ] \simeq 
	\epsilon^{R'} \sum_{n=0}^\infty \frac{1}{n!}  \prod_{i=1}^n \sum_{\ell_i} R'_{\ell_i} \frac{d\zeta_i}{\zeta_i}\frac{d\phi_i}{2\pi} \, \Theta(\zeta_i-\epsilon) \ \ , \qquad R'=\sum_\ell R'_\ell \ \ .
\end{equation}
The quantity $\epsilon$ in eq.~\eqref{eq:dZ-rescaled} acts as a cutoff for the integration over the variables $\zeta_i$.
Considering again the calculation of $\cNP{V}$, the factor $\exp[-R(v)]$ cancels out between numerator and denominator, and we obtain
\begin{equation} \label{eq:np-shift-NLL}
	\cNP{V} = 
	\frac{\sum_\ell \int d\eta^{(\ell)} \frac{d\phi}{2\pi} \int d\mathcal{Z}[\{R'_{\ell_i},k_i\}] \, h_V(\eta^{(\ell)},\phi,\{\tilde{p}\},\{k_i\}) \, \delta\Big(1-\frac{\Vsc{\{k_i\}}}{v}\Big)}{\int d\mathcal{Z}[\{R'_{\ell_i},k_i\}] \,  \delta\Big(1-\frac{\Vsc{\{k_i\}}}{v}\Big)} \ \ .
\end{equation} 
We know that, at NLL accuracy and keeping only soft contributions, we have\footnote{At NLL accuracy, $\Sigma_{\rm PT}(v)$ receives a factorised contribution due to hard-collinear virtual corrections. This would cancel between the numerator and denominator in eq.~\eqref{eq:np-shift-NLL}, so we can neglect it at this stage. It will become important for observables for which eq.~\eqref{eq:np-shift-NLL} is divergent for $R'\to 0$, and hence needs to be improved with the inclusion of sub-leading terms, see section~\ref{sec:subtraction}.}
\begin{equation}
	\Sigma_{\rm PT}(v) = e^{-R(v)} \, \mathcal{F}(R') \ \ ,
\end{equation}
where
\begin{equation}
	\mathcal{F}(R') = 
	\int d\mathcal{Z}[\{R'_{\ell_i},k_i\}] \, \Theta\bigg(1-\frac{\Vsc{\{k_i\}}}{v}\bigg) \ \ .
\end{equation}
One can also show (see appendix~\ref{sec:monte-carlo-determ}) that
\begin{equation}
	\int d\mathcal{Z}[\{R'_{\ell_i},k_i\}] \,  \delta\bigg(1-\frac{\Vsc{\{k_i\}}}{v}\bigg) = 
	R' \mathcal{F}(R') \ \ .
\end{equation}
We can then recast the denominator of eq.~\eqref{eq:np-shift-NLL} in terms of $\mathcal{F}(R')$ and obtain
\begin{equation} \label{eq:np-shift-NLL-final}
	\cNP{V} \mathcal{F}(R') = 
	\frac{1}{R'} \sum_\ell \int d\eta^{(\ell)} \, \frac{d\phi}{2\pi} \int d\mathcal{Z}[\{R'_{\ell_i},k_i\}] \, h_V(\eta^{(\ell)},\phi,\{\tilde{p}\},\{k_i\}) \, \delta\bigg(1-\frac{\Vsc{\{k_i\}}}{v}\bigg) \ \ .
\end{equation}
This expression is suitable for both analytic calculations and numerical determinations according to the method explained in appendix~\ref{sec:monte-carlo-determ}. The idea, developed originally in \cite{Banfi:2001bz}, is to label $k_1$ the emission such that $\zeta_1$ is the largest of the $\zeta_i$, neglect all emissions with $\zeta_i<\epsilon \zeta_1$ and  use the constraint on the observable to perform the integration over $\zeta_1$. This gives
\begin{multline} \label{eq:np-shift-NLL-numerix}
	\cNP{V} \mathcal{F}(R') =  \sum_\ell
	\int d\eta^{(\ell)} \, \frac{d\phi}{2\pi} \sum_{\ell_1=1,2}  \frac{R'_{\ell_1}}{R'}\int_0^{2\pi}\!\frac{d\phi_1}{2\pi}\left( \epsilon^{R'}\sum_{n=0}^\infty \frac{1}{n!}  \prod_{i=2}^{n+1}\sum_{\ell_i=1,2} R'_{\ell_i}\int^{1}_{\epsilon}\frac{d\zeta_i}{\zeta_i}\int_0^{2\pi}\frac{d\phi_i}{2\pi}\right) \times \\ 
	\times \left. \left(\frac{\Vsc{k_1,k_2,\dots,k_{n+1}}}{v}\right)^{-R'} h_V(\eta^{(\ell)},\phi,\{\tilde{p}\},k_1,k_2,\dots,k_{n+1})\right|_{\Vsc{k_1}=v} \ \ ,
\end{multline}
where now $\Vsc{k_1} = v$.

\subsection{The Milan factor} 

Due to the non-inclusive nature of event shapes, it was initially
suspected~\cite{Nason:1995np} that the leading hadronisation
correction is {\em not} universal, which cast doubts about the utility
of the dispersive approach in predicting the NP parameter
$\alpha_0$. Using the thrust variable,~\cite{Dokshitzer:1997iz}
considered the hadronisation corrections due to the decay of the NP
gluon to a gluon (or $q\bar{q}$) pair. The non-inclusiveness of the
thrust variable turned out to be encoded in a calculable
multiplicative constant, called the {\em Milan factor}, multiplying
the leading moment of the effective coupling.

In this section we describe, within our approach, how the Milan factor
arises. The Milan factor is historically given by the sum of two contributions. One is the
`inclusive' contribution, where one upgrades the NP corrections due to
one emission by dressing the ultra-soft gluon with its inclusive splittings, as well as virtual
corrections. This contribution, computed in~\cite{Dokshitzer:1997iz,Dokshitzer:1998pt}, is constructed in such a way as to give a
multiplicative factor times the contribution of a single ultra-soft
gluon. The remaining contribution is the so-called `non-inclusive' correction, arising from the difference of the equivalent of
eq.~\eqref{eq:diffstep} in the presence of two NP emissions
$(k_a,k_b)$, and of their inclusive contribution. 

The introduction of an inclusive contribution is just a convenience and, as shown in~\cite{Dokshitzer:1997iz,Dokshitzer:1998pt}, it is possible to obtain the Milan factor for the thrust and $C$-parameter by combining the contribution of single and double soft emission and the corresponding virtual corrections. We are not interested in computing the Milan factor here, but want to show that for certain observables the contribution of two ultra-soft emissions gives rise to the same $\cNP{V}$ as one ultra-soft emission. As the contribution of two ultra-soft emissions has a collinear divergence, we employ an inclusive subtraction as a technical convenience, in a similar way as subtraction terms are employed in fixed-order calculations. The so-regularised contribution of two ultra-soft emissions to the observable constraint reads
\begin{multline}
	\Theta\big(v-V(\{\tilde p\},k_a,k_b,\{k_i\})\big) - \Theta\big(v-V(\{\tilde p\},k_a+k_b,\{k_i\})\big) \\ \simeq 
	- \delta V^{\rm (n.i.)}_{\rm NP}(\{\tilde p\},k_a,k_b,\{k_i\}) \, \delta\big(v-V(\{\tilde p\},\{k_i\})\big) \ \ ,
\end{multline}
where 
\begin{equation}
  \label{eq:dV-NP2}
  \delta V^{\rm (n.i.)}_{\rm NP}(\{\tilde p\},k_a,k_b,\{k_i\}) \equiv 
  V(\{\tilde p\},k_a,k_b,\{k_i\})- V(\{\tilde p\},k_a+k_b,\{k_i\}) \ \ .
\end{equation}
We can now express the non-inclusive correction in our approach as follows
\begin{equation}
	\langle\delta V_{\rm NP}\rangle^{(\mathrm{n.i.})} \equiv 
	\frac{\int [dk_a] [dk_b] \mathcal{M}_{\rm NP}^2(k_a,k_b) \int dZ_{\rm sc}[\{k_i\}] \, \delta V^{\rm (n.i.)}_{\rm NP}(\{\tilde p\},k_a,k_b,\{k_i\}) \, \delta(v-V(\{\tilde p\},\{k_i\}))} {\int dZ_{\rm sc}[\{k_i\}] \, \delta(v-V(\{\tilde p\},\{k_i\}))}\ \ .
\end{equation}
The crucial assumption now is that the ultra-soft matrix-element
squared $ \mathcal{M}_{\rm NP}^2(k_a,k_b)$ has the same form as the
perturbative one, where the {\em physical}, i.e.\ infrared finite,
coupling is assumed to be a function of the invariant mass
$m^2 = (k_a+k_b)^2$. In particular, $ \mathcal{M}_{\rm NP}^2(k_a,k_b)$
is independent of the rapidity of the parent ultra-soft gluon,
$k = k_a +k_b$, while it depends on the azimuthal angle difference
$\phi_a-\phi_b$ solely through the dependence on the invariant
mass. The two-body phase space can be cast in terms of the following
standard variables \cite{Dokshitzer:1997iz}
	\begin{align}
	 [dk_a] [dk_b] = \frac{1}{2!} \sum_\ell \frac{1}{64 \pi^4} d\eta^{(\ell)} d z \frac{d^2 \vec{u}_a}{2\pi} \frac{d^2 \vec{u}_b}{2\pi} q^2 \, dm^2\, \delta\big(1 - u_a^2 - u_b^2 + 2 u_a u_b \cos(\phi_a-\phi_b) \big) \ ,
\end{align}
where
\begin{align}
	q^2 \equiv \frac{m^2}{z(1-z)}, \quad \vec{u}_a \equiv \frac{\vec
		\kappa_a}{z q} ,\quad \vec{u}_b \equiv \frac{\vec
		\kappa_b}{(1-z) q} \ .
\end{align}
The Dirac delta function is used above to keep the invariant mass fixed. One can easily use the delta function to integrate over the azimuthal angle difference $\phi_a - \phi_b$, and we find
\begin{align}
	\nonumber
	[dk_a] [dk_b] &= \frac{1}{2!}  \sum_\ell
	\frac{1}{64 \pi^4} d\eta^{(\ell)} d z\, \frac{d\phi}{2\pi} \, d m^2 \,q^2 \frac{u_a d u_a d u_b u_b}{\pi \sqrt{J}}\, \Theta(J) \ \ , \\
	J &\equiv \left((u_a+u_b)^2-1\right) \left(1 - (u_a-u_b)^2\right)\ ,
\end{align}
where $\phi$ is a residual azimuthal angle, i.e.\ $\phi_a$, $\phi_b$ or the azimuth of $k_a+k_b$, which the squared matrix-element does not depend upon. 
We do not need the explicit form of $ \mathcal{M}_{\rm NP}^2(k_a,k_b)$, but only its generic dependence on the phase space variables
\begin{align}\label{eq:ME}
	\mathcal{M}^2_{\text{NP}}(k_a,k_b) = \left(4 \pi \alpha_s(m^2)\right)^2 \frac{\mathcal{M}^2(u_a^2,u_b^2,z)}{ m^2 \left(z u_a^2 + (1-z)u_b^2\right) q^2} \ \ .
\end{align} 
 In terms of these variables we define the `inclusive' observable
\begin{equation}
\label{eq:dV-inclusive}
 V(\{\tilde p\},k_a+k_b,\{k_i\})\equiv \frac{\sqrt{\kappa^2+m^2}}{Q}h_V(\eta^{(\ell)},\phi,\{\tilde{p}\},\{k_i\}) + V(\{\tilde p\},\{k_i\}) \ \ ,
\end{equation}
where $\vec{\kappa} \equiv \vec{\kappa}_a + \vec{\kappa}_b$. 

In the
particular case in which $V(\{\tilde p\},k_a,k_b,\{k_i\})$ is additive, i.e.\ 
\begin{equation}
  \label{eq:dV-additive}
  V(\{\tilde p\},k_a,k_b,\{k_i\})=\frac{\kappa_a}{Q}h_V(\eta^{(\ell)}_a,\phi_a,\{\tilde{p}\},\{k_i\})+\frac{\kappa_b}{Q}h_V(\eta^{(\ell)}_b,\phi_b,\{\tilde{p}\},\{k_i\}) + V(\{\tilde p\},\{k_i\}) \ \ ,
\end{equation}
 one can integrate freely over $\eta^{(\ell)}$ and $\phi$. As $V(\{\tilde p\},k_a,k_b,\{k_i\})$ is a function of $\eta^{(\ell)}_a$ and $\eta^{(\ell)}_b$, then we must express the observable function $h_V$ in terms of $\eta^{(\ell)}$ which requires the following relations
\begin{align}
	\eta^{(\ell)}_a = \eta^{(\ell)} - \ln \frac{u_a}{ \sqrt{z u_a^2 + (1-z) u_b^2}} \ \ , \quad 	\eta^{(\ell)}_b = \eta^{(\ell)} - \ln \frac{u_b}{\sqrt{z u_a^2 + (1-z) u_b^2}} \ \ .
\end{align}
We notice that the relation between $\eta_a^{(\ell)}, \eta_b^{(\ell)}$
and $\eta^{(\ell)}$ corresponds to a boost that depends on
$u_a,u_b$ and $z$. Last, we need to discuss the integration limits on
$\eta^{(\ell)}$. As for a single massless gluon, the limits on the
individual rapidities are as follows
$0 < \eta_{a,b}^{(\ell)} < \infty $, therefore, we integrate the
functions $h_V(\eta^{(\ell)}_a,\phi_a=\phi,\{k_i\})$ and
$h_V(\eta^{(\ell)}_b,\phi_b=\phi,\{k_i\})$ over $\eta^{(\ell)}$ and
$\phi$
\begin{align}
	\int_{\ln \frac{u_a}{ \sqrt{z u_a^2 + (1-z) u_b^2}} }^\infty \!\!\! d\eta^{(\ell)}  \int \frac{d\phi}{2 \pi}\, h_V(\eta^{(\ell)}_a,\phi,\{k_i\}) = 	\int_0^\infty d\eta^{(\ell)}  \int \frac{d\phi}{2\pi}\, h_V(\eta^{(\ell)},\phi,\{k_i\}) \ ,
\end{align}
and thus obtain 
\begin{multline}
  \langle\delta V_{\rm NP}\rangle^{(\mathrm{n.i.})}=
  \frac{\sum_\ell \int d\eta^{(\ell)} \frac{d\phi}{2\pi} \int dZ_{\rm sc}[\{k_i\}] \, h_V(\eta^{(\ell)},\phi,\{\tilde{p}\},\{k_i\})\, \delta\Big(1-\frac{\Vsc{\{k_i\}}}{v}\Big)}{\int dZ_{\rm sc}[\{k_i\}] \, \delta\Big(1-\frac{\Vsc{\{k_i\}}}{v}\Big)} \times \\ \times
 \frac{1}{Q}\int [d(k_a,k_b)]\left(\kappa_a+\kappa_b-\sqrt{\kappa^2+m^2}\right)\mathcal{M}_{\rm NP}^2(k_a,k_b)  
\ \ .
\end{multline}
The `residual' two-body phase space is given by
\begin{align}
	[d(k_a,k_b)]   = \frac{1}{2!} \frac{1}{64 \pi^4} dz \, dm^2 \, q^2  \frac{u_a d u_a d u_b u_b}{\pi \sqrt{J}}\, \Theta(J) \ \ .
\end{align}
It has been shown for the first time
in~\cite{Dokshitzer:1997iz} that, in the dispersive approach of~\cite{Dokshitzer:1995qm}, we have
\begin{equation}
  \label{eq:Milan-ni}
  \int [d(k_a,k_b)]\left(\kappa_a+\kappa_b-\sqrt{\kappa^2+m^2}\right)\mathcal{M}_{\rm NP}^2(k_a,k_b) = \langle\kappa\rangle_{\rm NP} \, \mathcal{M^{\rm (n.i.)}}\ \ ,
\end{equation}
with $\mathcal{M}^{\rm (n.i.)}$ the non-inclusive part of the Milan
factor whose correct numerical form is given
in~\cite{Dokshitzer:1998pt,Dasgupta:1999mb}, and was later determined
analytically in~\cite{Smye:2001gq}. The analysis we presented for the
Milan factor confirms that, in our approach, any observable satisfying
eq.~\eqref{eq:dV-additive} exhibits a non-inclusive contribution to
the leading hadronisation correction proportional to the contribution
of a single ultra-soft gluon multiplied by the non-inclusive
contribution to the Milan factor, {\em viz.}
\begin{equation}
  \label{eq:dV-ni}
  \langle\delta V_{\rm NP}\rangle^{\mathrm{(n.i.)}} = \frac{\langle\kappa\rangle_{\rm NP} }{Q}\cNP{V} \, \mathcal{M^{\rm (n.i.)}} \ .
\end{equation}
Therefore, considering also the inclusive contribution to the Milan
factor, the total shift in our approach is given by
\begin{equation}
	\label{eq:dV-ni}
	\langle\delta V_{\rm NP}\rangle = \frac{\langle\kappa\rangle_{\rm NP} }{Q}\cNP{V} \, \mathcal{M} \ ,
      \end{equation}
where $\mathcal{M}$ is the full Milan factor.

\section{NP shifts for ``popular'' $e^+e^-$ event shapes}
\label{sec:observables}

The function $h_V(\eta^{(\ell)},\phi,\{ \tilde{p}\},\{k_i\})$ needs to be computed separately for each observable. In this section we perform the calculation for rIRC safe event shapes that have been routinely measured at LEP, namely thrust, $C$-parameter, heavy-jet mass, jet broadenings and thrust major. Among those, only the thrust major has not been computed so far. This is due to the fact that this observable does not admit an analytical treatment in the two-jet limit. Therefore, only a semi-numerical approach such as the one presented here is viable to compute the NP shift to its distribution. From the discussion below it will be evident that event shapes can be grouped into classes showing a similar structure for $h_V(\eta^{(\ell)},\phi,\{ \tilde{p}\},\{k_i\})$. However, it is not clear to us whether a general classification can be formulated and hence we leave this question for future work. In the following we derive the expressions for the NP shifts for the above observables, as provided by our method.

\subsection{Recovery of known results}

We first show that with our method we can compute the shift to all observables that have been studied analytically so far and obtain perfect agreement.

\paragraph{Thrust and $\mathbf{C}$-parameter:}

The function $h_V(\eta^{(\ell)},\phi,\{ \tilde{p}\},\{k_i\})$ for one minus the thrust, $1\!-\!T$, and $C$-parameter does not depend explicitly on the momenta of PT soft-collinear emissions
\begin{equation} \label{eq:hV-additive}
	h_V(\eta^{(\ell)},\phi,\{ \tilde{p}\},\{k_i\}) = 
	h_V(\eta^{(\ell)},\phi) \ \ .
\end{equation}
In this case, one has 
\begin{equation} \label{eq:ave-hV-additive}
	\cNP{V} = \sum_\ell
	\int d\eta^{(\ell)} \, \frac{d\phi}{2\pi} \, h_V(\eta^{(\ell)},\phi) \ \ .
\end{equation}
 The lower limit on $\eta^{(\ell)}$ can be set to zero because $p_{t\ell} \sim \sqrt{v} Q$ for such observables. Also,
 \begin{equation} \label{eq:T-C-hV}
 	h_{1\!-\!T}(\eta^{(\ell)},\phi) = e^{-\eta^{(\ell)}} \ \ , \qquad h_C(\eta^{(\ell)},\phi) = \frac{3}{\cosh\eta^{(\ell)}} \ \ ,
 \end{equation}
  which implies that the upper limit on $\eta^{(\ell)}$ can be set to infinity due to exponential damping. This gives
 \begin{equation} \label{eq:T-C-ave-hV}
 	\cNP{1\!-\!T} = 2 \ \ , \qquad \cNP{C} = 3\pi \ \ .
 \end{equation}

\paragraph{Heavy-jet mass:} In the case of the heavy-jet mass there is an interplay between NP and PT radiation since, in the presence of perturbative emissions, a non-zero NP correction to the heavy-jet mass arises only when the ultra-soft emission is in the heavier hemisphere. This is due to the fact that the contribution to the invariant mass of either hemisphere due to a NP emission is of order $\kappa$, which is much less than the contribution of all PT emissions, which is of order $\rho_{\ell}$. As a consequence, in the presence of multiple soft and collinear PT emissions, a NP emission can never determine which hemisphere is heavier. Therefore
\begin{equation} \label{eq:f-mh}
	h_{\rho_H}(\eta^{(1)},\phi,\{\tilde{p}\},k_1,\dots,k_n) = 
	\Theta(\rho_1-\rho_2) \, e^{-\eta^{(1)}}  , \:\:\: h_{\rho_H}(\eta^{(2)},\phi,\{\tilde{p}\},k_1,\dots,k_n) = 
	\Theta(\rho_2-\rho_1) \, e^{-\eta^{(2)}} \ \ ,
\end{equation}
where $\rho_1$ and $\rho_2$ are the invariant masses of the two hemispheres. This gives 
\begin{align} \label{eq:ave-h-mh}
	\nonumber
	\cNP{\rho_H} &= 
	\frac{1}{R' \mathcal{F}_{\rho_H}(R')} \int_0^\infty \, d\eta^{(1)} \, e^{-\eta^{(1)}} \int d\mathcal{Z}[\{R'_{\ell_i},k_i\}] \, \times \\
	&\quad \; \times \left[\Theta\Big(\rho_1(\{\tilde{p}\},\{k_i\}) - \rho_2(\{\tilde{p}\},\{k_i\})\Big) \, \delta\bigg(1-\frac{\rho_1(\{\tilde{p}\},\{k_i\})}{\rho_H}\bigg)\right] + 1 \leftrightarrow 2  \ \ ,
\end{align}
which implies 
\begin{equation}
	\cNP{\rho_H} = 1 \ \ .
\end{equation}
   
\paragraph{Total and wide-jet broadening:}  
The interplay between PT and NP effects is more complicated for the total and wide-jet broadening than it is for the heavy-jet mass. To better understand the issues involved, we consider an ultra-soft emission $k$ in the hemisphere containing leg $\ell$. The broadening of the corresponding hemisphere (which we denote by $B_\ell$) is then given by
\begin{equation} \label{eq:Bell}
	2 B_{\ell}(\{\tilde p\},k,\{k_i\})Q = 
	k_t + \sum_{i\in \mathcal{H}_\ell} k_{ti}+ \tilde p_{t,\ell} \ \ .
\end{equation}
From the definition of the thrust axis, we have
\begin{equation} \label{eq:ptell-T}
  \tilde p_{t,\ell} = \left|\sum
    _{i\in \mathcal{H}_\ell} \vec k_{ti}+\vec k_t\right| = \left|\vec
    p_{t,\ell}-\vec k_t\right| \ \ ,
\end{equation}
with $\vec p_{t,\ell}$ defined in eq.~\eqref{eq:ptell}.  The
hadronisation contribution to $B_\ell$ is then given by
\begin{equation} \label{eq:dBell}
	2 \delta B_{\ell}(\{\tilde p\},k,\{k_i\})Q 
	= k_t + \left|\vec p_{t,\ell} - \vec k_t\right| - p_{t,\ell} \ \ .
\end{equation}
We remark that $k_t$ is the transverse momentum of the NP emission with respect to the thrust axis and not with respect to its emitter. In fact, $k_t$ is different from $\kappa$ and depends also on $\eta^{(\ell)}$, $\phi$ and the recoiled momentum $p_{t,\ell}$. More precisely, if we express all quantities in eq.~\eqref{eq:dBell} in terms of
$\kappa$, $\eta^{(\ell)}$ and $\phi$, we find 
\begin{equation}
  \label{eq:kt-pt-ktT}
  \begin{split}
    k_t& = \left| \vec \kappa+z^{(\ell)} \vec p_t\right| = \kappa\, \sqrt{1+2 e^{\eta^{(\ell)}} \frac{p_{t,\ell}}{Q}\cos\phi+\left(e^{\eta^{(\ell)}} \frac{p_{t,\ell}}{Q}\right)^2} \ \ , \\
  \left|\vec p_{t,\ell}-\vec k_t\right| -p_{t,\ell}& =\left|(1-z^{(\ell)})\vec p_{t,\ell}-\vec \kappa\right|- p_{t,\ell}\simeq -z^{(\ell)} p_{t,\ell}-\kappa\cos\phi  = -\kappa \, e^{\eta^{(\ell)}} \frac{p_{t,\ell}}{Q} +\dots \ \ ,
  \end{split}
\end{equation}
where in the last line we have omitted a term that vanishes after
integration over the azimuthal angle $\phi$.  This gives, for the
total broadening
\begin{equation}
    \label{eq:h-BT}
    h_{B_T}(\eta^{(\ell)},\phi, \{\tilde{p}\}, k_1,\dots,k_n)=\frac{1}{2}\left[\sqrt{1+2 e^{\eta^{(\ell)}}\frac{p_{t,\ell}}{Q}\cos\phi+e^{2\eta^{(\ell)}} \left(\frac{ p_{t,\ell}}{Q}\right)^2}-e^{\eta^{(\ell)}}\frac{p_{t,\ell}}{Q}\right] \ \ .
  \end{equation}
For the wide-jet broadening, where the ultra-soft emission gives a contribution only if it is emitted in the broader hemisphere, we get 
\begin{align}\label{eq:h-BW}
	h_{B_W}(\eta^{(1)},\phi,\{\tilde{p}\}, k_1,\dots,k_n) = \, &\frac{1}{2}\left[\sqrt{1+2e^{\eta^{(1)}}\frac{p_{t,1}}{Q}\cos\phi+e^{2\eta^{(1)}} \left(\frac{ p_{t,1}}{Q}\right)^2}-e^{\eta^{(1)}}\frac{p_{t,1}}{Q}\right]\Theta(B_1-B_2) \ , \\ 
    h_{B_W}(\eta^{(2)},\phi,\{\tilde{p}\}, k_1,\dots,k_n) = \, 
    &\frac{1}{2}\left[\sqrt{1+2e^{\eta^{(2)}}\frac{p_{t,2}}{Q}\cos\phi+e^{2\eta^{(2)}} \left(\frac{ p_{t,2}}{Q}\right)^2}-e^{\eta^{(2)}}\frac{p_{t,2}}{Q}\right]\Theta(B_2-B_1) \ .                                                             
\end{align}
Notice that when integrating $h_{B_T}$ and $h_{B_W}$ we can take the limits on $\eta^{(\ell)}$ to be between $0$ and infinity since the integral is convergent. Thus we obtain the following integral, with $p>0$,
\begin{equation}
  \label{eq:B-eta-phi-integral}
  \int_0^{\infty}\!\! d\eta \int \frac{d\phi}{2\pi} \left[\sqrt{1+2 e^{\eta} \frac{p}{Q}\cos\phi+\left(e^{\eta} \frac{p}{Q}\right)^2}-e^{\eta} \frac{p}{Q}\right]=\ln\frac{Q}{p}+\eta_0^{(B)}+\mathcal{O}\left(\frac{p}{Q}\right) \ \ ,
\end{equation}
where
\begin{equation}
\label{eq:Q0B}
 \eta_0^{(B)}= \int_0^{\infty}\frac{du}{u}\int \frac{d\phi}{2\pi}  \left[\sqrt{1+2 u\cos\phi+u^2}-u-\Theta(1-u)\right] =-0.6137056 \ \ .
 \end{equation}
Since $p$ is the transverse recoil induced by perturbative emissions,  all terms that vanish as a power of $p/Q$  give contributions that are at most suppressed by powers of $B_T,B_W$ and hence can be neglected. With this approximation, our final result for $\cNP{B_T}$ and $\cNP{B_W}$ reads 
\begin{equation} \label{eq:hB-almost-final}
    \begin{split}
		\cNP{B_T} = \,
		&\frac{1}{R' \, \mathcal{F}_{B_T}(R')} \int d\mathcal{Z}[\{R'_{\ell_i},k_i\}] \,  \delta\bigg(1-\frac{B_{T,\rm sc}(\{\tilde{p}\},\{k_i\})}{B_T}\bigg) \, \times \\ 
		&\times \int\prod_{\ell}d^2p_{t,\ell} \, \delta^{(2)}\Bigg(\vec p_{t,\ell}+\sum_{i\in \mathcal{H}_{\ell}}\vec k_{t,i}\Bigg) \, \frac12 \sum_{\ell} \ln\frac{Qe^{\eta_0^{(B)}}}{p_{t,\ell}} \ \ , \\
		\cNP{B_W} = \,
		&\frac{1}{R' \, \mathcal{F}_{B_W}(R')} \int d\mathcal{Z}[\{R'_{\ell_i},k_i\}] \, \delta\bigg(1-\frac{B_{W,\rm sc}(\{\tilde{p}\},\{k_i\})}{B_W}\bigg) \, \times \\ 
		&\times\left[\int\prod_{\ell}d^2p_{t,1} \, \delta^{(2)}\Bigg(\vec p_{t,1}+\sum_{i\in \mathcal{H}_{1}}\vec k_{t,i}\Bigg) \frac12 \ln\frac{Qe^{\eta_0^{(B)}}}{p_{t,1}} \, \Theta(B_1-B_2) + 1 \leftrightarrow 2\right] \ \ .	
    \end{split}
\end{equation}
Remarkably, the fact that the hard legs are displaced from the thrust axis has profound implications for the NP shift to the jet broadenings. In fact, from eq.~\eqref{eq:B-eta-phi-integral}, one can see that the rapidity of an ultra-soft emission collinear to leg $\ell$ is effectively cut at $\ln (Q/p_{t,\ell})$, which is far away from the collinear limit $\ln(Q/\kappa)$. This is consistent with our treatment of NP corrections, which requires that the ultra-soft emission is at large angles.

To explicitly compute $\cNP{B_T}$ and $\cNP{B_W}$, we further manipulate eq.~\eqref{eq:hB-almost-final} by rescaling $p_{t,\ell}$ with respect to $B_T,B_W$ as follows
\begin{equation}
\vec x_\ell \equiv \frac{\vec p_{t,\ell}}{B Q} \ \ , \qquad B=B_T,B_W \ \ ,
\end{equation}
and by introducing the two-dimensional vectors $\vec\zeta_i\equiv \zeta_i(\cos\phi_i,\sin\phi_i)$. This gives
\begin{equation}
  \label{eq:hB-final}
  \begin{split}
  \cNP{B_T} & =\ln\frac{1}{B_T}+\eta_0^{(B)}+\chi_T(R') \ \ , \\ 
  \cNP{B_W} &=\frac 12\left(\ln\frac{1}{B_W}+\eta_0^{(B)}\right)+\chi_W(R') \ \ ,
  \end{split}
  \end{equation}
where
\begin{equation} \label{eq:chiB}
	\begin{split}
		\chi_T(R') = \,
		&\frac{1}{R' \, \mathcal{F}_{B_T}(R')} \int d\mathcal{Z}[\{R'_{\ell_i},k_i\}] \, \delta\bigg(1-\frac{B_{T,\rm sc}(\{\tilde{p}\},\{k_i\})}{B_T}\bigg) \, \times \\ 
		&\times \int\prod_{\ell}d^2 \vec x_{\ell} \, \delta^{(2)}\Bigg(\vec x_{\ell}+\sum_{i\in \mathcal{H}_{\ell}}\vec \zeta_{i}\Bigg) \, \frac12 \sum_{\ell} \ln\frac{1}{|\vec x_{\ell}|} \ \ , \\
		\chi_W(R') = \,
		&\frac{1}{R' \, \mathcal{F}_{B_W}(R')} \int d\mathcal{Z}[\{R'_{\ell_i},k_i\}] \, \delta\bigg(1-\frac{B_{W,\rm sc}(\{ \tilde{p}\},\{k_i\})}{B_W}\bigg) \, \times \\ 
		&\times \left[\int d^2 \vec x_{1} \, \delta^{(2)}\Bigg(\vec x_{1}+\sum_{i\in \mathcal{H}_{1}}\vec \zeta_{i}\Bigg) \, \frac12 \ln\frac{1}{|\vec x_{1}|}\Theta(B_1-B_2) + 1 \leftrightarrow 2\right] \ \ .  
   \end{split}
 \end{equation}
 As has been noted already in~\cite{Dokshitzer:1998qp}, for small $B_T,B_W$ the shift depends logarithmically on the event shape's value, with corrections given by $\eta_0^{(B)}$ and the functions $\chi_T(R')$ and $\chi_W(R')$. The latter have been computed analytically in appendices~\ref{subsec:NP-total} and \ref{subsec:NP-wide} respectively and agree with the results of~\cite{Dokshitzer:1998qp}.
 
\begin{figure}[htbp]
	\centering
	\includegraphics{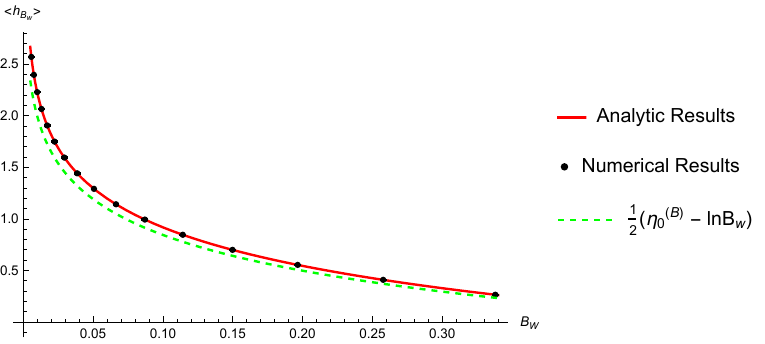}
	\caption{Comparison of the numerical and analytic calculation of the NP shift for $B_W$ (with $\alpha_s = 0.118$).}
	\label{fig:NPshift-Bw}
\end{figure}

\begin{figure}[htbp]
	\centering
	\includegraphics{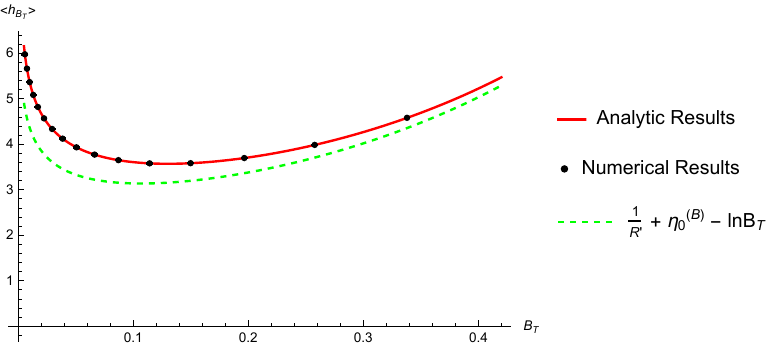}
	\caption{Comparison of the numerical and analytic calculation of the NP shift for $B_T$ (with $\alpha_s = 0.118$).}
	\label{fig:NPshift-Bt}
\end{figure}

We also compute the NP shift fully numerically and compare with the analytic calculation. This is shown in Figs.~\ref{fig:NPshift-Bw} and \ref{fig:NPshift-Bt}.
We observe a perfect agreement between the analytic calculation and
the output of our numerical program. We also plot the limiting
behaviour of the shift for $R' \to 0$ (see eqs.~\eqref{eq:hBW-leading}
and \eqref{eq:hBT-leading} for $B_W$ and $B_T$ respectively) given by the green-dashed curves. The
logarithmic piece and the constant $\eta_0^{(B)}$ capture most of the
shift for all values of $B_W$ and for small values of $B_T$. We notice
however a fundamental difference between the shift of $B_W$ and of
$B_T$. The former is a smooth function of $R'$ and decreases for large
$B_W$. In contrast, the shift for $B_T$ has a minimum and increases
for large $B_T$. This is the symptom of an unphysical behaviour. In
fact, the shift for $B_T$ develops a $1/R'$ singularity as can be seen
from the explicit form of the function $\chi_T(R')$ in
eq.~\eqref{eq:chiT}. From Fig.~\ref{fig:NPshift-Bt}, one can see that
this singular behaviour dominates the shift at large values of $B_T$,
or equivalently small values of $R'$.

This behaviour is also reflected in numerical issues when calculating $\cNP{B_T}$ with the procedure described in appendix~\ref{sec:monte-carlo-determ}. In fact, it may transpire, especially when $R'$ is small, that only one hemisphere is populated by a small number of emissions with
$k_t\sim B_T Q$, whereas all emissions in the other hemisphere (which we can refer to as the ``empty'' hemisphere) fall below the cutoff of the Monte Carlo integration. As a consequence, one would find a zero
value of $p_{t,\ell}$ in the empty hemisphere and eq.~\eqref{eq:chiB}, which requires the calculation of $\ln(Q/p_{t,\ell})$, would give a floating point exception. This issue is ultimately due to the fact that the Monte Carlo procedure of appendix~\ref{sec:monte-carlo-determ} assumes that only perturbative emissions with comparable transverse momenta contribute to the shift, which is clearly not the case for $p_{t,\ell}$ of the empty hemisphere. In order to obtain finite numerical predictions we are forced to decrease the cutoff $\epsilon$ more and more as $R'$ approaches zero. We reiterate that this problem is of physical and not technical nature. Its ultimate solution requires upgrading the probability of soft-gluon emissions in the empty hemisphere as was done in \cite{Dokshitzer:1998qp}. This leads to a finite value for $\cNP{B_T}$ even for $R'\to 0$. The generalisation of this procedure to an arbitrary event shape will be discussed in section~\ref{sec:subtraction}. 

\subsection{NP shift for the thrust major}

\begin{figure}[htbp]
	\centering
	\includegraphics{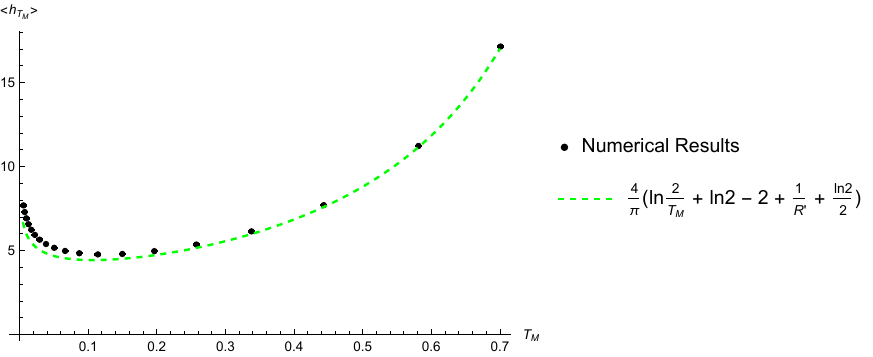}
	\caption{The numerical calculation of the NP shift for $T_M$, compared to its limiting behaviour for $R'\to 0$ (with $\alpha_s = 0.118$).}
	\label{fig:hTM}
\end{figure}

Another observable that has a similar behaviour to the jet broadenings is the thrust major. In this case, only PT soft and collinear emissions determine the thrust-major axis $\vec n_M$, which we conventionally set to give the $y$-direction for all momenta. In fact, when varying the trial direction for the thrust-major axis, one finds a set of local maxima which are the candidates for the magnitude of $T_M$. In the presence of soft and collinear PT emissions that contribute to the observable in the two-jet region, all these local maxima are of order $T_M$ as all emissions have comparable transverse momenta. The addition of an ultra-soft emission, with $\kappa \ll T_M Q$, will not change the hierarchy of the local maxima and thus will not change the value of $T_M$ up to sub-leading power corrections. This is similar to the argument that is used to compute NP corrections to the heavy-jet mass. Ultimately this implies that the thrust-major axis is not altered by a NP emission. This gives
\begin{equation} \label{eq:TM-np}
	T_M(\{\tilde p\},k,k_1,\dots,k_n) Q = 
	\sum_i^n |k_{yi}|+|k_y|+|\tilde p_{y,1}|+|\tilde p_{y,2}| \ \ .
\end{equation}
Using the transverse momentum component $p_{y,\ell}$ defined in eq.~(\ref{eq:ptell}), the change in the thrust major due to a NP gluon $k$ collinear to leg $\ell$ is given by 
\begin{equation} \label{eq:dTM-np}
	\delta T_M(\{\tilde p\},k,\{k_i\}) Q =  |k_y| + |\tilde p_{y,\ell}| - | p_{y,\ell}| \ \ .
\end{equation}
We now need to recast the quantities in square bracket in terms of the transverse momentum with respect to the emitter $\kappa_y=\kappa\sin\phi$. 
Using the symmetry properties of the $\phi$-integral, we obtain that $\delta T_M$ depends only on $|p_{y,\ell}|$. 
This gives
\begin{equation} \label{eq:ky-np}
	\begin{split}
		|k_y| &= 
		\big|\kappa_y+z^{(\ell)} |p_{y,\ell}|\big| = 
		\kappa \left|\sin\phi+e^{\eta^{(\ell)}} \frac{|p_{y,\ell}|}{Q}\right| \ \ , \\
		|\tilde p_{y,\ell}|-|p_{y,\ell}| &= 
		\big|(1-z^{(\ell)})|p_{y,\ell}|-\kappa_y\big|-|p_{y,\ell}| = \sqrt{\left[(1-z^{(\ell)})|p_{y,\ell}|-\kappa_y\right]^2}-|p_{y,\ell}| \\
		&\simeq -z^{(\ell)}|p_{y,\ell}|-\kappa\sin\phi = 
		-\kappa \, e^{\eta^{(\ell)}} \frac{|p_{y,\ell}|}{Q} + \dots
  \end{split}
\end{equation}
In the last equality we have dropped a term that vanishes upon integration over $\phi$. We then obtain 
\begin{equation} \label{eq:dTM-final}
	\delta T_M(\{\tilde p\},k,\{k_i\}) Q \simeq
	\kappa \left(\,\left|\sin\phi + e^{\eta^{(\ell)}} \frac{|p_{y,\ell}|}{Q}\right| - e^{\eta^{(\ell)}}\frac{|p_{y,\ell}|}{Q}\right) \ \ .
\end{equation}
Therefore, the $\eta$ and $\phi$ integrals now give, with $p>0$
\begin{equation} \label{eq:TM-eta-phi-integrals}
	\int_0^\infty \!\! d\eta \int_0^\pi \frac{d\phi}{\pi} \left(\,\left|\sin\phi+e^{\eta}\frac{p}{Q}\right|-e^{\eta}\frac{p}{Q}\right) =
	\frac{2}{\pi}\left(\ln\frac{Q}{p} + \ln 2 - 2\right) + \mathcal{O}\left(\frac{p}{Q}\right) \ \ .
\end{equation}  
The constant $\ln (2e^{-2})$ is the same as found for the thrust minor \cite{Banfi:2001sp}. This is expected because, in both cases, it is harder emissions that fix the thrust-major axis and here we are considering the magnitude of one of the two components of $\vec p_{t,\ell}$. Inserting the above expression into the general formula for $\cNP{V}$ gives
\begin{equation} \label{eq:cTM}
	\begin{split}
		\cNP{T_M} = \,
		&\frac{1}{R' \, \mathcal{F}_{T_M}(R')} \, \frac{2}{\pi} \, \int d\mathcal{Z}[\{R'_{\ell_i},k_i\}] \, \delta\bigg(1-\frac{T_{M,\rm sc}(\{\tilde{p}\},\{k_i\})}{T_M}\bigg) \, \times \\ 
		&\times \int_{-\infty}^\infty \prod_{\ell}d p_{y,\ell} \, \delta\Bigg(p_{y,\ell} + \sum_{i\in \mathcal{H}_{\ell}}k_{y,i}\Bigg) \sum_{\ell}\ln\frac{2Qe^{-2}}{|p_{y,\ell}|} \ \ .
    \end{split}
\end{equation}
In a similar way as for the broadenings, we can introduce the rescaled variables $x_{\ell}=2p_{y,\ell}/(T_M Q)$ and the two-dimensional vectors $\vec\zeta_i\equiv \zeta_i(\cos\phi_i,\sin\phi_i)$ and obtain
\begin{equation} \label{eq:cTM-rescaled}
	\begin{split}
		\cNP{T_M} = 
		\frac{4}{\pi}\left(\ln\frac{2}{T_M} + \ln2 - 2 \right) + \chi_M(R') \ \ ,
    \end{split}
\end{equation}
where
\begin{equation} \label{eq:chiM}
	\begin{split}
		\chi_M(R') = \,
		&\frac{1}{R' \, \mathcal{F}_{T_M}(R')} \, \frac{2}{\pi} \, \int d\mathcal{Z}[\{R'_{\ell_i},k_i\}] \, \delta\bigg(1-\frac{T_{M,\rm sc}(\{\tilde{p}\},\{k_i\})}{T_M}\bigg) \, \times \\ 
		& \times\int_{-\infty}^\infty \prod_{\ell}d x_{\ell} \, \delta\Bigg(x_{\ell}+\sum_{i\in \mathcal{H}_{\ell}}\zeta_i\sin\phi_i\Bigg) \sum_{\ell}\ln\frac{1}{|x_{\ell}|} \ \ .
	\end{split}
      \end{equation}
We first observe that, similar to the broadenings, the NP shift for the thrust major has an explicit logarithmic dependence on $T_M$.       
The function $\chi_M(R')$ can only be computed numerically with the procedure outlined in appendix~\ref{sec:monte-carlo-determ}. This calculation constitutes the first new result of this paper. The
resulting value of $\cNP{T_M}$ is plotted as a function
of $T_M$ in Fig.~\ref{fig:hTM}, and compared to its limiting behaviour for $R'\to 0$ (which has been computed analytically in eq.~\eqref{eq:chiM-leading-full}). 

As was the case for $B_T$, the shift for $T_M$ has a minimum and increases for large $T_M$, or equivalently small values of $R'$, where a $1/R'$ singular behaviour dominates the shift. The calculation of $\chi_{M}(R')$ has the same problem as that of $ \chi_{T}(R')$. In fact, when $R'\to 0$, there will be one emission with $k_t\sim T_M Q $ in one hemisphere and all emissions in the other hemisphere will have much smaller transverse momenta. From eq.~\eqref{eq:chiM-leading-full} we find that in this limit $\chi_M(R)=\frac{4}{\pi}\left(1/R' + \frac{\ln 2}{2}\right)+\mathcal{O}(R')$. This implies that a naive implementation of eq.~\eqref{eq:cTM} will result in $p_{y,\ell}$ of the empty hemisphere to be equal to zero,
and hence in a floating point exception, unless one decreases the cutoff for the integration over the $\zeta_i$. In the next section, we discuss how to deal with this problem in full generality and devise a procedure to obtain finite NP shifts down to $R'=0$.

\begin{figure}[htbp]
	\centering
	\includegraphics{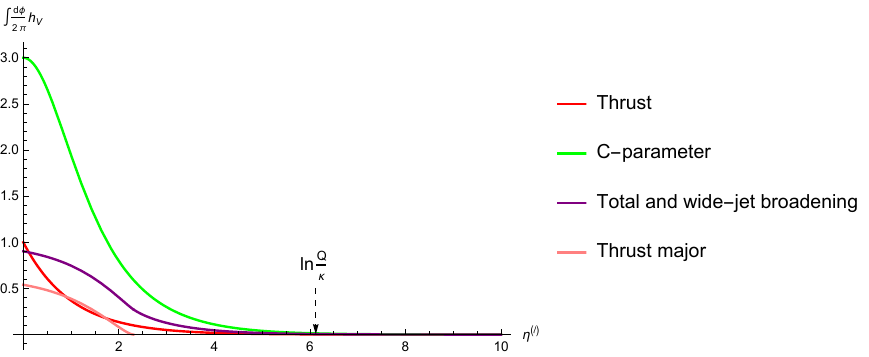}
	\caption{Plot of $h_V(\eta^{(\ell)},\{\tilde{p}\},\{k_i\})$ (after integrating over azimuth) for the various event-shape variables considered (with $Q = 91.2\,\rm{GeV}$, $p_{t,\ell}/Q \sim 0.1$ and $\kappa \sim \Lambda_{\rm QCD}$ for illustration) to show the collinear suppression in each case.} 
	\label{fig:rapidity_damping}
\end{figure}

We observe that, for soft and collinear PT emissions, the scaling of $\Vsc{k_i}$ with respect to $\eta^{(\ell)}$ is very different for the different observables. In particular, there is a rapidity suppression for thrust, $C$-parameter and heavy-jet mass, but no rapidity dependence for the jet broadenings and thrust major. This is not the case for a NP emission \emph{in the presence of multiple soft and collinear PT emissions}. In this case the contribution to $\delta V_{\rm NP}$ of the collinear regions $\eta^{(\ell)} \to \infty$ is always suppressed for the event shapes considered here, see Fig.~\ref{fig:rapidity_damping}. Also, the suppression is such that the rapidity integral in eq.~\eqref{eq:np-shift-NLL} is convergent.

\section{General treatment of divergent $\mathbf{\cNP{V}}$ for $\mathbf{R'\to 0}$}
\label{sec:subtraction}

The analytic expressions for $\cNP{B_W}$ and $\cNP{B_T}$ as a function
of $R'$ can be found in eqs.~\eqref{eq:hBW-analytic} and
\eqref{eq:hBT-analytic} respectively. There one can see that, while
$\cNP{B_W}$ is finite for all values of $R'$, $\cNP{B_T}$ has a
divergence for $R'\to 0$. As already anticipated, this divergence
originates from the fact that, when $R'\to 0$, one hemisphere contains
a few emissions with $k_t\sim B_T Q$, while emissions in the other
hemisphere have much lower transverse momenta. Mathematically, this corresponds to the situation where $R'$ is much smaller than all higher derivatives of $R(v)$, which can no longer be neglected. Numerically, this
creates a problem for a Monte Carlo implementation of $\cNP{B_T}$ which
assumes all transverse momenta of perturbative emissions to be of the
same order. For $B_T$, the problem is solved in
\cite{Dokshitzer:1998qp} by retaining higher derivatives of the radiator, e.g.\ $R^{\prime\prime}$. While this \emph{improved}
evaluation can be carried out fully analytically for $\cNP{B_T}$, the
same is not true for $\cNP{T_M}$ because the direction of the
thrust-major axis depends on all perturbative emissions.

Therefore, we need to devise a procedure to compute $\cNP{V}$ that is suitable for all observables and gives a finite result for $R'\to 0$. What we propose is to add and subtract to $\cNP{V}$ a counterterm that displays the appropriate $1/R'$ behaviour. The counterterm is designed in such a way as to cancel the divergence of $\cNP{V}$ for $R'\to 0$ at the
\emph{integrand} level. It must also be simple enough to be computed fully analytically for all values of $V$. This procedure ensures that $\cNP{V}$ is finite for
all values of $R'$.

\subsection{Counterterm for $\langle h_{B_T} \rangle$}
\label{subsec:BTsubtraction}

To see how a subtraction procedure might work, we consider the case of the total broadening for which we have full analytic control. In the limit $R' \to 0$, one of the hemispheres will contain a single emission, which we denote $k_1$, with a transverse momentum of order  $B_T Q$. In the limit $R'\to 0$, there exists $\delta$ with $\epsilon \ll \delta \ll 1$ such that  all other emissions have transverse momenta less than $\delta B_T Q$, and thus 
\begin{align} \label{eq:B-1emsn}
	B_{T,{\rm sc}}(\{\tilde{p}\},\{k_i\})\simeq 
	B_{T,{\rm sc}}(\{\tilde{p}\}, k_1) \ \ .
\end{align}
Using eq.~\eqref{eq:dZ-rescaled}, invoking the symmetry between the two hemispheres defined by the thrust axis ($\mathcal{H}_1\leftrightarrow\mathcal{H}_2$ symmetry) and the fact that $R' = R'_1 + R'_2$ (where we will eventually set $R'_1 = R'_2 = R'/2$ due to the $\mathcal{H}_1 \leftrightarrow \mathcal{H}_2$ symmetry), we write $\chi_T(R')$, defined in eq.~\eqref{eq:chiB}, in the limit $R' \to 0$ as
\begin{align} \label{eq:chiT-smallR'-start}
	\nonumber
	\chi_T(R') \simeq \, &\frac{1}{2}\int_{-\infty}^{\infty}d^2\vec{x}_2\, \left(\lim_{\epsilon \to 0}\,\epsilon^{R'_2}\,\sum_{n=0}^{\infty}\frac{\left(R'_2\right)^n}{n!}\,\prod_{i=2}^{n+1}\,\int_{\epsilon}^{\delta}\frac{d\zeta_i}{\zeta_i}\,\int_0^{2\pi}\frac{d\phi_i}{2\pi}\right) \times \\
	&\times \, \delta^{(2)}\Bigg(\vec{x}_2+\sum_{i\in \mathcal{H}_2}\vec{\zeta}_{i}\Bigg) \, \ln\frac{1}{|\vec{x}_2|}  \ \ .
\end{align}
We rescale $\zeta_i$ and introduce a two-dimensional Fourier transform, noting that we must apply a cut on large values of the transverse momenta of the recoil in $\mathcal{H}_2$ due to the small transverse momenta of emissions in this hemisphere. We achieve this by imposing an upper-bound $|\vec{x}_2| < x_{\rm max}$ (where $x_{\rm max} \sim \delta$) and a corresponding lower-bound on the conjugate parameter $|\vec{b}_2| > b_{\rm min}$ (where $b_{\rm min} \sim 1/\delta$). We obtain
\begin{align}
	\nonumber
	\chi_T(R') \simeq \,
	&\frac12\int_{0} ^{x_{\rm max}} d^2\vec{x}_2 \int_{b_{\rm min}}^{\infty} d^2\vec{b}_2 \left(\lim_{\epsilon \to 0}\, (\epsilon \delta)^{R'_2}\,\sum_{n=0}^{\infty}\frac{\left(R'_2\right)^n}{n!}\,\prod_{i=2}^{n+1}\,\int_{\epsilon}^{1}\frac{d\zeta_i}{\zeta_i}\,\int_0^{2\pi}\frac{d\phi_i}{2\pi}\right) \times \\
	&\times \exp\left(- i \vec{b}_2\cdot\vec{x}_2 - i \delta \sum_{i \in \mathcal{H}_2} \vec{b}_2 \cdot \vec{\zeta_i}\right) \, \ln\frac{1}{x_2} \ \ ,
\end{align}
where $x_2 \equiv |\vec{x}_2|$. This may be written in a simplified exponential form to give
\begin{equation}\label{eq:hT-smallR'-factorised}
	\chi_T(R') \simeq 
	\frac12 \, \delta^{R'_2} \int_0^{x_{\rm max}} x_2 \, dx_2 \int_{b_{\rm min}}^{\infty} b_2 \, db_2 \, J_0(b_2x_2) \, \exp\left(R'_2 \int_0^1\frac{d\zeta}{\zeta}\,\big[J_0(\delta b_2\zeta)\,-1\big]\right) \, \ln\frac{1}{x_2} \ \ .
\end{equation}
As emissions in $\mathcal{H}_2$ do not contribute to the observable, as per eq.~\eqref{eq:B-1emsn}, we notice that there is no exponential damping factor in the $\zeta$-integral. Using the fact that, for large $b$,
\begin{equation} \label{eq:radiator-large-b}
	\int_0^1 \frac{d\zeta}{\zeta} \left[J_0(\delta b_2\zeta)-1\right] \simeq 
	- \ln b_2 - \ln \delta + \mathcal{O}(1) \ \ ,
\end{equation}
we obtain
\begin{equation}
	\chi_T(R')\simeq 
	\frac12 \int_0^{x_{\rm max}} x_2 \, dx_2 \int_{b_{\rm min}}^{\infty} b_2 \, db_2 \, J_0(b_2x_2) \, b_2^{- R'_2} \, \ln\frac{1}{x_2}  \ \ .
\end{equation}
Performing the $x_2$- and $b_2$-integrations and extracting the singular $R' \to 0$ behaviour (setting $R'_1 = R'_2 = R'/2$) we obtain
\begin{align} \label{eq:chiT-smallR'-result}
	 \chi_T(R') \simeq \frac{1}{R'} \ \ .
\end{align}
We note that we do not control terms of $\mathcal{O}(1)$ in this calculation as they may arise from the interplay of unknown terms of order $R'$ multiplied by the leading $1/R'$. From the full analytical evaluation of $\cNP{B_T}$ performed in appendix~\ref{subsec:NP-total}, the leading $R' \to 0$ behaviour is presented in eq.~\eqref{eq:hBT-leading} and we find that the uncontrolled $\mathcal{O}(1)$ term is in fact equal to zero in this limit. Therefore in the limit $R' \to 0$
\begin{equation} 
	\cNP{B_T} = \ln\frac{1}{B_T} + \eta_0 + \frac{1}{R'} + \mathcal{O}(R') \ \ .
\end{equation}
At this point we face essentially two possibilities. One is to
relax this assumption and try computing $\cNP{V}$ from scratch, a
framework that we do not adopt here because of its limited
scope. The other is to devise a {\em local} counterterm, which we call
$\cNPct{V}$, that exhibits the same singularity for $R'\to 0$ as
$\cNP{V}$. We then subtract (add) this counterterm from (to) the full
shift. The subtraction of the singularity for $R'\to 0$,
within $\cNP{V} - \cNPct{V}$ which will be carried out via a general Monte
Carlo procedure, takes place at the integrand level and
thus we obtain a finite result as $R'\to 0$. The counterterm is
constructed so as to have full analytic control; in particular we must
be able to obtain all constant terms appearing in the shift of any
observable as $R' \to
0$. 

\begin{figure}[htbp]
	\centering
	\includegraphics{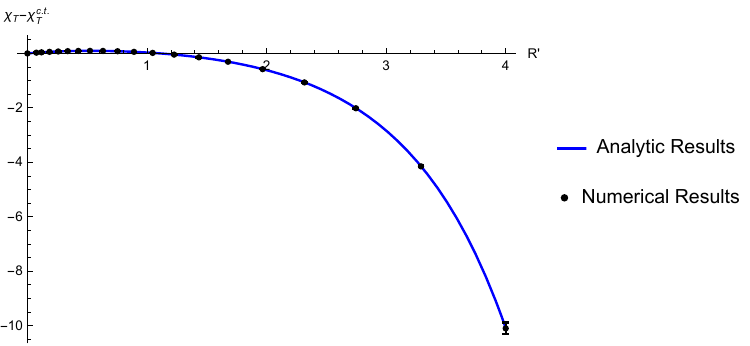}
	\caption{Comparison of the numerical and analytic calculation of $\chi_T(R') - \chi^{\textrm{c.t.}}_T(R')$ and demonstration that the contributions of $\chi_T(R')$ and $\chi^{\textrm{c.t.}}_T(R')$ cancel perfectly in the limit $R'\to 0$.}
  \label{fig:chiT-chiTbar}
\end{figure}

Let us return to the case of the total broadening. Fully generally, for $R'\to 0$ one emission, emitted in hemisphere $\mathcal{H}_{\ell}$, has a value of transverse momentum that is much larger than all of the other emissions. In this situation $B_T$ is determined by this emission and the hemisphere that does not contain this emission, $\mathcal{H}_{\bar{\ell}}$, is automatically the hemisphere with the smaller broadening. Motivated by the above discussion, a good counterterm is then constructed as follows 
\begin{align} \label{eq:BT-counterterm}
	\nonumber
 	\cNPct{B_T} = \,
 	&\frac{1}{R' \mathcal{F}_{B_T}(R')} \int d^2\vec{p}_{t,\ell} \, d^2\vec{p}_{t,\bar{\ell}} \int d\mathcal{Z}[\{R'_{\ell_i},k_i\} ] \, \delta\bigg(1-\frac{\max_{i}\{k_{ti}\}}{B_T Q}\bigg) \frac 12 \ln\frac{Q}{p_{t,\bar{\ell}}} \, \times \\ \nonumber 
 	&\times \delta^{(2)}\Big(\vec{p}_{t,\ell}+\vec{k}_{t,\max}\Big) \, \delta^{(2)}\Bigg(\vec{p}_{t,\bar{\ell}}+\sum_{i\in \mathcal{H}_{\bar{\ell}}}\vec k_{t,i}\Bigg) \, \Theta\Bigg(\max_{i}\{k_{ti}\} - \frac12 p_{t,\bar{\ell}} - \frac12\sum_{i\in \mathcal{H}_{\bar{\ell}}} k_{t,i}\Bigg) \\ \nonumber
 	&\!\!\!\!\!\!\!= \frac{1}{R' \mathcal{F}_{B_T}(R')} \int d^2\vec{p}_{t,\bar{\ell}} \int d\mathcal{Z}[\{R'_{\ell_i},k_i\} ] \, \delta\bigg(1-\frac{\max_{i}\{k_{ti}\}}{B_T Q}\bigg) \frac 12 \ln\frac{Q}{p_{t,\bar{\ell}}} \, \times \\ 
 	&\times \delta^{(2)}\Bigg(\vec{p}_{t,\bar{\ell}}+\sum_{i\in \mathcal{H}_{\bar{\ell}}}\vec k_{t,i}\Bigg) \, \Theta\Bigg(\max_{i}\{k_{ti}\} - \frac12 p_{t,\bar{\ell}} - \frac12\sum_{i\in \mathcal{H}_{\bar{\ell}}} k_{t,i}\Bigg) \ \ ,
\end{align}
where $\big|\vec{k}_{t,\max}\big| = \max_i\{k_{ti}\}$. 
In a similar way as for $\cNP{B_T}$, we can introduce the rescaled variables $\vec{x}_{\ell}=\vec{p}_{t,\ell}/(B_T Q)$  and the two-dimensional vectors $\vec\zeta_i\equiv \zeta_i(\cos\phi_i,\sin\phi_i)$ and obtain (see appendix~\ref{subsec:ct-BT})
\begin{equation} \label{eq:hBTct-rescaled}
	\cNPct{B_T} = \frac12 f_T(R')\ln\frac{1}{B_T} + \chi^{\textrm{c.t.}}_T(R') \ \ ,
\end{equation}
where
\begin{equation} \label{eq:f_T(R')}
	f_T(R') \equiv 
	\frac{e^{\gamma_E \frac{R'}{2}}}{2^{\frac{R'}{2}} \, \sigma\big(\frac{R'}{2}\big)} \, \frac{\Gamma(1+R')}{\Gamma\big(1+\frac{R'}{2}\big)} \ \ ,
\end{equation}
with $f_T(0) = 1$, the function $\sigma(R'/2)$ defined in eq.~\eqref{eq:sigmaR'} and 
	\begin{align} \label{eq:chibarT}
		\nonumber
		\chi^{\textrm{c.t.}}_T(R') = \,
		&\frac{1}{R' \mathcal{F}_{B_T}(R')} \int d^2\vec{x}_{\bar{\ell}} \int d\mathcal{Z}[\{R'_{\ell_i},k_i\} ] \, \delta\bigg(1-\max_{i}\{\zeta_i\}\bigg) \frac 12 \ln\frac{1}{x_{\bar{\ell}}} \, \times \\ 
		&\times \delta^{(2)}\Bigg(\vec{x}_{\bar{\ell}}+\sum_{i\in \mathcal{H}_{\bar{\ell}}} \vec{\zeta}_{i}\Bigg) \, \Theta\Bigg(1 - \frac12 x_{\bar{\ell}} - \frac12\sum_{i\in \mathcal{H}_{\bar{\ell}}} \zeta_{i}\Bigg) \ \ , 
\end{align}
where $x_{\bar{\ell}} \equiv |\vec{x}_{\bar{\ell}}|$.
Such counterterm has the same $1/R'$ singularity as $\cNP{B_T}$ and is constructed so that it may be computed analytically, as is performed in appendix~\ref{subsec:ct-BT}. Therefore, the combination $\cNP{B_T} - \cNPct{B_T}$ is free of singularities for all values of $R'$ with the cancellation of singularities occurring \emph{locally}, i.e.\ for each Monte Carlo configuration used to compute $\cNP{B_T} - \cNPct{B_T}$. What we obtain is
\begin{equation} \label{eq:hBT-hBTct}
	\cNP{B_T} - \cNPct{B_T} 
	= \bigg[1 - \frac12 f_T(R')\bigg] \ln\frac{1}{B_T} + \eta_0^{(B)} + \chi_T(R') - \chi^{\textrm{c.t.}}_T(R') \ \ .
\end{equation}
The function $\chi^{\textrm{c.t.}}_T(R')$ is computed analytically in appendix~\ref{subsec:ct-BT} and its expression can be found in eq.~\eqref{eq:chibarTNLL}. We may also rewrite $\chi^{\textrm{c.t.}}_T(R')$ in a way that is suitable for Monte Carlo integration, following the strategy described in appendix~\ref{sec:monte-carlo-determ}. We find
\begin{align} \label{eq:chiBT-chibarBT}
	\nonumber
	&\chi_T(R')-\chi^{\textrm{c.t.}}_T(R') = \,
	\frac{1}{2 \, \mathcal{F}_{B_T}(R')} \sum_{\ell_1=1,2	}  \frac{R'_{\ell_1}}{R'} \int_0^{2\pi} \! \frac{d\phi_1}{2\pi} \left( \epsilon^{R'}\sum_{n=0}^\infty \frac{1}{n!}  \prod_{i=2}^{n+1}\sum_{\ell_i=1,2} R'_{\ell_i}\int^{1}_{\epsilon}\frac{d\zeta_i}{\zeta_i}\int_0^{2\pi}\frac{d\phi_i}{2\pi}\right) \times \\ \nonumber
	&\times \Bigg[\left(\frac{B_{T, \rm sc}\big(\{\tilde{p}\},k_1,\dots,k_{n+1}\big)}{B_T}\right)^{-R'} \ln\left[\left(\frac{B_{T, \rm sc}\big(\{\tilde{p}\},k_1,\dots,k_{n+1}\big)}{B_T}\right)^{2}\frac{1}{\left|\sum_{i \in \mathcal{H}_{\ell_1}}\vec{\zeta}_i\right|\left|\sum_{i \notin \mathcal{H}_{\ell_1}}\vec{\zeta}_i\right|}\right] \\
	& \qquad - \ln\frac{1}{\left|\sum_{i \notin \mathcal{H}_{\ell_1}}\vec{\zeta}_i\right|} \, \Theta\Bigg(1 - \frac12\Bigg|\sum_{i \notin \mathcal{H}_{\ell_1}}\vec{\zeta}_i\Bigg| - \frac12\sum_{i \notin \mathcal{H}_{\ell_1}}\zeta_i\Bigg)\Bigg]	\ \ ,
\end{align} 
where, as explained in appendix~\ref{sec:monte-carlo-determ}, now $B_{T, \rm sc}(\{\tilde{p}\},k_1) = B_T$.
We observe that $\chi_T(R')-\chi^{\textrm{c.t.}}_T(R')$ is finite for all values of $R'$. When the calculation involves the recoil due to emissions in the same hemisphere as emission $k_1$, i.e.\ hemisphere $\mathcal{H}_{\ell_1}$, the counterterm $\chi^{\textrm{c.t.}}_T(R')$ has no effect whatsoever. When the calculation involves the recoil due to emissions in the other hemisphere, the counterterm makes sure that the Monte Carlo gives a finite result even for $R'\to 0$. From a numerical point of view, this regime is problematic when we have no emissions other than $k_1$ above the cutoff $\epsilon$ and are required to compute $\ln \left|\sum_{i \notin \mathcal{H}_{\ell_1}}\vec{\zeta}_i\right|$. However, in this limit $B_{T, \rm sc}\big(\{\tilde{p}\},k_1,\dots,k_{n+1}\big)/B_T \to 1$ and the theta-constraint of the counterterm will be trivially satisfied, thus the contributions of $\chi_T(R')$ and $\chi^{\textrm{c.t.}}_T(R')$ will cancel perfectly as shown in Fig.~\ref{fig:chiT-chiTbar}.

\begin{figure}[htbp] 
	\centering
	\includegraphics{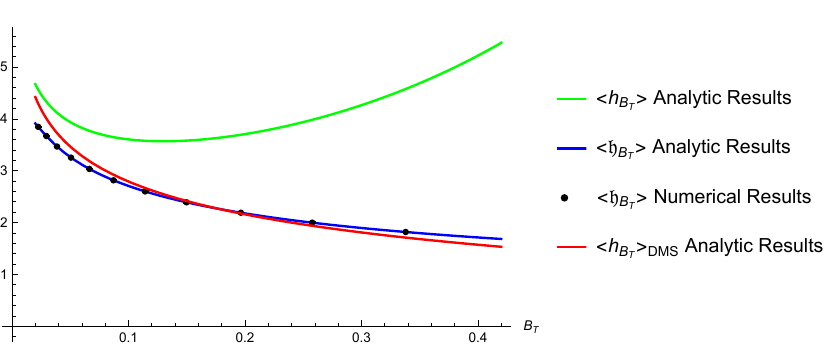}
	\caption{Plot of our numerical and analytical results for $\cNPtilde{B_T}$ and comparisons with the analytic expressions for $\cNP{B_T}$ and $\cNPDMS{B_T}$ in \cite{Dokshitzer:1998qp} (with $\alpha_s = 0.118$).}
	\label{fig:shift-BT-comparison}		
\end{figure}

We now describe how to compute the counterterm in eq.~\eqref{eq:BT-counterterm}. If we do this naively we of course obtain a $1/R'$ divergence from eq.~\eqref{eq:chibarT}. This is because in devising the measure $d\mathcal{Z}[\{R'_{\ell_i},k_i\} ]$ we have neglected all higher derivatives of the radiator. In particular, the second derivative regularises the $1/R'$ divergence. As a result the shift now behaves as $1/\sqrt{\alpha_s}$ and the product of such a contribution with a finite correction of order $\alpha_s$, which is beyond our nominal accuracy, gives a $\sqrt{\alpha_s}$ contribution. Therefore, in the improved version of $\chi^{\textrm{c.t.}}_T$ we can account for all contributions up to order $\sqrt{\alpha_s}$ excluded. In order to do so, we also need to upgrade the expression for $\chi^{\textrm{c.t.}}_T$ to take into account hard-collinear real and virtual corrections. The exact procedure to perform this upgrade is technically involved and is explained in appendix~\ref{sec:NP-counterterms}. The outcome of this procedure is the following improved counterterm
\begin{equation} \label{eq:hBT-ct-imp}
	\cNPctimp{B_T} = \frac12 f_T(R')\ln\frac{1}{B_T} + \chi^{\textrm{c.t.$\,$imp}}_T(R',R^{\prime\prime},R^{(3)}) \ \ ,
\end{equation}
with
\begin{align} \label{eq:chiBT-ct-imp}
	\nonumber
	\chi^{\textrm{c.t.$\,$imp}}_T(R',R^{\prime\prime},R^{(3)}) = \,
	&\frac12 f_T(R') 
	\Bigg[-2 - \rho\bigg(\frac{R'}{2}\bigg) + \psi\bigg(1+\frac{R'}{2}\bigg) + \gamma_E \\ 
	&+ \frac{1}{\sigma\big(\frac{R'}{2}\big)} \left\{\frac{1}{1+\frac{R'}{2}}\,{}_2 F_1\bigg(1,1;2+\frac{R'}{2};-1\bigg) + H_T\Big(R^\prime,R^{\prime\prime},R^{(3)}\Big)\right\}\Bigg] \ \ ,
\end{align}
where
\begin{align} \label{eq:HT}
	\nonumber
	H_T\Big(R^\prime,R^{\prime\prime},R^{(3)}\Big) &= \frac{2\sqrt{\pi} \, s}{R'} \, e^{s^2} \, {\rm Erfc}(s)  + \frac{2}{3}\frac{R^{(3)}}{\left(R^{\prime\prime}\right)^2} \, \frac{\sqrt{\pi}}{8}\frac{d^3}{ds^3} e^{s^2} \, {\rm Erfc}(s)  \\ \nonumber
	&\quad + \left(\gamma_E -\ln2 + \psi\bigg(1+\frac{R'}{2}\bigg)\right) \left(\sqrt{\pi} \, s \, e^{s^2} \, {\rm Erfc}(s) - 1\right) \\
	&\quad - \frac{3C_F\alpha_s}{\pi \, R^{\prime\prime}} \left(\sqrt{\pi} \, s \, e^{s^2} {\rm Erfc}(s) - 1 \right) \ \ , \qquad s \equiv \frac{R'}{2\sqrt{R^{\prime\prime}}} \ \ .
\end{align}
The final expression for the shift for the total broadening is then
\begin{equation}
	\langle\delta B_T\rangle = \frac{\langle \kappa\rangle_{\rm NP}}{Q}\cNPtilde{B_T} \, \mathcal{M} \ \ ,
\end{equation}
where 
\begin{equation} \label{eq:htilde-BT}
	\cNPtilde{B_T} \equiv 
	\cNP{B_T} - \cNPct{B_T} + \cNPctimp{B_T} \ \ .  
\end{equation}
The behaviour of the shift for the total broadening can be appreciated by considering two separate regimes, $R'\gg \sqrt{R^{\prime\prime}}$ and $R'\ll \sqrt{R^{\prime\prime}}$.
As such, we observe
\begin{equation} \label{eq:erfc-two-regimes}
	e^{s^2} \, {\rm Erfc}(s) = 
	\left\{
	\begin{split}
		& \frac{1}{\sqrt{\pi}s} \left(1 - \frac{1}{2s^2} + \frac{3}{s^4} + \dots \right) \qquad\;\;\;\,\,,\; {\rm for}\,\, s \gg 1 \\
		& 1 - \frac{2s}{\sqrt{\pi}} + s^2 - \frac{4s^3}{3\sqrt{\pi}} + \dots \qquad\qquad ,\;{\rm for}\,\,  s \ll 1 \ \ ,
	\end{split}
	\right .
\end{equation}
which gives:
\begin{itemize}
\item $R'\gg \sqrt{R^{\prime\prime}}$: this regime describes the small-$B_T$ region where many emissions populate both hemispheres and we have
\begin{equation} \label{eq:BT-R'R''}
	\cNPtilde{B_T} = 
    \ln\frac{1}{B_T} + \eta_0^{(B)} + \chi_T(R') + \mathcal{O}(\alpha_s) \ \ .
\end{equation}
We see that $\cNPctimp{B_T}$ cancels almost completely against $\cNPct{B_T}$, leaving a finite contribution of order $\alpha_s$ which is beyond our control. Indeed, the precise form of the latter crucially depends on the choice of counterterm, arising from $\cNPctimp{B_T}$.
\item $R'\ll \sqrt{R^{\prime\prime}}$: this regime describes the large-$B_T$ region where the hemisphere that does not contain the emission setting the broadening is almost empty and we obtain 
\begin{equation} \label{eq:BT-R''R'}
	\cNPtilde{B_T} = 
	\frac12\ln\frac{1}{B_T} + \eta_0^{(B)} + \frac12\left[\frac{\pi}{2\sqrt{C_F \,\alpha_s}} + \frac34 - \frac{2\pi\beta_0}{3C_F}\right] + \mathcal{O}(\sqrt{\alpha_s}) \ \ .
\end{equation}
We re-emphasise that, as anticipated, we do not control terms of order $\sqrt{\alpha_s}$ as they could arise from the product of terms of order $\alpha_s$ beyond our nominal accuracy, multiplied by the prefactor $1/\sqrt{\alpha_s}$. 
\end{itemize}
The procedure outlined above to obtain the shift for $B_T$ is similar to the one originally proposed in \cite{Dokshitzer:1998qp}. There, a counterterm was introduced at the level of the integral transforms that were employed to compute the shift. Our procedure is designed to handle a generic observable, in a way that is amenable to an efficient numerical implementation. We plot our results for $\cNPtilde{B_T}$ in Fig.~\ref{fig:shift-BT-comparison} and compare them both to $\cNP{B_T}$ and the analytic expressions in \cite{Dokshitzer:1998qp} (which we refer to as $\cNPDMS{B_T}$).

We first remark that in the region of large $B_T$, corresponding to small $R'$, we have specifically checked that $\cNPtilde{B_T}$ and $\cNPDMS{B_T}$ agree up to corrections of order $\sqrt{\alpha_s}$, which are beyond our nominal accuracy. In the region of small $B_T$, corresponding to large $R'$, we note that our result differs both from \cite{Dokshitzer:1998qp} and the analytic behaviour of $\cNP{B_T}$. This deviation is due to the residual $\mathcal{O}(\alpha_s)$ term in eq~\eqref{eq:BT-R'R''}. This occurs because our method relies on cancelling the $1/R'$ singularity \emph{locally} at the integrand level, which unavoidably leaves a residual $\mathcal{O}(\alpha_s)$ contribution in our final shift. It would be desirable to devise a counterterm which automatically switches off these spurious contributions for $B_T \to 0$, but we leave this for future work. 

At the phenomenological level, the results that we present are valid up to values of $B_T$ that are not too small, which contain the range in which simultaneous fits of $\alpha_s$ and $\alpha_0$ take place and is thus appropriate for the study performed in this paper.

\subsection{Counterterm for $\langle h_{T_M} \rangle$}
\label{subsec:TMsubtraction}
Once we have validated our procedure we are in a position to compute the shift for the thrust major. Fully generally, for $R'\to 0$, one emission, emitted in hemisphere $\mathcal{H}_{\ell}$, has a value of transverse momentum that is much larger than all of the other emissions. In this situation $T_M$ is determined by this emission only, and it is this emission that sets the thrust-major axis. This reduces enormously the complexity of the calculation of the shift for the thrust major and makes it amenable to an analytic calculation. 

\begin{figure}[htbp]
	\centering
	\includegraphics{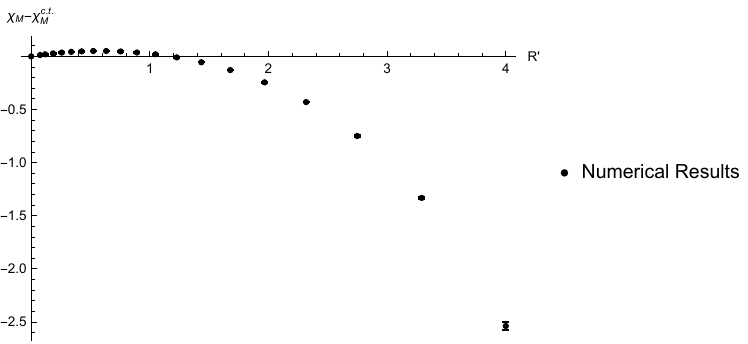}
	\caption{Plot of the numerical determination of $\chi_M(R') - \chi^{\textrm{c.t.}}_M(R')$ and demonstration that the contributions of $\chi_M(R')$ and $\chi^{\textrm{c.t.}}_M(R')$ cancel perfectly in the limit $R'\to 0$.}
	\label{fig:chiM-chiMbar}
\end{figure}

A suitable counterterm that may be computed analytically is constructed as follows 
\begin{align} \label{eq:TM-counterterm}
	\nonumber
	\cNPct{T_M} &= 
	\frac{1}{R' \mathcal{F}_{T_M}(R')} \int dp_{y,\ell} \, dp_{y,\bar{\ell}} \int d\mathcal{Z}[\{R'_{\ell_i},k_i\} ] \, \delta\bigg(1-\frac{2\max_{i}\{k_{ti}\}}{T_M Q}\bigg) \frac{2}{\pi} \ln\frac{Q}{|p_{y,\bar{\ell}}|} \, \times \\ \nonumber
	&\quad\times \delta\Big(p_{y,\ell}+k_{t,\max}\Big) \, \delta\Bigg(p_{y,\bar{\ell}}+\sum_{i\in \mathcal{H}_{\bar{\ell}}} k_{y,i}\Bigg) \, \Theta\Bigg(\max_{i}\{k_{ti}\} - |p_{y,\bar{\ell}}| - \sum_{i\in \mathcal{H}_{\bar{\ell}}} k_{t,i}\Bigg) \\ \nonumber
	&= \frac{1}{R' \mathcal{F}_{T_M}(R')} \int \, dp_{y,\bar{\ell}} \int d\mathcal{Z}[\{R'_{\ell_i},k_i\} ] \, \delta\bigg(1-\frac{2\max_{i}\{k_{ti}\}}{T_M Q}\bigg) \frac{2}{\pi} \ln\frac{Q}{|p_{y,\bar{\ell}}|} \, \times \\ 
	&\quad\times \delta\Bigg(p_{y,\bar{\ell}}+\sum_{i\in \mathcal{H}_{\bar{\ell}}} k_{y,i}\Bigg) \, \Theta\Bigg(\max_{i}\{k_{ti}\} - |p_{y,\bar{\ell}}| - \sum_{i\in \mathcal{H}_{\bar{\ell}}} k_{t,i}\Bigg) \ \ ,  
\end{align}
where again $\big|\vec{k}_{t,\max}\big| = \max_i\{k_{ti}\}$ and the $y$-direction is along $\vec{k}_{t,\max}$.
In a similar way as for $\cNPct{B_T}$, we can introduce the rescaled variables $x_{\ell}=2p_{y,\ell}/(T_M Q)$ and the two-dimensional vectors $\vec\zeta_i\equiv \zeta_i(\cos\phi_i,\sin\phi_i)$ and obtain
\begin{equation} \label{eq:hTMct-rescaled}
	\cNPct{T_M} = 
	\frac{2}{\pi}f_M(R')\ln\frac{2}{T_M} + \chi^{\textrm{c.t.}}_M(R') \ \ ,
\end{equation}
where
\begin{equation} \label{eq:f_M(R')}
	f_M(R') \equiv 
	\frac{1}{\mathcal{F}_{T_M}(R')} \, \rho_1\bigg(\frac{R'}{2}\bigg) \frac{e^{- \gamma_E \frac{R'}{2}}}{\Gamma\big(1+\frac{R'}{2}\big)} \ \ ,
\end{equation}
with $f_M(0) = 1$, the function $\rho_1(R'/2)$ defined in eq.~\eqref{eq:b2integral1} and 
\begin{align} \label{eq:chibarM}
		\nonumber
		\chi^{\textrm{c.t.}}_M(R') = \,
		&\frac{1}{R' \mathcal{F}_{T_M}(R')} \int dx_{\bar{\ell}} \int d\mathcal{Z}[\{R'_{\ell_i},k_i\} ] \, \delta\bigg(1-\max_{i}\{\zeta_i\}\bigg) \frac{2}{\pi} \ln\frac{1}{|x_{\bar{\ell}}|} \, \times \\ 
		&\times \delta\Bigg(x_{\bar{\ell}}+\sum_{i\in \mathcal{H}_{\bar{\ell}}} \zeta_{i}\sin\phi_{i,\max}\Bigg) \, \Theta\Bigg(1 - |x_{\bar{\ell}}| - \sum_{i\in \mathcal{H}_{\bar{\ell}}} \zeta_{i}\Bigg) \ \ , 
\end{align}
where $\phi_{i,\max}$ denotes the angle between $\vec{k}_{t,i}$ and $\vec{k}_{t,\max}$. Such counterterm has the same $1/R'$ singularity as $\cNP{T_M}$ and is constructed so that it may be computed analytically, as is performed in appendix~\ref{subsec:ct-TM}. Therefore, the combination $\cNP{T_M} - \cNPct{T_M}$ is 
free of singularities for all values of $R'$ with the cancellation of singularities occurring \emph{locally}, i.e.\ for each Monte Carlo configuration used to compute $\cNP{T_M} - \cNPct{T_M}$. What we obtain is
\begin{equation} \label{eq:hTM-hTMct}
	\cNP{B_T} - \cNPct{B_T} 
	= \frac{4}{\pi}\bigg[1 - \frac12 f_M(R')\bigg] \ln\frac{2}{T_M} + \frac{4}{\pi}\left(\ln 2 - 2\right) + \chi_M(R') - \chi^{\textrm{c.t.}}_M(R') \ \ .
\end{equation}
The function $\chi^{\textrm{c.t.}}_M(R')$ is computed analytically in appendix~\ref{subsec:ct-TM} and its expression can be found in eq.~\eqref{eq:chibarMNLL}. We may also rewrite $\chi^{\textrm{c.t.}}_M(R')$ in a way that is suitable for Monte Carlo integration, following the strategy described in appendix~\ref{sec:monte-carlo-determ}. We find 
\begin{align} \label{eq:chiTM-chibarTM}
	\nonumber
	&\chi_M(R')-\chi^{\textrm{c.t.}}_M(R') = \,
	\frac{2}{\pi \, \mathcal{F}_{T_M}(R')} \sum_{\ell_1=1,2}  \frac{R'_{\ell_1}}{R'} \int_0^{2\pi} \! \frac{d\phi_1}{2\pi} \left( \epsilon^{R'}\sum_{n=0}^\infty \frac{1}{n!}  \prod_{i=2}^{n+1}\sum_{\ell_i=1,2} R'_{\ell_i}\int^{1}_{\epsilon}\frac{d\zeta_i}{\zeta_i}\int_0^{2\pi}\frac{d\phi_i}{2\pi}\right) \times \\ \nonumber
	&\times \Bigg[\left(\frac{T_{M, \rm sc}\big(\{\tilde{p}\},k_1,\dots,k_{n+1}\big)}{T_M}\right)^{-R'} \Bigg\{\ln\left(\frac{T_{M, \rm sc}\big(\{\tilde{p}\},k_1,\dots,k_{n+1}\big)}{T_M}\right)^{2} + \ln\frac{1}{\left|\sum_{i \in \mathcal{H}_{\ell_1}}\zeta_i\sin\phi_i\right|} \\
	&\qquad+ \ln\frac{1}{\left|\sum_{i \notin \mathcal{H}_{\ell_1}}\zeta_i\sin\phi_i\right|}\Bigg\} - \ln\frac{1}{\left|\sum_{i \notin \mathcal{H}_{\ell_1}}\zeta_i\sin\phi_{i,\max}\right|} \, \Theta\Bigg(1 - \Bigg|\sum_{i \notin \mathcal{H}_{\ell_1}}\zeta_i\sin\phi_{i,\max}\Bigg| - \sum_{i \notin \mathcal{H}_{\ell_1}}\zeta_i\Bigg)\Bigg]	\ \ ,
\end{align}
where, as explained in appendix~\ref{sec:monte-carlo-determ}, now $T_{M, \rm sc}(\{\tilde{p}\},k_1) = T_M$. We observe that $\chi_M(R')-\chi^{\textrm{c.t.}}_M(R')$ is finite for all values of $R'$. When the calculation involves the recoil due to emissions in the same hemisphere as emission $k_1$, i.e.\ hemisphere $\mathcal{H}_{\ell_1}$, the counterterm $\chi^{\textrm{c.t.}}_M(R')$ has no effect whatsoever. When the calculation involves the recoil due to emissions in the other hemisphere, the counterterm makes sure that the Monte Carlo gives a finite result even for $R'\to 0$. From a numerical point of view, without a counterterm this regime would be problematic when we have no emissions other than $k_1$ above the cutoff $\epsilon$ and are required to compute $\ln \left|\sum_{i \notin \mathcal{H}_{\ell_1}}\zeta_i\sin\phi_i\right|$. However, in this limit $T_{M, \rm sc}\big(\{\tilde{p}\},k_1,\dots,k_{n+1}\big)/T_M \to 1$, $\phi_i \to \phi_{i,\max}$ and the theta-constraint of the counterterm will be trivially satisfied, thus the contributions of $\chi_M(R')$ and $\chi^{\textrm{c.t.}}_M(R')$ will cancel perfectly as shown in Fig.~\ref{fig:chiM-chiMbar}.

\begin{figure}[htbp] 
	\centering
	\includegraphics{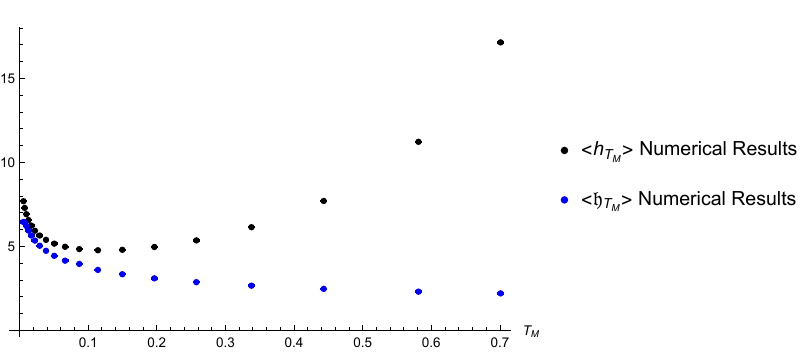}
	\caption{Plot of our numerical results for $\cNPtilde{T_M}$ and $\cNP{T_M}$ (with $\alpha_s = 0.118$).}
	\label{fig:shift-TM-full}
\end{figure}

Performing a similar upgrade as for the total broadening we obtain the needed, improved expression for $\cNPctimp{T_M}$, computed in appendix~\ref{subsec:ct-TM} and repeated here 
\begin{equation} 
	\cNPctimp{T_M} = 
	\frac{2}{\pi} f_M(R')\ln\frac{2}{T_M} + \chi^{\textrm{c.t.$\,$imp}}_M(R',R^{\prime\prime},R^{(3)}) \ \ ,
\end{equation}
with
\begin{align}
	\nonumber
	\chi^{\textrm{c.t.$\,$imp}}_M(R',R^{\prime\prime},R^{(3)}) = \, 
	&\frac{2}{\pi \, \mathcal{F}_{T_M}(R')} \, \frac{e^{-\gamma_E\frac{R'}{2}}}{\Gamma\big(1+\frac{R'}{2}\big)} \Bigg[\bigg\{\gamma_E + \psi\bigg(1+\frac{R'}{2}\bigg)\bigg\}\rho_1\bigg(\frac{R'}{2}\bigg) + \rho_3\bigg(\frac{R'}{2}\bigg) \\ 
	&+ \frac{1}{1+\frac{R'}{2}}\,{}_2 F_1\bigg(1,1;2+\frac{R'}{2};-1\bigg) + H_M\Big(R^\prime,R^{\prime\prime},R^{(3)}\Big)\Bigg] \ \ ,
\end{align} 
where
\begin{align}\label{eq:HM}
	\nonumber
	H_M\Big(R^\prime,R^{\prime\prime},R^{(3)}\Big) &= \frac{2\sqrt{\pi} \, s}{R'} \,e^{s^2} \, {\rm Erfc}(s)  + \frac{2}{3}\frac{R^{(3)}}{\left(R^{\prime\prime}\right)^2} \, \frac{\sqrt{\pi}}{8}\frac{d^3}{ds^3} e^{s^2} \, {\rm Erfc}(s)  \\ \nonumber
	&\quad + \left(\gamma_E + \psi\bigg(1+\frac{R'}{2}\bigg)\right) \left(\sqrt{\pi} \, s \, e^{s^2} \, {\rm Erfc}(s) - 1\right) \\
	&\quad - \frac{3C_F\alpha_s}{\pi\, R^{\prime\prime}} \left(\sqrt{\pi} \, s \, e^{s^2}  {\rm Erfc}(s) - 1 \right) \ \ , \qquad s \equiv \frac{R'}{2 \sqrt{R^{\prime\prime}}} \ \ .
\end{align}
The final expression of the shift for the thrust major is given by 
\begin{equation}
	\langle\delta T_M\rangle = 
	\frac{\langle \kappa\rangle_{\rm NP}}{Q}\cNPtilde{T_M} \, \mathcal{M} \ \ ,
\end{equation}
where 
\begin{equation} \label{eq:htilde-TM}
	\cNPtilde{T_M}\equiv  
	\cNP{T_M} - \cNPct{T_M} + \cNPctimp{T_M} \ \ .
\end{equation}
The behaviour of the shift for the thrust major can be appreciated by considering two separate regimes using eq.~\eqref{eq:erfc-two-regimes}:
\begin{itemize}
	\item $R'\gg \sqrt{R^{\prime\prime}}$: this regime describes the small-$T_M$ region, where many emissions populate both hemispheres and we have
	\begin{equation} \label{eq:TM-R'R''}
		\cNPtilde{T_M} = 
			\frac{4}{\pi}\left(\ln\frac{2}{T_M} + \ln2 - 2 \right) + \chi_M(R') + \mathcal{O}(\alpha_s) \ \ .
	\end{equation}
	We see that $\cNPctimp{T_M}$ cancels almost completely against $\cNPct{T_M}$, leaving a finite contribution of order $\alpha_s$ which is beyond our control. Indeed, the precise form of the latter crucially depends on the choice of counterterm, arising from $\cNPctimp{T_M}$.
	\item $R'\ll \sqrt{R^{\prime\prime}}$: this regime describes the large-$T_M$ region where the hemisphere that does not contain the emission that sets the thrust-major axis is almost empty and we obtain 
	\begin{equation} \label{eq:TM-R''R'}
		\cNPtilde{T_M} = 
		 \frac 4\pi\left[\frac{1}{2}\ln\frac{1}{T_M}+2\ln 2 -2+ \frac{1}{2}\left(\frac{\pi}{2\sqrt{C_F\,\alpha_s}}+\frac{3}{4}-\frac{2\pi\beta_0}{3C_F}\right) + \frac{\ln2}{2}\right] + \mathcal{O}(\sqrt{\alpha_s}) \ \ .
	\end{equation}
Note that we do not control terms of order $\sqrt{\alpha_s}$ as they could arise from the product of unknown terms of order $\alpha_s$ multiplied by the prefactor $1/\sqrt{\alpha_s}$. 
\end{itemize}
The function $\cNPtilde{T_M}$ is plotted in Fig.~\ref{fig:shift-TM-full}. 
As for the total broadening we notice that the 1/$R'$ divergence has been replaced with a constant value for large values of $T_M$, while at very small values of $T_M$ discrepancies between $\cNPtilde{T_M}$ and $\cNP{T_M}$ will be of order $\alpha_s$ and beyond our control. 

We conclude by commenting on the fact that the exact behaviour of the shift at small values of $T_M$ (the black dots in Fig.~\ref{fig:shift-TM-full}) is perfectly under control. The fact that $\cNPtilde{T_M}$ does not tend to $\cNP{T_M}$ for $T_M \to 0$ is due to a residual $\mathcal{O}(\alpha_s)$ term, as was explained in the case for the total broadening. 

\section{Mean values}
\label{sec:means}

With the formalism we have developed so far we can also compute the non-perturbative corrections to the mean values of event shapes. The mean value of an event shape is defined as
\begin{equation} \label{eq:mean}
	\langle v \rangle = 
	\int_0^{v_{\max}} \!\!\!\! dv \, v\frac{d}{dv}\Sigma(v) \ \ .
\end{equation}
Substituting $\Sigma(v) = \Sigma_{\rm PT}(v) + \Sigma_{\rm NP}(v)$, from eq.~\eqref{eq:Sigmatot}, into the above equation, we obtain 
\begin{equation} \label{eq:mean-separated}
	\langle v \rangle  = 
	\langle v \rangle_{\rm PT} + \langle v \rangle_{\rm NP} \ \ , 
\end{equation}
where
\begin{equation} \label{eq:mean-PT-NP}
	\begin{split}
		\langle v \rangle_{\rm PT} &\equiv 
		\int_0^{v_{\max}} \!\!\!\! dv \, v\frac{d}{dv}\Sigma_{\rm PT}(v) \ \ , \\ 
		\langle v \rangle_{\rm NP} &\equiv \int_0^{v_{\max}} \!\!\!\! dv \, v\frac{d}{dv}\delta\Sigma_{\rm NP}(v) \ \ ,
\end{split}
\end{equation}
with $\delta \Sigma_{\rm NP}$ given in eq.~\eqref{eq:SigmaNP}. The perturbative element of the mean value, $\langle v \rangle_{\rm PT}$, can be computed as an expansion in powers of $\alpha_s$. The non-perturbative element, $\langle v \rangle_{\rm NP}$, can be simplified by performing an integration by parts
\begin{equation} \label{eq:mean-NP-simp}
	\langle v \rangle_{\rm NP} = 
	v \, \delta\Sigma_{\rm NP}(v) \bigg|_0^{v_{\max}} - \int_0^{v_{\max}} \!\!\!\! dv \, \delta\Sigma_{\rm NP}(v) \ \ .
\end{equation}
We can now use the fact that $v_{\max}$ is the same for both perturbative and non-perturbative distributions to set the first term on the right-hand side of the above equality to zero. In the second term we can substitute the expression for $\delta\Sigma_{\rm NP}(v)$
in eq.~\eqref{eq:SigmaNP} and approximate the resulting expression in terms of $\delta V_{\rm NP}$ defined in eq.~\eqref{eq:dV-NP}
\begin{equation} \label{eq:mean-NP-expanded}
	\langle v \rangle_{\rm NP}\simeq 
	\int [dk] \, M^2_{\rm NP}(k) \, \int dZ[\{k_i\}] \, \delta V_{\rm NP}(\{\tilde p\},k,\{k_i\}) \, \Theta\Big(v_{\max} - V(\{\tilde p\},\{k_i\})\Big) \ \ .
\end{equation}
Using the assumption in eq.~\eqref{eq:M2NP}, inserting the expression for $\delta V_{\rm NP}(\{\tilde p\},k,\{k_i\} )$ in eq.~\eqref{eq:deltaV} and assuming that the ultra-soft emission $k$ is accompanied by perturbative emissions that are soft, collinear and widely separated in angle, we obtain 
\begin{equation} \label{eq:mean-NP-simplified}
	\langle v \rangle_{\rm NP} \simeq 
	\mathcal{M}\frac{\langle \kappa \rangle_{\rm NP}}{Q} \sum_{\ell} \int \!d\eta^{(\ell)} \, \frac{d\phi}{2\pi} \int dZ_{\rm sc}[\{k_i\}] \, h_V\big(\eta^{(\ell)},\phi,\{\tilde{p}\},\{k_i\}\big) \, \Theta\Big(v_{\max} - V_{\rm sc}(\{\tilde{p}\},\{k_i\})\Big) \ \ .
\end{equation}
We can now compute the NP corrections to the mean values of the event-shape variables considered in section \ref{sec:observables}.

\paragraph{Thrust, $C$-parameter, heavy-jet mass.} In these cases $h_V(\eta^{(\ell)},\phi,\{k_i\} )=h_V(\eta^{(\ell)},\phi)$, which gives
\begin{equation} \label{eq:mean-NP-simplified-additive}
	\langle v \rangle_{\rm NP} = 
	\mathcal{M}\frac{\langle \kappa \rangle_{\rm NP}}{Q} \sum_{\ell} \int\!d\eta^{(\ell)} \, \frac{d\phi}{2\pi} \, h_V(\eta^{(\ell)},\phi) \int dZ_{\rm sc}[\{k_i\}] \, \Theta\Big(v_{\max} - V_{\rm sc}(\{\tilde{p}\},\{k_i\})\Big) = 
	\mathcal{M}\frac{\langle \kappa \rangle_{\rm NP}}{Q}\cNP{V} \ \ .
\end{equation}

\paragraph{Jet broadenings and thrust major.}
Starting from eq.~\eqref{eq:mean-NP-simplified}, the NP corrections to the mean values of the jet broadenings and thrust major (unlike the shift for the thrust major distribution) may be computed fully analytically. Since no novel numerical procedure is required, we leave the details of the calculation to appendix~\ref{sec:NP-means}. Here we report the final results: 
\begin{align} 
	\label{eq:BW-mean-value-result}
	\langle B_W \rangle_{\rm NP} &\simeq 
	\mathcal{M}\frac{\langle \kappa \rangle_{\rm NP}}{Q} \, \frac12 \left[\eta_0^{(B)} + \frac{\pi}{2\sqrt{2C_F \, \alpha_s(Q)}} + \frac34 - \frac{\pi\beta_0}{3 C_F}\right] + \mathcal{O}(\sqrt{\alpha_s}) \ \ , \\
	\label{eq:BT-mean-value-result}
	\langle B_T \rangle_{\rm NP} &\simeq 
	\mathcal{M}\frac{\langle \kappa \rangle_{\rm NP}}{Q} \, \left[\eta_0^{(B)} + \frac{\pi}{2\sqrt{C_F \, \alpha_s(Q)}} + \frac34 - \frac{2\pi\beta_0}{3 C_F}\right] + \mathcal{O}(\sqrt{\alpha_s}) \ \ , \\
	\label{eq:TM-mean-value-result}
	\langle T_M \rangle_{\rm NP} &\simeq
	\mathcal{M}\frac{\langle \kappa \rangle_{\rm NP}}{Q} \, \frac{4}{\pi}\left[2\ln 2 - 2 + \frac{\pi}{2\sqrt{C_F \, \alpha_s(Q)}} + \frac34 - \frac{2\pi\beta_0}{3 C_F} + \frac{\ln 2}{2}\right] + \mathcal{O}(\sqrt{\alpha_s}) \ \ .		
\end{align}
We note that we do not control terms of order $\sqrt{\alpha_s}$ and to this accuracy eqs.~\eqref{eq:BW-mean-value-result} and \eqref{eq:BT-mean-value-result} agree with the corresponding results in \cite{Dokshitzer:1998qp}. Comparing results explicitly, the sole difference is the replacement of the $\sqrt{\alpha_s(Q)}$ term in the denominator of eqs.~\eqref{eq:BW-mean-value-result} and \eqref{eq:BT-mean-value-result} with $\sqrt{\alpha_{s, \rm CMW}(\bar{Q})}$, with $\bar{Q} = Qe^{-3/4}$, in \cite{Dokshitzer:1998qp}. This is highlighted in \cite{Dokshitzer:1998qp} and results in a difference of order $\sqrt{\alpha_s}$ which is not formally under control. 

\section{Phenomenology}
\label{sec:pheno}

\subsection{Mean Values}
\label{subsec:means}
We take the non-perturbative corrections to the mean, $\langle v \rangle_{\rm NP}$, computed in section~\ref{sec:means} for the various event-shape observables of interest and add these to $\langle v \rangle_{\rm PT}$, the perturbative element of the mean value. This provides the full mean value, $\langle v \rangle = \langle v \rangle_{\rm PT} + \langle v \rangle_{\rm NP}$, for which we perform simultaneous fits for $\alpha_s(m_Z)$ and $\alpha_0(2 \,\mathrm{GeV})$ to data. The PT element of the mean value is given by
\begin{equation} \label{eq:mean-PT}
	\langle v \rangle_{\rm PT} = 
	\mathcal{A}_V\left(\frac{\alpha_s}{2\pi}\right) + \left(\mathcal{B}_V-2\mathcal{A}_V\right)\left(\frac{\alpha_s}{2\pi}\right)^2 \ \ ,
\end{equation}
with $\mathcal{A}_V$ and $\mathcal{B}_V$ the $\mathcal{O}\left(\alpha_s\right)$ and $\mathcal{O}\left(\alpha_s^2\right)$ fixed-order perturbative coefficients. $\mathcal{A}_V$ and $\mathcal{B}_V$ are known analytically for $C$ \cite{Ellis:1980wv} and have otherwise been determined using the program \textsc{EVENT2} \cite{Catani:1996jh}. The values used have been set out in Table~\ref{table:PT-coefficients}.
\begin{table}[h!]
	\centering
	\begin{tabular}{| c | c | c |} 
		\hline
		Observable & $\mathcal{A}_V$ & $\mathcal{B}_V$ \\ [0.5ex] 
		\hline
		$1-T$ & $2.103$ & $44.892$ \\
		\hline
		$C$ & $8.638$ & $164.120$ \\
		\hline
		$\rho_H$ & $2.103$ & $23.208$ \\
		\hline
		$B_W$ & $4.067$ & $-10.555$ \\
		\hline
		$B_T$ & $4.067$ & $63.784$ \\ 
		\hline
		$T_M$ & $8.134$ & $59.459$ \\ 
		\hline
	\end{tabular}
	\caption{Fixed-order perturbative coefficients of the event-shape observables taken from \cite{Ellis:1980wv} for $C$ and otherwise determined using \textsc{EVENT2} \cite{Catani:1996jh}. These are used in performing the simultaneous fits for $\alpha_s$ and $\alpha_0$.}
	\label{table:PT-coefficients}
\end{table}

\begin{figure}[h!]
\centering
\includegraphics[width=0.65\textwidth]{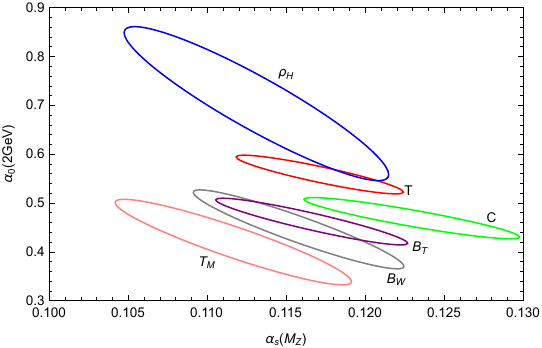}
\caption{95\% confidence level contours for the fitted values of $\alpha_s$ and $\alpha_0$ to experimental data for the mean values of event-shape observables.}
\label{fig:contour-mean}
\end{figure}

\begin{figure}
  \begin{subfigure}{\linewidth}
  \includegraphics[width=.5\linewidth]{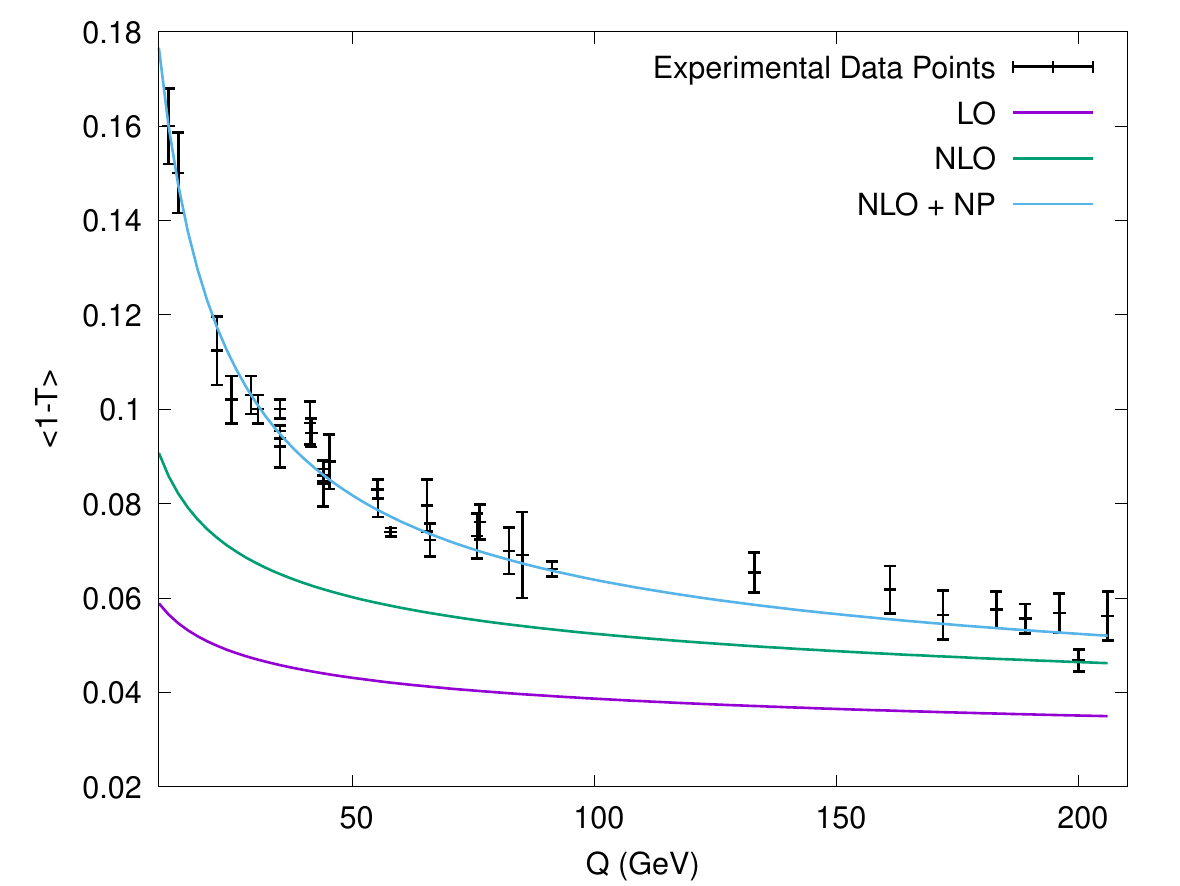}\hfill
  \includegraphics[width=.5\linewidth]{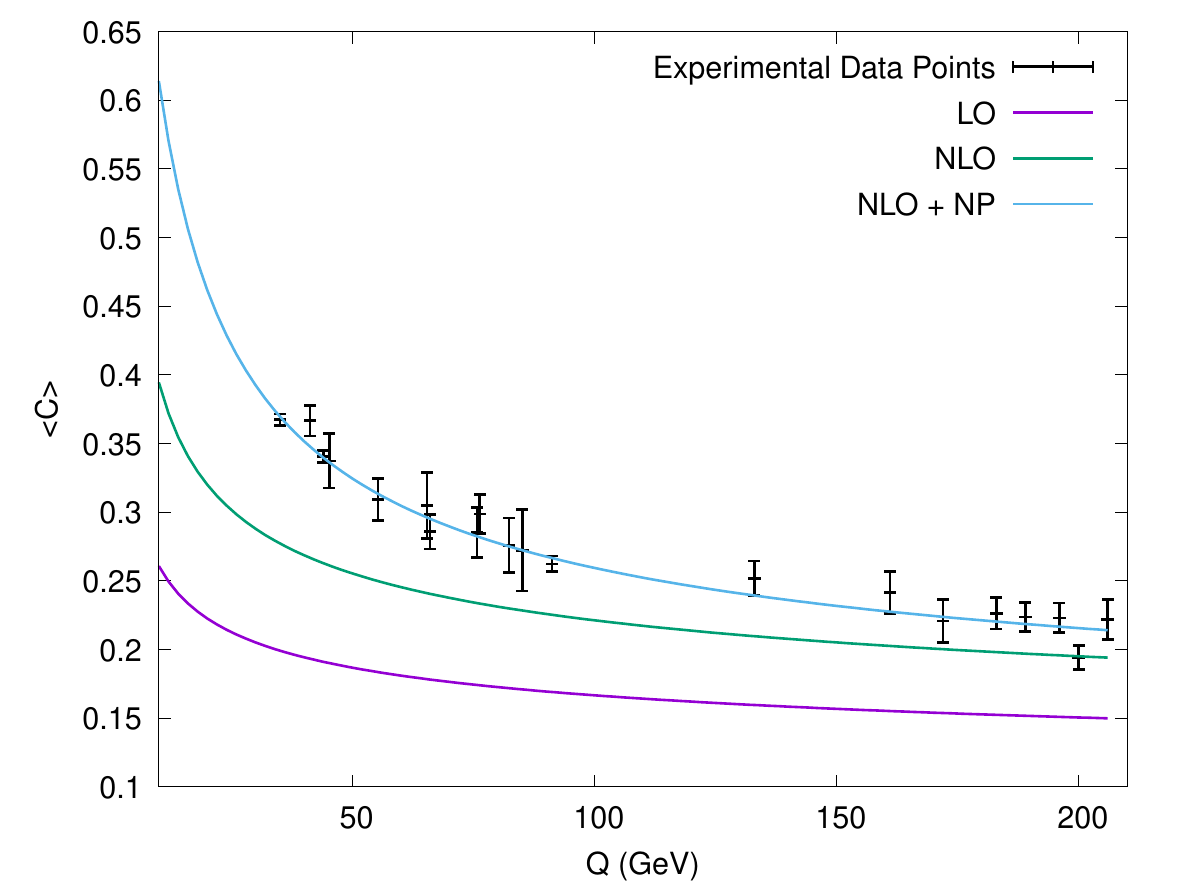}
  \end{subfigure}\par\medskip
  \begin{subfigure}{\linewidth}
  \includegraphics[width=.5\linewidth]{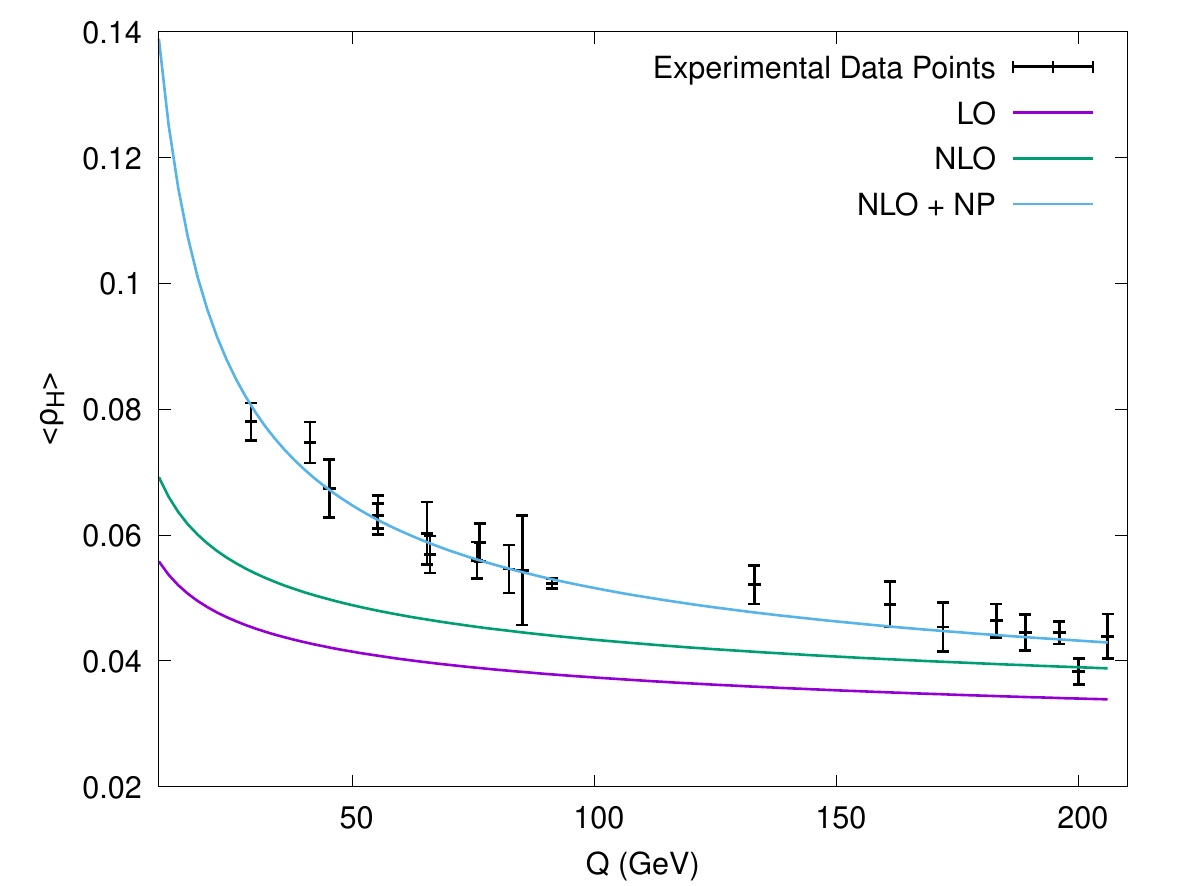}\hfill
  \includegraphics[width=.5\linewidth]{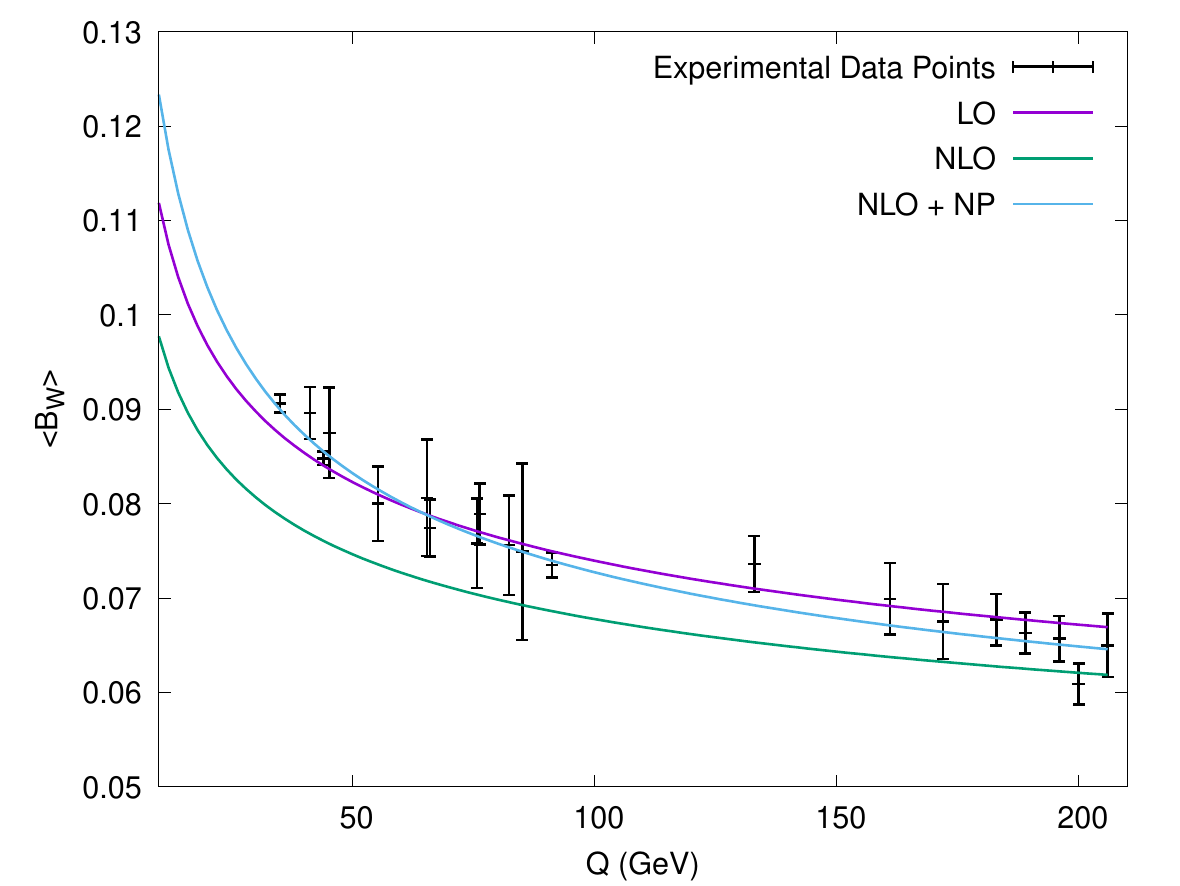}\hfill
  \end{subfigure}\par\medskip
  \begin{subfigure}{\linewidth}
  \includegraphics[width=.5\linewidth]{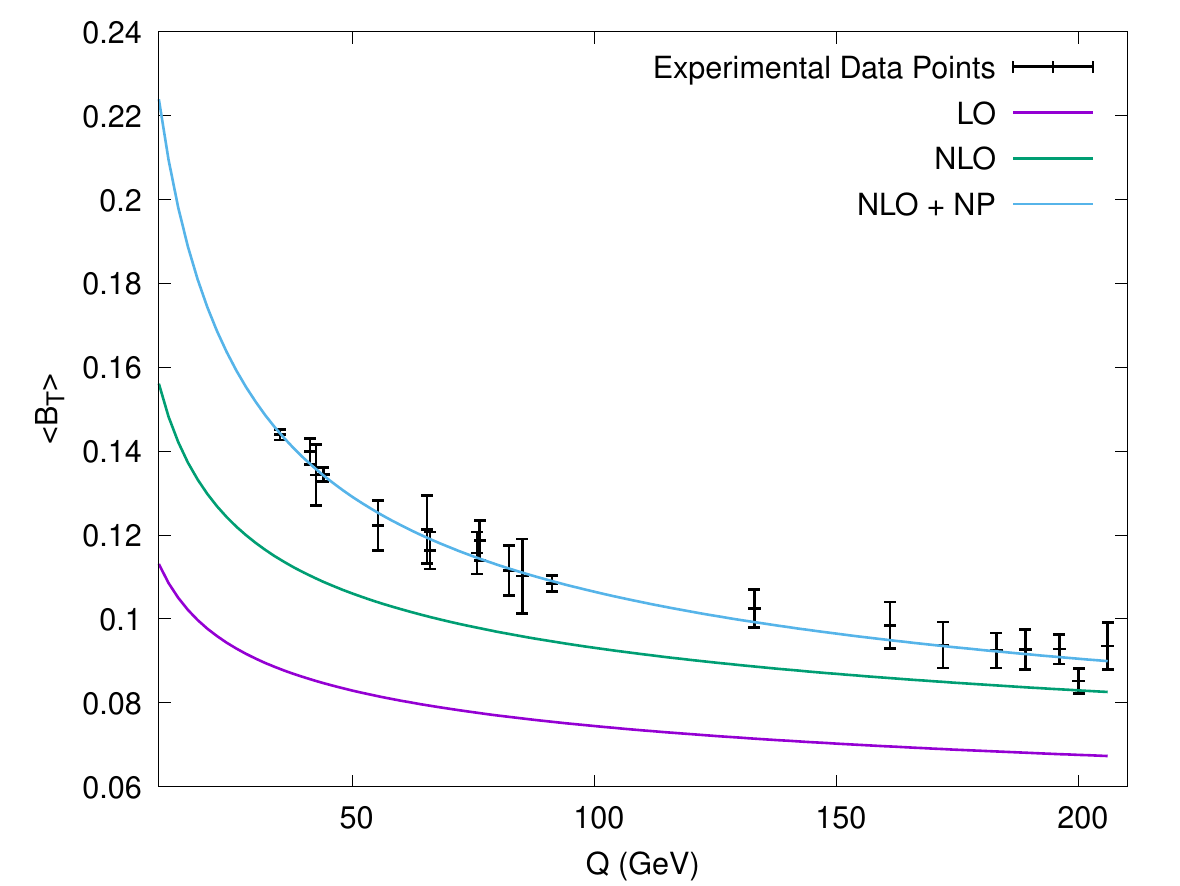}\hfill
  \includegraphics[width=.5\linewidth]{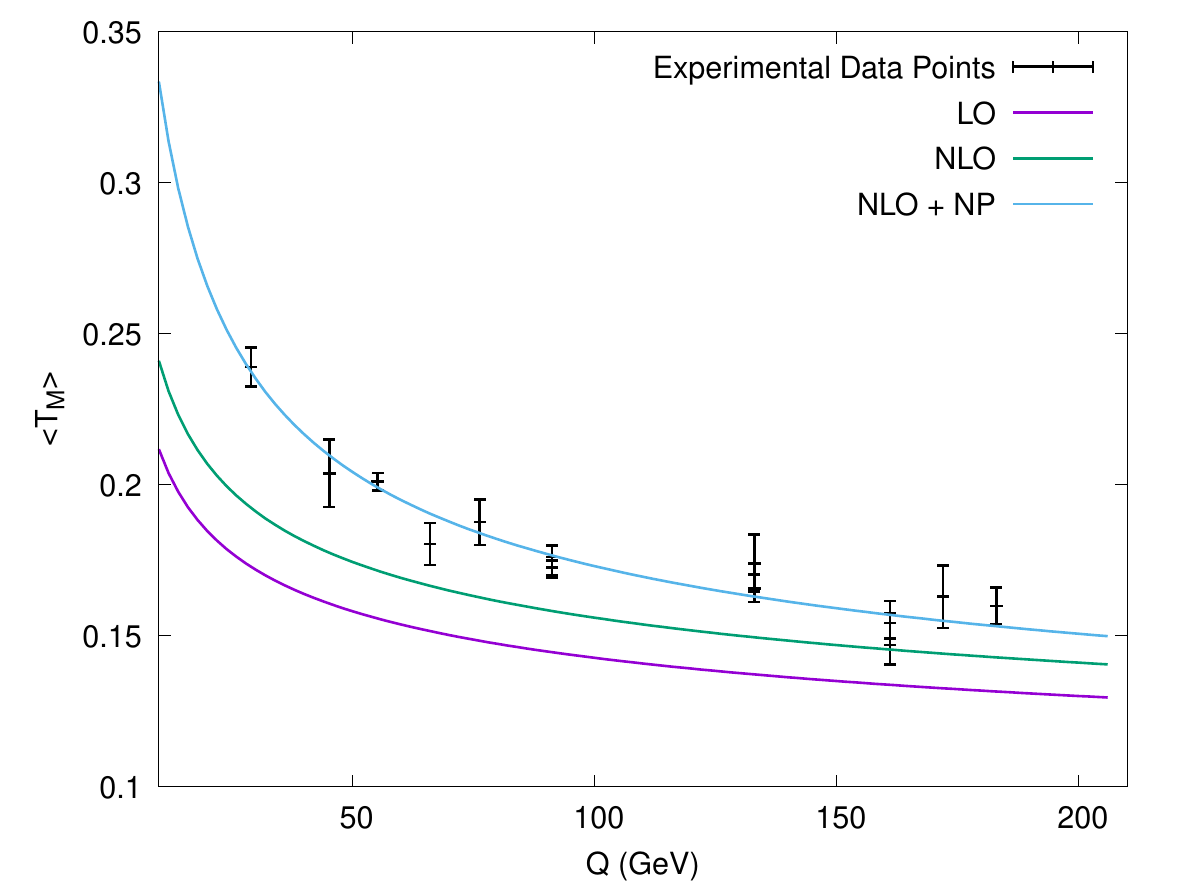}\hfill
  \end{subfigure}
  \caption{Plots of the energy dependence of the mean values of event-shape observables using the fitted values of $\alpha_s$ and $\alpha_0$ from Table~\ref{table:dist-fits} and compared with experimental data.}
\label{fig:mean-energy_dependence}
\end{figure}

The latest ALEPH QCD publication \cite{ALEPH:2003obs} provides references to the data \cite{TASSO:1990cdg,TASSO:1983cre,TASSO:1989kdk,PLUTO:1981inb,MovillaFernandez:1997fr,Biebel:1999zt,CELLO:1989okb,Bender:1984fp,Petersen:1987bq,AMY:1989feg,TOPAZ:1992wgt,DELPHI:2003yqh,L3:2000shd} used to perform the simultaneous fits for the means of $1-T$, $C$, $\rho_H$, $B_W$ and $B_T$. Where possible we have identified the exact same data, however it has not been possible to obtain 13 of the data points used for the fits for $1-T$ and 22 of the data points used for the fits for $\rho_H$. For $T_M$, as this observable was not considered in \cite{ALEPH:2003obs}, we have identified all available data for $T_M$ that we were able to find in the literature \cite{ALEPH:website,Petersen:1987bq,AMY:1989feg,DELPHI:2003yqh,DELPHI:1999vbd,L3:1992nwf,OPAL:1996fae,OPAL:1997asf}. We note that there is considerably less data available for $T_M$ than was used in the fits for the other event-shape observables in \cite{ALEPH:2003obs}. With this data we have performed the simultaneous fits for $\alpha_s$ and $\alpha_0$ with the results set out in Table~\ref{table:mean-fits}. Using the results from Table~\ref{table:mean-fits} we have plotted the 95\% CL contours for the event-shape means in Fig.~\ref{fig:contour-mean} and the energy dependence of the mean values of the event-shape observables, comparing LO perturbative, NLO perturbative and NLO+NP predictions with data in Fig.~\ref{fig:mean-energy_dependence}. 
\begin{table}[h!]
	\centering
	\begin{tabular}{| c | c | c | c |} 
		\hline
		Observable & $\alpha_s(m_Z)$ & $\alpha_0(2\,\mathrm{GeV})$ & $\chi^2/\rm{d.o.f}$ \\ [0.5ex]
		\hline
		$1-T$ & $0.1171\pm0.0022$ & $0.558\pm0.016$ & $57.3/(35-2)$ \\ 
		\hline
		$C$ & $0.1230\pm0.0028$ & $0.469\pm0.017$ & $15.3/(20-2)$ \\
		\hline
		$\rho_H$ & $0.1131\pm0.0034$ & $0.703\pm0.064$ & $15.4/(20-2)$ \\
		\hline
		$B_W$ & $0.1158\pm0.0027$ & $0.446\pm0.033$ & $10.8/(20-2)$ \\
		\hline
		$B_T$ & $0.1166\pm0.0025$ & $0.462\pm0.020$ & $7.6/(20-2)$ \\
		\hline
		$T_M$ & $0.1117\pm0.0031$ & $0.420\pm0.036$ & $9.2/(13-2)$ \\
		\hline
	\end{tabular}
	\caption{Results for the simultaneous fits for $\alpha_s$ and $\alpha_0$ to experimental data for the mean values of event-shape observables.}
	\label{table:mean-fits}
\end{table} 

For $1-T$ (despite the 13 data points that we have been unable to obtain), $C$ and $B_T$ (despite the $\mathcal{O}(\sqrt{\alpha_s})$ difference between our result for $\langle B_T \rangle$ and that of \cite{Dokshitzer:1998qp}) our simultaneous fitted values for $\alpha_s$ and $\alpha_0$ lie within the 95\% confidence level contours deduced from the fitted values and associated errors calculated by the ALEPH collaboration \cite{ALEPH:2003obs}. For $\rho_H$ and $B_W$ however, our simultaneous fitted values for $\alpha_s$ and $\alpha_0$ do not lie within the 95\% confidence level contours deduced from \cite{ALEPH:2003obs}. 

For $B_W$, this is due to the $\mathcal{O}(\sqrt{\alpha_s})$ difference between our result for $\langle B_W \rangle_{\rm NP}$ and that of \cite{Dokshitzer:1998qp}. This has been confirmed by reperforming the simultaneous fitted values for $\alpha_s$ and $\alpha_0$ but using the formula for $\langle B_W \rangle_{\rm NP}$ in \cite{Dokshitzer:1998qp}. Doing so we obtain $\alpha_s = 0.1157\pm0.0028$, $\alpha_0 = 0.476\pm0.038$ and $\chi^2 = 10.8/(20-2)$ which does indeed lie within the deduced ALEPH 95\% confidence level contour. 

For $\rho_H$, this is due to the 22 data points (at low centre-of-mass energies) that we have been unable to find in public repositories. This has resulted in a noticeably large 95\% confidence level contour in Fig.~\ref{fig:contour-mean} which is particularly driven by the uncertainty in $\alpha_0$. We note that the $\alpha_s$ and $\alpha_0$ values for $\langle \rho_H \rangle$ computed in \cite{ALEPH:2003obs} lie within our large 95\% confidence level contour. It is our belief that including the 22 data points would both reduce the size of our contour and also produce simultaneous fitted values for $\alpha_s$ and $\alpha_0$ that lie within the deduced ALEPH 95\% confidence level contours.      

Values of $\alpha_s$ between 0.1117 and 0.1230 are compatible with the Particle Data Group's world average value of 0.118 \cite{Workman:2022ynf}. The values of $\alpha_0$ lie close to 0.5 apart from for $\rho_H$ which is noticeably higher. The large $\alpha_0$ value for $\rho_H$ is in part due to the low centre-of-mass energy data points that have not been obtained (cf.~\cite{ALEPH:2003obs} which included these data points and found $\alpha_0 = 0.627$). This may also be due to the effect of hadron masses, as discussed in \cite{Salam:2001bd}, which we neglect in this preliminary study.

In Fig.~\ref{fig:contour-mean}, the $95\%$ confidence level contours for all 6 observables show a strong negative correlation between the fitted values for $\alpha_s$ and $\alpha_0$. We note that the contours for $1-T$, $C$, $B_W$ and $B_T$ lie close together with the contour for $T_M$ lying a little below that for $B_W$. We stress that far fewer data points were identified for $T_M$ than for the other observables and that, for all event shapes, very few measurements were available at low centre-of-mass energies.

\clearpage
\subsection{Distributions}
\label{subsec:distributions}
For the event-shape observables considered in section \ref{sec:observables}, we performed an NLL resummation using the \textsc{CAESAR} program \cite{Banfi:2004yd}. These NLL resummed distributions were then matched with fixed order distributions at NLO, obtained from \textsc{EVENT2}, using the log-R scheme \cite{Catani:1992ua}. To the resummed, matched distributions (NLL+NLO) we then apply the NP shifts that we have computed in sections~\ref{sec:observables} and \ref{sec:subtraction}. 

The latest ALEPH QCD publication \cite{ALEPH:2003obs} provides experimental data covering centre-of-mass energies between $91.2\,\rm{GeV}$ and $209\,\rm{GeV}$, as well as the fit ranges that were employed for the simultaneous fits for $\alpha_s(m_Z)$ and $\alpha_0(2\,\mathrm{GeV})$ for each of the event-shape observables. As a test of our method we have attempted to repeat these simultaneous fits. Where possible we have used the same fit ranges, however for the $C$-parameter and jet broadenings it was necessary to slightly reduce the upper bound of the fit ranges. It was noted in \cite{ALEPH:2003obs} that the fit ranges for all observables have been selected in the central region of three-jet production. 

Using our method, we obtained simultaneous fit values for $\alpha_s$ and $\alpha_0$ for $1-T$, $C$ and $B_T$ that are consistent with the one-sigma confidence level contours calculated by the ALEPH collaboration \cite{ALEPH:2003obs}. For $\rho_H$ and $B_W$ it was not possible to obtain reliable and consistent fit values for $\alpha_s$ and $\alpha_0$. $\rho_H$ and $B_W$ exhibit a known property whereby the resummed, matched distributions (NLL+NLO) must be squeezed to smaller observable values to ensure that the non-perturbative shift is positive over the whole fit region and thus enabling an accurate fit to experimental data. This requires a significant reduction in the value of $\alpha_s$ from the Particle Data Group's world average value of 0.118 \cite{Workman:2022ynf}. In addition, fit ranges in the central region of three-jet production cover a region where this squeeze requirement is particularly pronounced for $\rho_H$ and $B_W$. 

\begin{figure}[h!]
	\centering
	\includegraphics[width=0.65\textwidth]{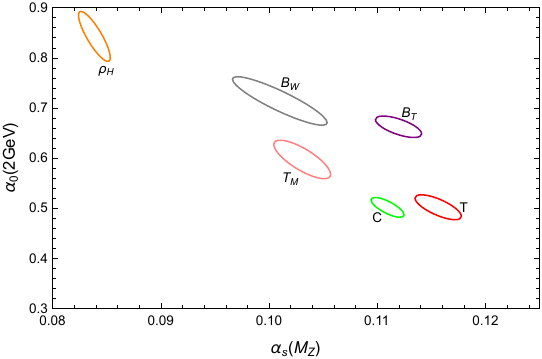}
	\caption{95\% confidence level contours for the fitted values of $\alpha_s$ and $\alpha_0$ to experimental data for the distributions of event-shape observables.}
	\label{fig:contour-dist}
\end{figure}

\begin{figure}
	\begin{subfigure}{\linewidth}
		\includegraphics[width=.5\linewidth]{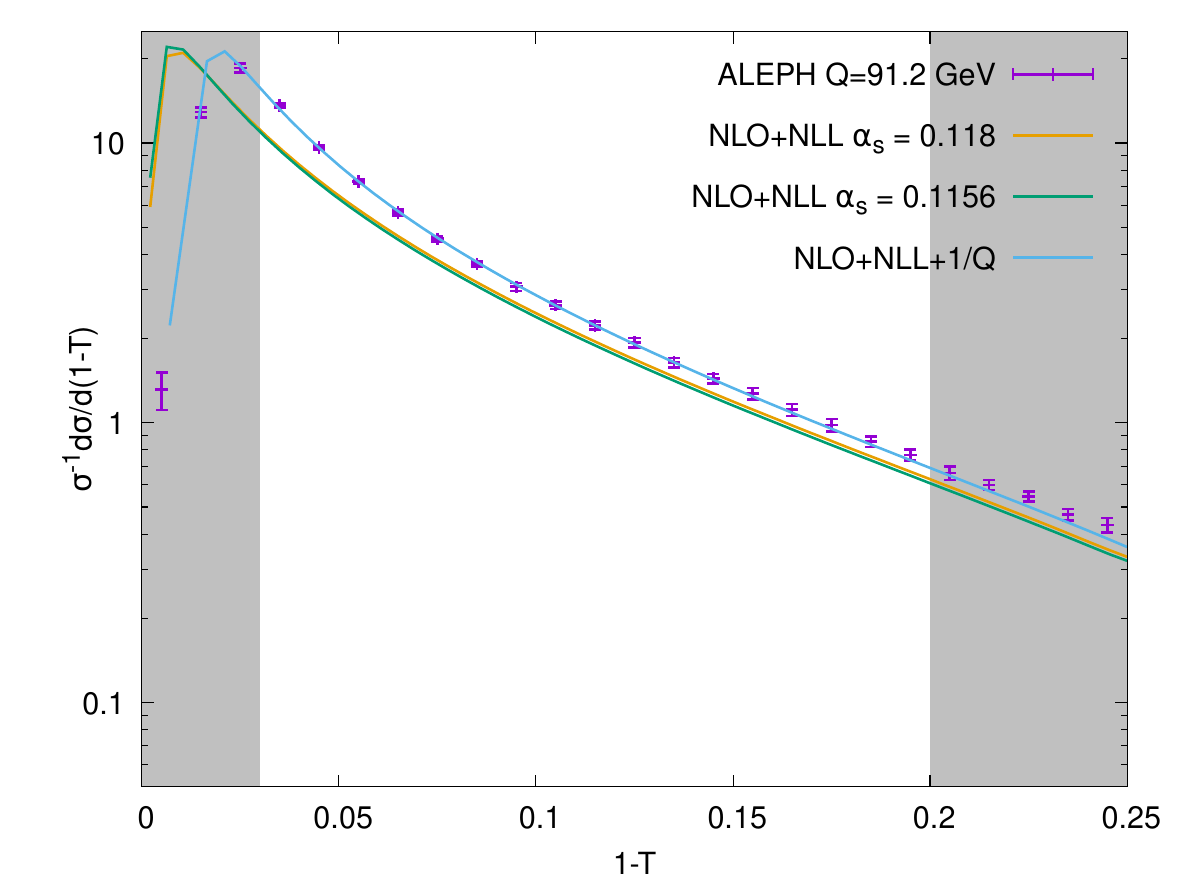}\hfill
		\includegraphics[width=.5\linewidth]{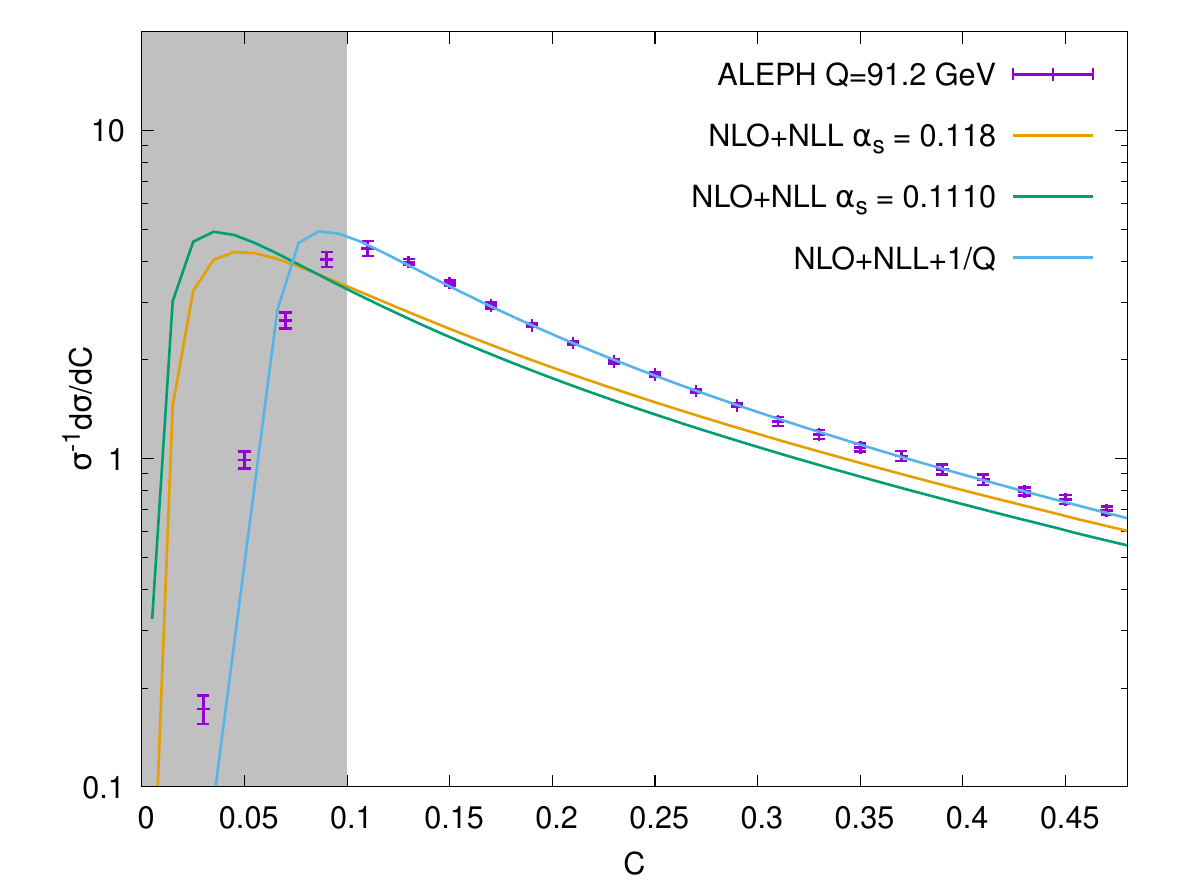}
	\end{subfigure}\par\medskip
	\begin{subfigure}{\linewidth}
		  \includegraphics[width=.5\linewidth]{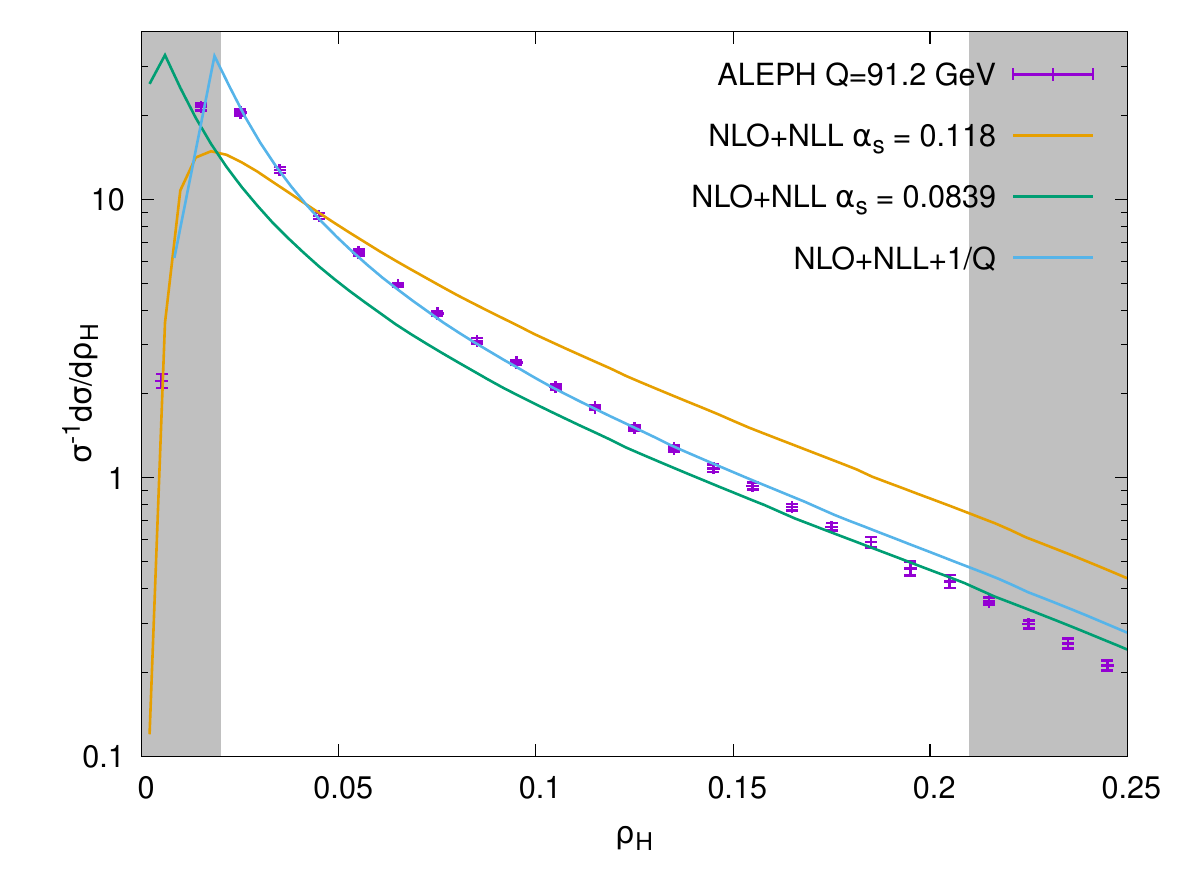}\hfill
		  \includegraphics[width=.5\linewidth]{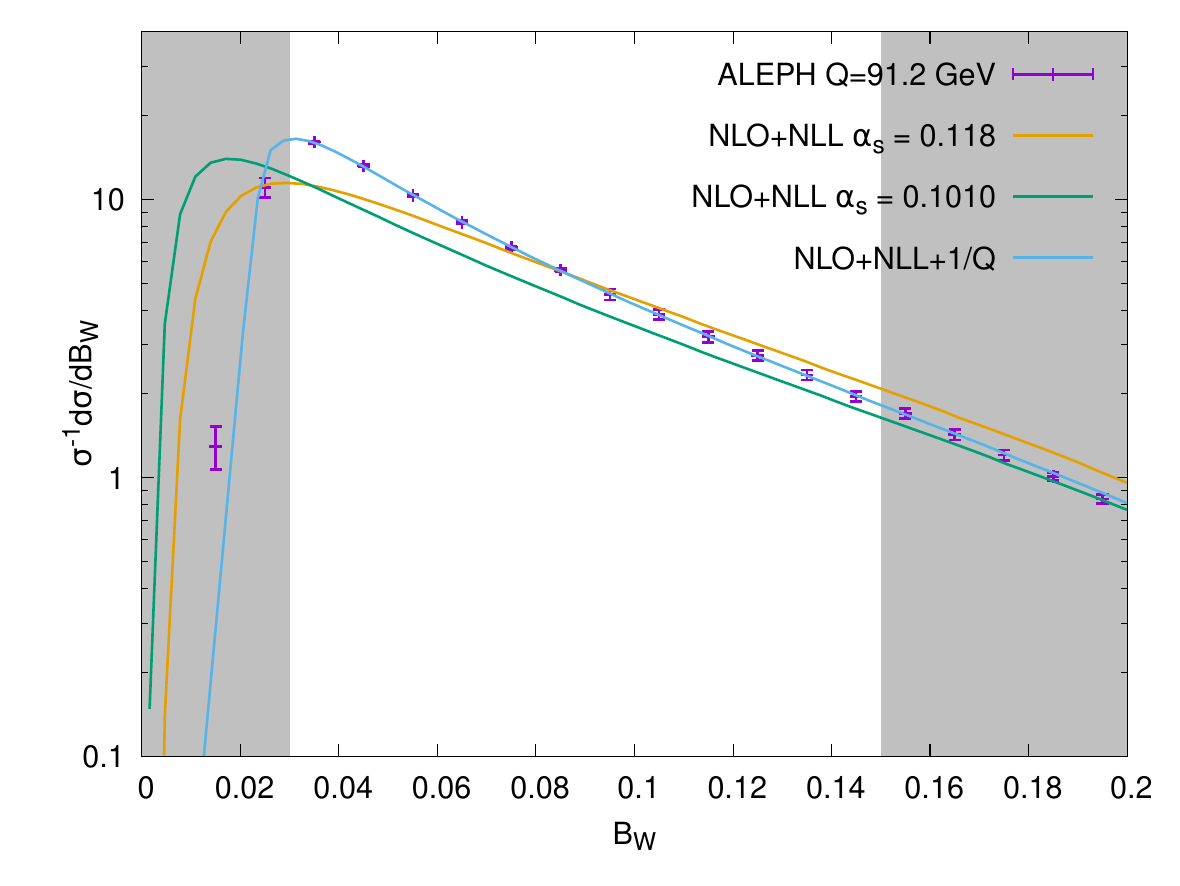}\hfill
	\end{subfigure}\par\medskip
	\begin{subfigure}{\linewidth}
		\includegraphics[width=.5\linewidth]{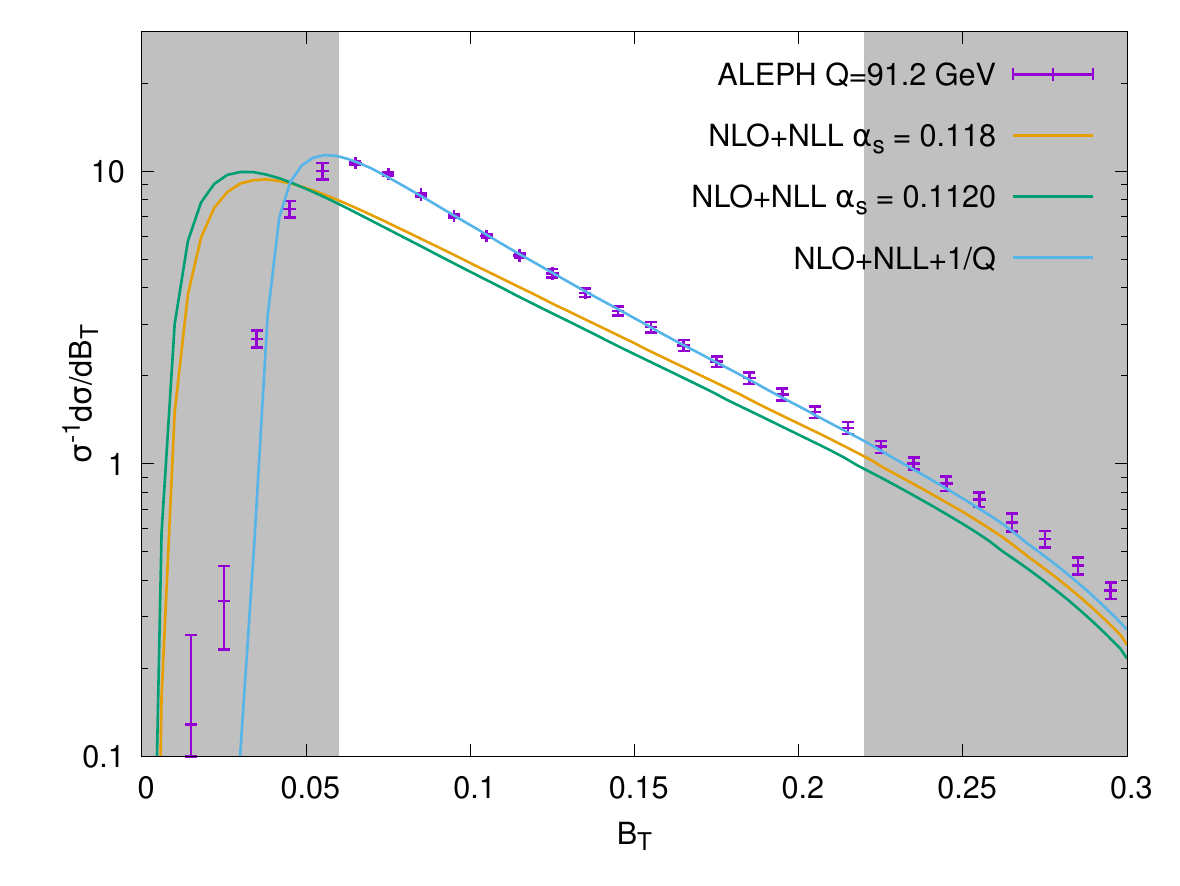}\hfill
		\includegraphics[width=.5\linewidth]{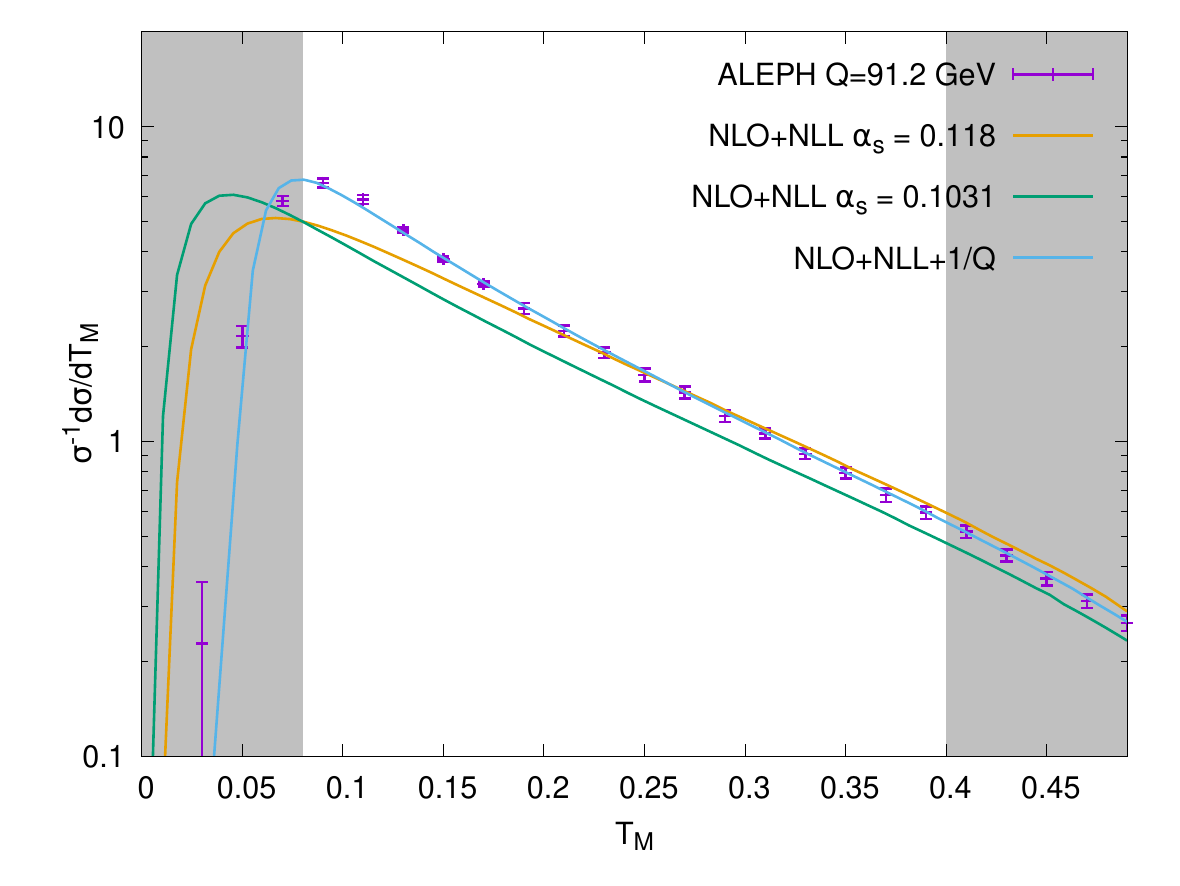}\hfill
	\end{subfigure}
\caption{Plots of the resummed, matched (NLO+NLL) and also fitted distributions (NLO+NLL+$1/Q$) of event-shape observables using both central values and the fitted values of $\alpha_s$ and $\alpha_0$ from Table~\ref{table:dist-fits} and compared with experimental data. The grey shading indicates the regions excluded from the fit range.}
\label{fig:dist-matched_distributions}
\end{figure}

While the fit ranges in \cite{ALEPH:2003obs} have been selected in the central region of three-jet production, the NP shifts are computed in the two-jet region. As a result we have performed a new simultaneous fit for $\alpha_s$ and $\alpha_0$ where we have extended the lower bound of the fit ranges in \cite{ALEPH:2003obs} to the two-jet peak. This criterion has been used to determine  appropriate fit ranges for the thrust major. The fit ranges that we have used in this study are set out in Table~\ref{table:dist-fit-range}. Using the experimental data in \cite{ALEPH:2003obs} and the fit ranges detailed in Table~\ref{table:dist-fit-range} we have performed simultaneous fits for $\alpha_s$ and $\alpha_0$ with the result set out in Table~\ref{table:dist-fits}.
\begin{table}[h!]
	\centering
	\begin{tabular}{ c | c | c | c | c | c | c |} 
		\cline{2-7}
		& \multicolumn{6}{|c|}{Event Shape} \\
		\cline{2-7}
		& $1-T$ & $C$ & $\rho_H$ & $B_W$ & $B_T$ & $T_M$ \\ 
		\hline
		\multicolumn{1}{|c|}{$91.2\,\rm{GeV}$} & 0.03 - 0.20 & 0.10 - 0.50 & 0.02 - 0.21 & 0.03 - 0.15 & 0.06 - 0.22 & 0.08 - 0.40 \\
		\hline
		\multicolumn{1}{|c|}{$133\,\rm{GeV}$} & 0.02 - 0.20 & 0.08 - 0.50 & 0.02 - 0.25 & 0.03 - 0.20 & 0.05 - 0.30 & 0.08 - 0.40 \\
		\hline
		\multicolumn{1}{|c|}{$161\,\rm{GeV}$} & 0.02 - 0.20 & 0.08 - 0.50 & 0.02 - 0.25 & 0.03 - 0.20 & 0.05 - 0.30 & 0.08 - 0.40 \\
		\hline
		\multicolumn{1}{|c|}{$172\,\rm{GeV}$} & 0.02 - 0.20 & 0.08 - 0.50 & 0.02 - 0.25 & 0.03 - 0.20 & 0.05 - 0.30 & 0.08 - 0.40 \\
		\hline
		\multicolumn{1}{|c|}{$183\,\rm{GeV}$} & 0.02 - 0.20 & 0.08 - 0.50 & 0.02 - 0.20 & 0.03 - 0.20 & 0.05 - 0.25 & 0.08 - 0.40 \\
		\hline
		\multicolumn{1}{|c|}{$189\,\rm{GeV}$} & 0.02 - 0.16 & 0.08 - 0.50 & 0.02 - 0.16 & 0.03 - 0.20 & 0.05 - 0.20 & 0.08 - 0.40 \\
		\hline
		\multicolumn{1}{|c|}{$200\,\rm{GeV}$} & 0.02 - 0.16 & 0.08 - 0.50 & 0.02 - 0.16 & 0.03 - 0.20 & 0.05 - 0.20 & 0.08 - 0.40 \\
		\hline
		\multicolumn{1}{|c|}{$206\,\rm{GeV}$} & 0.02 - 0.16 & 0.08 - 0.50 & 0.02 - 0.16 & 0.03 - 0.20 & 0.05 - 0.20 & 0.08 - 0.40 \\
		\hline
	\end{tabular}
	\caption{Fit ranges for the simultaneous fits for $\alpha_s$ and $\alpha_0$ to ALEPH data from $91.2\,\rm{GeV}$ to $206\,\rm{GeV}$ for the distributions of event-shape observables.}
	\label{table:dist-fit-range}
\end{table} 
\begin{table}[h!]
	\centering
	\begin{tabular}{| c | c | c | c |} 
		\hline
		Event Shape &$\alpha_s(m_Z)$ & $\alpha_0(2\,\mathrm{GeV})$ & $\chi^2/\rm{d.o.f}$ \\ 
		\hline
		$1-T$ & $0.1156\pm0.0009$ & $0.5020\pm0.0102$ & $54.9/(56-2)$ \\ 
		\hline
		$C$ & $0.1110\pm0.0006$ & $0.5018\pm0.0081$ & $56.0/(69-2)$ \\
		\hline
		$\rho_H$ & $0.0839\pm0.0006$ & $0.8424\pm0.0203$ & $137.7/(61-2)$ \\
		\hline
		$B_W$ & $0.1010\pm0.0018$ & $0.7138\pm0.0197$ & $52.1/(61-2)$ \\
		\hline
		$B_T$ & $0.1120\pm0.0009$ & $0.6624\pm0.0087$ & $77.4/(72-2)$ \\ 
		\hline
		$T_M$ & $0.1031\pm0.0011$ & $0.5973\pm0.0157$ & $45.6/(51-2)$ \\ 
		\hline
	\end{tabular}
	\caption{Results for the simultaneous fits for $\alpha_s$ and $\alpha_0$ to experimental data for the distributions of event-shape observables.}
	\label{table:dist-fits}
\end{table} 
\newline
Using the results from Table~\ref{table:dist-fits} we have plotted the 95\% CL contours for the event-shape distributions in Fig.~\ref{fig:contour-dist} and the comparison between the resummed, matched distributions (NLO+NLL) with $\alpha_s = 0.118$, our fitted distributions (NLO+NLL and NLO+NLL+$1/Q$) using the results in Table~\ref{table:dist-fits} and experimental data at a centre-of-mass energy of $91.2\,\rm{GeV}$ in Fig.~\ref{fig:dist-matched_distributions}.

As for the 95\% confidence level contours for the mean values, the $95\%$ confidence level contours for the distributions of each of the observables in Fig.~\ref{fig:contour-dist} show a strong negative correlation between the fitted values of $\alpha_s$ and $\alpha_0$. However, we notice that the 95\% confidence level contours for the distributions lie noticeably further apart from one another than was the case for the mean values in Fig.~\ref{fig:contour-mean}. 

In Fig.~\ref{fig:dist-matched_distributions} we observe that $T_M$ exhibits the same squeeze requirement as for $\rho_H$ and $B_W$. It is clear therefore that an extension of our calculation to the three-jet region is particularly needed for these observables. Nevertheless, we have still performed the simultaneous fits with these observables. We find that this squeeze requirement leads to noticeably smaller fitted values of $\alpha_s$ for $\rho_H$, $B_W$ and $T_M$ than for $T$, $C$ and $B_T$. In particular, for $T_M$ we observe a fitted value of $\alpha_s$ that is comparable with that for $B_W$ and a fitted value of $\alpha_0$ located between that of $B_T$ and the grouping of $C$ and $1-T$. For $\rho_H$, this effect is even more pronounced with a 95\% confidence level contour located far away from that for $B_W$ and $T_M$. As for $\langle \rho_H \rangle$, this may also be due to the effect of hadron masses, as discussed in \cite{Salam:2001bd}, which we neglect in this preliminary study.

\clearpage
\section{Conclusions}
\label{sec:the-end}

In this paper we have presented a general method to compute the leading non-perturbative corrections to event-shape distributions in the two-jet region. The power of our method relies on a numerical algorithm suitable to treat any observable, including those leading to logarithmic enhancement in the linear power correction. Indeed, prior to our work, only few examples of such observables were studied, specifically those who could be handled analytically. The crucial point is that leading power-suppressed corrections can be modelled in terms of the emission of an ultra-soft gluon, accompanied by an ensemble of perturbative soft and collinear emissions. The only incalculable quantity is a moment, known as $\alpha_0$, of the soft effective coupling which determines ultra-soft emission probability. The integration over the remaining kinematic variables, i.e.\ rapidity and azimuth, is assumed to follow the perturbative soft emission matrix element squared. Therefore, the ultra-soft gluon can be considered as a ``special'' emission, which together with an arbitrary number of soft and collinear emissions, can be simulated numerically with the ARES method. In fact, similar contributions appear when computing NNLL corrections to two-jet event-shape distributions.

First, we have validated the method to reproduce leading hadronisation corrections to all known event-shape distributions. Then, we have been able for the first time to compute the hadronisation correction to the distribution of the thrust major, which does not allow an analytic treatment, but for which data has existed for a long time. In order to do so, we had to tackle the problem of unphysical divergences of hadronisation corrections occurring for large values of recoil-sensitive event shapes, e.g.\ the total broadening and the thrust major itself. This is done by performing a subtraction procedure, so that all numerical integrations are finite, and unphysical divergences are treated fully analytically in a general way. 

Finally, we have performed new simultaneous fits of the strong coupling $\alpha_s$ and $\alpha_0$ using a selection of experimental data obtained by the ALEPH collaboration. We have been able to obtain consistent results for all event shapes for which hadronisation corrections were known. Moreover, we have performed a similar fit for the thrust major distributions and mean values. The resulting values of $\alpha_s$ and $\alpha_0$ are in the same ballpark as for other event shapes.

The method we have devised is just our own first attempt to improve the phenomenology of event-shape distributions and means at LEP. A further step is to extend our calculation to the three-jet region, along the lines of what was done in~\cite{Caola:2022vea,Nason:2023asn}. Such extension is particularly needed for the heavy-jet mass, wide-jet broadening and thrust major. Another direction of improvement is more sophisticated procedures to subtract unphysical divergences for recoil-sensitive event shapes. Last, we remark that although here for simplicity we have restricted our method to selected event shapes, our procedure could be extended to the two-jet rate, which is particularly important for precise $\alpha_s$ determinations. This requires including information on the rapidity of subsequent soft and collinear emissions, in a similar way as is done for NLL~\cite{Banfi:2001bz} and NNLL~\cite{Banfi:2016zlc} resummations and of course including the interplay between PT and NP emissions. We remark that for jet rates, the treatment of multiple ultra-soft emissions does not lead to the Milan factor and needs to be computed along the lines of~\cite{Dasgupta:2009tm}. We leave both issues for future work.

\acknowledgments{The work of AB has been funded by the Science Technology and Facilities Council (STFC) under grant number ST/T00102X/1. The work of BKE has been funded by the European Research Council (ERC) under the European Union's Horizon 2020 research and innovation program (grant agreement No.~788223). RW acknowledges the hospitality of the CERN Theory Group while part of this work was performed.}

\appendix
\section{Monte Carlo determination of non-perturbative shifts}
\label{sec:monte-carlo-determ}
This appendix describes the Monte Carlo procedure that we adopt to compute the various non-perturbative shifts. As an illustrative example, let us first consider the denominator of eq.~\eqref{eq:np-shift-NLL}
\begin{equation}
	\int d\mathcal{Z}[\{R'_{\ell_i},k_i\}] \, \delta\bigg(1-\frac{\Vsc{\{k_i\}}}{v}\bigg) \ \ .
\end{equation}
Here, the contribution with zero emissions does not satisfy the observable constraint $\delta(v-\Vsc{\{k_i\}})$. We therefore select emission $k_1$, such that $\Vsc{k_1}$ is the largest of all $\Vsc{k_i}$, and neglect all emissions such that $\Vsc{k_i}<\epsilon \Vsc{k_1}$. This gives
\begin{multline} \label{eq:np-den-MC}
	\int d\mathcal{Z}[\{R'_{\ell_i},k_i\}] \, \delta\bigg(1-\frac{\Vsc{\{k_i\}}}{v}\bigg) = \, 
	\epsilon^{R'} \sum_{\ell_1=1,2} R'_{\ell_1} \int_{0}^{\infty} \frac{d\zeta_1}{\zeta_1} \, \zeta_1^{R'} \int_0^{2\pi} \frac{d\phi_1}{2\pi} \, \times \\ 
	\times \left(\sum_{n=0}^{\infty} \frac{1}{n!} \prod_{i=2}^{n+1} \sum_{\ell_i=1,2} R'_{\ell_i} \int_{\epsilon \zeta_1}^{\zeta_1} \frac{d\zeta_i}{\zeta_i} \int_0^{2\pi} \frac{d\phi_i}{2\pi}\right) \, \delta\bigg(1-\frac{\Vsc{ k_1,\dots,k_{n+1}}}{v}\bigg) \ \ .
\end{multline}
We introduce the rescaling $\zeta_i \to \zeta_1 \zeta_i$ and rescale the corresponding momenta $k_i$ such that $\Vsc{k_i} \to \zeta_1\Vsc{k_i}$. As $V$ is rIRC safe we have the rescaling
\begin{equation} \label{eq:rescaling}
	\Vsc{k_1,\dots,k_{n+1}} \to 
	\zeta_1 \Vsc{k_1,\dots,k_{n+1}} \ \ .
\end{equation}
We can therefore write eq.~\eqref{eq:np-den-MC} as
\begin{multline}
	\int d\mathcal{Z}[\{R'_{\ell_i},k_i\}] \, \delta\bigg(1-\frac{\Vsc{\{k_i\}}}{v}\bigg) = \, 
	\epsilon^{R'} \sum_{\ell_1=1,2} R'_{\ell_1} \int_{0}^{\infty} \frac{d\zeta_1}{\zeta_1} \, \zeta_1^{R'} \int_0^{2\pi} \frac{d\phi_1}{2\pi} \, \times \\ 
	\times \left(\sum_{n=0}^{\infty} \frac{1}{n!} \prod_{i=2}^{n+1} \sum_{\ell_i=1,2} R'_{\ell_i} \int_{\epsilon}^{1} \frac{d\zeta_i}{\zeta_i} \int_0^{2\pi} \frac{d\phi_i}{2\pi}\right) \, \delta\bigg(1-\zeta_1\frac{\Vsc{ k_1,\dots,k_{n+1}}}{v}\bigg) \ \ .
\end{multline}
The $\zeta_1$-integration can now be performed using the delta-function constraint to obtain
\begin{equation} \label{eq:np-den-MC-rescaled-final}
	\int d\mathcal{Z}[\{R'_{\ell_i},k_i\}] \,  \delta\bigg(1-\frac{\Vsc{\{k_i\}}}{v}\bigg) =
	R'\mathcal{F}(R') \ \ ,
\end{equation}
where $\mathcal{F}(R')$ is the NLL multiple-emission function, whose expression in terms of $\{k_i\}$ reads
\begin{align} \label{eq:F-MC-NLL}
	\nonumber
	\mathcal{F}(R') =
	&\sum_{\ell_1=1,2}\frac{R'_{\ell_1}}{R'}\int_0^{2\pi} \frac{d\phi_1}{2\pi} \, \epsilon^{R'} \sum_{n=0}^{\infty}\frac{1}{n!} \prod_{i=2}^{n+1} \sum_{\ell_i=1,2} R'_{\ell_i}\int_{\epsilon}^{1} \frac{d\zeta_i}{\zeta_i}\int_0^{2\pi} \frac{d\phi_i}{2\pi} \, \times \\
	&\times \exp\left(-R'\ln\frac{\Vsc{k_1,\dots,k_{n+1}}}{v}\right) \ \ .
\end{align}
The above expression can be evaluated numerically using a Monte Carlo procedure. One can apply the same approach to the numerator of eq.~\eqref{eq:np-shift-NLL} and we find
\begin{multline} 
	\cNP{V} \mathcal{F}(R') =  \sum_\ell
	\int d\eta^{(\ell)} \, \frac{d\phi}{2\pi} \sum_{\ell_1=1,2}  \frac{R'_{\ell_1}}{R'}\int_0^{2\pi}\!\frac{d\phi_1}{2\pi}\left( \epsilon^{R'}\sum_{n=0}^\infty \frac{1}{n!}  \prod_{i=2}^{n+1}\sum_{\ell_i=1,2} R'_{\ell_i}\int^{1}_{\epsilon}\frac{d\zeta_i}{\zeta_i}\int_0^{2\pi}\frac{d\phi_i}{2\pi}\right) \times \\ 
	\times \left. \left(\frac{\Vsc{k_1,k_2,\dots,k_{n+1}}}{v}\right)^{-R'} h_V(\eta^{(\ell)},\phi,\{\tilde{p}\},k_1,k_2,\dots,k_{n+1})\right|_{\Vsc{k_1}=v} \ \ ,
\end{multline}
where now $\Vsc{k_1} = v$.

\section{Analytical determination of non-perturbative shifts}
\label{sec:NP-broadenings}
This appendix presents analytical computations of the non-perturbative shifts for the wide-jet and total broadenings within our framework. For the thrust major, the analytical computation of the non-perturbative shift in the limit $R' \to 0$ is also presented.

\subsection{Wide-Jet Broadening}
\label{subsec:NP-wide}
For the wide-jet broadening we wish to analytically compute $\cNP{B_W}$, thus enabling us to determine $\langle\delta B_{\rm{W, \, NP}}\rangle$ via eq.~\eqref{eq:cV}. We start with eqs.~\eqref{eq:hB-final} and \eqref{eq:chiB} which we repeat here for clarity
\begin{equation}
	\cNP{B_W} = 
	\frac12 \left(\ln\frac{1}{B_W} + \eta_0^{(B)}\right) + \chi_W(R') \ \ ,
\end{equation}
with
\begin{align}
	\nonumber
	\chi_W(R') = \, 
	&\frac{1}{R' \, \mathcal{F}_{B_W}(R')} \int d\mathcal{Z}[\{R'_{\ell_i},k_i\}] \, \delta\bigg(1-\frac{B_{W,\rm sc}(\{\tilde{p}\},\{k_i\})}{B_W}\bigg) \times \\
	&\times \left[\int d^2 \vec{x}_1 \, \delta^{(2)}\Bigg(\vec x_1 + \sum_{i\in \mathcal{H}_1} \vec\zeta_{i}\Bigg) \, \frac12 \ln\frac{1}{|\vec x_{1}|} \, \Theta(B_1 - B_2) + 1\leftrightarrow 2 \right] \ \ .
\end{align}
Using eq.~\eqref{eq:dZ-rescaled}, invoking the symmetry between the two hemispheres defined by the thrust axis ($\mathcal{H}_1\leftrightarrow\mathcal{H}_2$ symmetry) and performing the rescaling $\zeta_i \to 2\zeta_i$, we write $\chi_W(R')$ as,\footnote{We highlight that $R'_{\ell_i}$ is the radiator corresponding to hemisphere $\mathcal{H}_{\ell_i}$ and that $R' = R'_1 + R'_2$ (where we will eventually set $R'_1 = R'_2 = R'/2$ due to the $\mathcal{H}_1 \leftrightarrow \mathcal{H}_2$ symmetry).}
\begin{align}
	\nonumber
	\chi_W(R') = \, 
	&\frac{1}{R' \, \mathcal{F}_{B_W}(R')} \, \epsilon^{R'} \, 2^{R'} \, \sum_{n=0}^{\infty}\frac{1}{n!} \, \prod_{i=1}^{n} \, \sum_{\ell_i = 1,2} R'_{\ell_i} \int_{\epsilon}^{\infty}\frac{d\zeta_i}{\zeta_i} \, \int_0^{2\pi}\frac{d\phi_i}{2\pi} \, \times \\ \nonumber
	&\times \Bigg[\int\prod_{\ell}d^2 \vec{x}_{\ell} \, \delta^{(2)}\Bigg(\vec x_{\ell} + \sum_{i\in \mathcal{H}_{\ell}} \vec\zeta_{i}\Bigg) \ln\frac{1}{2x_1} \, \Theta\Bigg(\sum_{i \in \mathcal{H}_1}\zeta_i + x_1 - \sum_{i \in \mathcal{H}_2}\zeta_i - x_2\Bigg) \, \times \\
	&\qquad \times \delta\Bigg(1 - \sum_{i \in \mathcal{H}_1} \zeta_i - x_1\Bigg)\Bigg] \ \ ,
\end{align}
where $x_\ell \equiv |\vec x_{\ell}|$. Introducing the Fourier and Mellin transforms of the kinematic and observable constraints respectively, this may be written as
\begin{align}
	\nonumber
	\chi_W(R') = 
	&\frac{2^{R'}}{R' \, \mathcal{F}_{B_W}(R')} \, \int_{\mathcal{C}} \frac{d\nu}{2\pi i\nu} \, \int_{\mathcal{C}} \frac{d\mu}{2\pi i} \, e^{\mu} \, \int\frac{d^2\vec{x}_1 \, d^2\vec{b}_1}{(2\pi)^2} \, \ln\frac{1}{2x_1} \, e^{-\left(\mu - \nu\right) x_1} \, \,e^{-i\vec{b}_1\cdot\vec{x}_1}  \times \\ \nonumber 
	&\times\int\frac{d^2\vec{x}_2 \, d^2\vec{b}_2}{(2\pi)^2} \, e^{-\nu x_2} \, e^{-i\vec{b}_2\cdot\vec{x}_2} \, \exp\left(R'_1 \int_0^{\infty}\frac{d\zeta}{\zeta} \, \int_0^{2\pi} \frac{d\phi}{2\pi} \, \left[e^{-i\vec{b}_1\cdot\vec{\zeta}} \, e^{-\left(\mu-\nu\right)\zeta} - \Theta(1-\zeta)\right]\right) \times \\ 
	&\times \exp\left(R'_2 \int_0^{\infty}\frac{d\zeta}{\zeta} \, \int_0^{2\pi}\frac{d\phi}{2\pi} \, \left[e^{-i\vec{b}_2\cdot\vec{\zeta}} \, e^{-\nu\zeta} - \Theta(1-\zeta)\right]\right) \ \ .
\end{align}
We make use of the following integral
\begin{equation} \label{eq:radiatorstandint}
	\int_0^{\infty}\frac{d\zeta}{\zeta} \, \int_0^{2\pi}\frac{d\phi}{2\pi} \, \left[e^{-i\vec{b}\cdot\vec{\zeta}} \, e^{-\nu\zeta} - \Theta(1-\zeta)\right] = 
	-\gamma_E - \ln\nu - \ln\left[\frac{1+\sqrt{1+\frac{b^2}{\nu^2}}}{2}\right] \ \ ,
\end{equation}
and find
\begin{align}
	\nonumber
	\chi_W(R') = 
	&\frac{2^{R'} e^{-\gamma_E R'}}{R' \, \mathcal{F}_{B_W}(R')} \, \int_{\mathcal{C}} \frac{d\nu}{2\pi i\nu} \, \int_{\mathcal{C}} \frac{d\mu}{2\pi i} \, e^{\mu} \, \int\frac{d^2\vec{x}_1 \, d^2\vec{b}_1}{(2\pi)^2} \, \ln\frac{1}{2x_1} \, e^{-\left(\mu - \nu\right) x_1} \, e^{-i\vec{b}_1\cdot\vec{x}_1} \int\frac{d^2\vec{x}_2 \, d^2\vec{b}_2}{(2\pi)^2} \times \\
	&\times e^{-\nu x_2} \, e^{-i\vec{b}_2\cdot\vec{x}_2} \, \left(\mu-\nu\right)^{-R'_1} \, \left(\frac{1+\sqrt{1+\frac{b_1^2}{\left(\mu-\nu\right)^2}}}{2}\right)^{-R'_1} \, \nu^{-R'_2} \, \left(\frac{1+\sqrt{1+\frac{b_2^2}{\nu^2}}}{2}\right)^{-R'_2} \ \ .
\end{align}
It is convenient to redefine $\mu \to \mu + \nu$, rescale $b_1 \to \mu b_1$, $b_2 \to \nu b_2$, $x_1 \to x_1/\mu$ and $x_2 \to x_2/\nu$, after which we perform the $x_1$- and $x_2$-integrals directly and find 
\begin{align} \label{eq:BW-NP-factorised}
	\nonumber
	\chi_W(R') = \,
	&\frac{2^{R'} e^{-\gamma_E R'}}{R' \, \mathcal{F}_{B_W}(R')} \, \int_{\mathcal{C}} \frac{d\nu}{2\pi i\nu} \, e^{\nu} \, \nu^{-R'_2} \, \int_{\mathcal{C}} \frac{d\mu}{2\pi i} \, e^{\mu} \, \mu^{-R'_1} \, \int_1^{\infty} \frac{dy_2}{y_2^2} \left(\frac{1+y_2}{2}\right)^{-R'_2} \times \\
	&\times \int_1^{\infty} \frac{dy_1}{y_1^2} \left(\frac{1+y_1}{2}\right)^{-R'_1} \left[-2 + \gamma_E + y_1 - \ln\frac{1+y_1}{y_1^2} + \ln\mu\right] \ \ ,
\end{align}
where $y_i \equiv \sqrt{1+b_i^2}$. To evaluate the $y_i$-integrals we make use of the following auxiliary functions taken from~\cite{Dokshitzer:1998qp}:\footnote{To compare notation, the function denoted in this paper by $\sigma(a)$ is equivalent to $\left[\lambda(a)\right]^{-a}$ in \cite{Dokshitzer:1998qp}.}
\begin{align} 
	\label{eq:sigmaR'}
	\sigma(a) &\equiv 
	\int_1^{\infty} \frac{dy}{y^2} \left(\frac{1+y}{2}\right)^{-a} 
	= \frac{2^{a}}{1+a} \, {}_2 F_1(a,1+a;2+a;-1) \ \ , \\
	\label{eq:chiR'}
	\chi(a) &\equiv 
	-\frac{1}{a} + \frac{1}{\sigma(a)} \int_1^{\infty}\frac{dy}{y} \left(\frac{1+y}{2}\right)^{-a} = 
	\frac{2}{a} \left(\frac{1}{\sigma(a)} - 1\right) \ \ , \\
	\label{eq:rhoR'}
	\rho(a) &\equiv 
	\frac{1}{\sigma(a)} \int_1^{\infty}\frac{dy}{y^2} \left(\frac{1+y}{2}\right)^{-a} \, \ln\frac{1+y}{y^2} \ \ .
\end{align}
For the wide-jet broadening, we know that
\begin{equation}
	\mathcal{F}_{B_W}(R') = 
	\left[2^{\frac{R'}{2}} \, \sigma\bigg(\frac{R'}{2}\bigg) \, \frac{e^{-\gamma_E \frac{R'}{2}}}{\Gamma\big(1+\frac{R'}{2}\big)} \right]^2 \ \ ,
\end{equation}
therefore, setting $R'_1 = R'_2 = R'/2$ we find
\begin{equation}
	\chi_W(R') = 
	\frac{1}{2}\left[-2 + \chi\bigg(\frac{R'}{2}\bigg) - \rho\bigg(\frac{R'}{2}\bigg) + \psi\bigg(1+\frac{R'}{2}\bigg) + \gamma_E \right] \ \ ,
\end{equation}
with the polygamma function $\psi(a) = d\ln\Gamma(a)/da$. In agreement with \cite{Dokshitzer:1998qp} we obtain
\begin{equation} \label{eq:hBW-analytic}
	\cNP{B_W} = 
	\frac12 \left[\ln\frac{1}{B_W} + \eta_0^{(B)} - 2 + \chi\bigg(\frac{R'}{2}\bigg) - \rho\bigg(\frac{R'}{2}\bigg) + \psi\bigg(1+\frac{R'}{2}\bigg) + \gamma_E \right] \ \ .
\end{equation} 
It is important to inspect the leading $R' \to 0$ behaviour of $\cNP{B_W}$. To do so we require the following limits
\begin{align} \label{eq:limitsato0}
	\lim_{a \to 0} \sigma(a) = 1, \quad 
	\lim_{a \to 0} \rho(a) = -2 +  2\ln 2, \quad 
	\lim_{a \to 0} \chi(a) = 2\ln 2 \ \ ,
\end{align}
such that
\begin{equation} \label{eq:hBW-leading}
	\lim_{R' \to 0} \cNP{B_W} =
	\frac{1}{2}\left(\ln\frac{1}{B_W}+\eta_0^{(B)}\right) \ \ .
\end{equation}

\subsection{Total Broadening}
\label{subsec:NP-total}
The analytical determination of the non-perturbative shift for the total broadening follows a similar process to that for the wide-jet broadening in appendix~\ref{subsec:NP-wide}. As before, we start with eqs.~\eqref{eq:hB-final} and \eqref{eq:chiB} which we repeat here for clarity
\begin{equation}
	\cNP{B_T} = 
	\ln\frac{1}{B_T} + \eta_0^{(B)} + \chi_T(R') \ \ ,
\end{equation}
with
\begin{align}
	\nonumber
	 \chi_T(R') = \,
	 &\frac{1}{R' \, \mathcal{F}_{B_T}(R')} \int d\mathcal{Z}[\{R'_{\ell_i},k_i\}] \,  \delta\bigg(1-\frac{B_{T,\rm sc}(\{\tilde{p}\},\{k_i\})}{B_T}\bigg) \, \times \\
	 &\times \int\prod_{\ell}d^2 \vec{x}_{\ell} \, \delta^{(2)}\Bigg(\vec x_{\ell} + \sum_{i\in \mathcal{H}_{\ell}}\vec\zeta_{i}\Bigg) \, \frac12 \, \sum_{\ell} \ln\frac{1}{|\vec x_{\ell}|} \ \ .
\end{align}
We follow similar steps as for the wide-jet broadening in appendix~\ref{subsec:NP-wide}. Using eq.~\eqref{eq:dZ-rescaled}, performing the rescaling $\zeta_i \to 2\zeta_i$, introducing the Fourier and Mellin transforms of the kinematic and observable constraints respectively and utilising the $\mathcal{H}_1 \leftrightarrow \mathcal{H}_2$ symmetry, we write $\chi_T(R')$ as
\begin{align}\label{eq:unimpciT}
	\nonumber
	\chi_T(R') = 
	&\frac{2^{R'}}{R' \, \mathcal{F}_{B_T}(R')} \, \int_{\mathcal{C}} \frac{d\nu}{2\pi i} \, e^\nu \, \int\frac{d^2\vec{x}_1 \, d^2\vec{b}_1}{(2\pi)^2} \, e^{-\nu x_1} \, e^{-i\vec{b}_1\cdot\vec{x}_1} \, \ln\frac{1}{2x_1} \, \int\frac{d^2\vec{x}_2 \, d^2\vec{b}_2}{(2\pi)^2} \, e^{-\nu x_2} \, e^{-i\vec{b}_2\cdot\vec{x}_2} \, \times \\ \nonumber 
	&\times \exp\left(R'_1 \int_0^{\infty}\frac{d\zeta}{\zeta} \, \int_0^{2\pi}\frac{d\phi}{2\pi} \, \left[e^{-i\vec{b}_1\cdot\vec{\zeta}} \, e^{-\nu \zeta} - \Theta(1-\zeta)\right]\right) \times \\ 
	&\times \exp\left(R'_2 \int_0^{\infty}\frac{d\zeta}{\zeta} \, \int_0^{2\pi}\frac{d\phi}{2\pi} \, \left[e^{-i\vec{b}_2\cdot\vec{\zeta}} \, e^{-\nu\zeta} - \Theta(1-\zeta)\right]\right) \ \ ,
\end{align}
where $x_\ell \equiv |\vec x_{\ell}|$. Using eq.~\eqref{eq:radiatorstandint} we find
\begin{align}
	\nonumber
	\chi_T(R') = \,
	&\frac{2^{R'} \, e^{-\gamma_E R'}}{R' \, \mathcal{F}_{B_T}(R')} \, \int_{\mathcal{C}} \frac{d\nu}{2\pi i} \, e^{\nu} \, \nu^{-R'} \, \int\frac{d^2 \vec{x}_1 \, d^2 \vec{b}_1}{(2\pi)^2} \, e^{-\nu x_1} \, e^{-i\vec{b}_1\cdot\vec{x}_1} \, \int\frac{d^2 \vec{x}_2 \, d^2 \vec{b}_2}{(2\pi)^2} \, e^{-\nu x_2} \, e^{-i\vec{b}_2\cdot\vec{x}_2} \, \times \\ 
	&\times \, \ln\frac{1}{2x_1} \, \left(\frac{1+\sqrt{1+\frac{b_1^2}{\nu^2}}}{2}\right)^{-R'_1} \, \left(\frac{1+\sqrt{1+\frac{b_2^2}{\nu^2}}}{2}\right)^{-R'_2} \ \ .
\end{align}
We rescale $b_i \to \nu b_i$ and $x_i \to x_i/\nu$, after which we perform the $x_1$- and $x_2$-integrals directly and find 
\begin{align} \label{eq:BT-NP-factorised}
	\nonumber
	\chi_T(R') = \,
	&\frac{2^{R'} e^{-\gamma_E R'}}{R' \, \mathcal{F}_{B_T}(R')} \, \int_{\mathcal{C}} \frac{d\nu}{2\pi i} \, e^{\nu} \, \nu^{-R'_2} \, \int_1^{\infty} \frac{dy_2}{y_2^2} \left(\frac{1+y_2}{2}\right)^{-R'_2} \times \\
	&\times \int_1^{\infty} \frac{dy_1}{y_1^2} \left(\frac{1+y_1}{2}\right)^{-R'_1} \left[-2 + \gamma_E + y_1 - \ln\frac{1+y_1}{y_1^2} + \ln\nu\right] \ \ ,
\end{align}
where $y_i \equiv \sqrt{1+b_i^2}$. We note that compared to the equivalent expression for $B_W$, eq.~\eqref{eq:BW-NP-factorised}, the key difference is that for $B_T$ there is only one Laplace variable whereas for $B_W$ there are two. We will see below that this will have the effect of leaving a residual $1/R'$ singularity which is not cancelled. For the total broadening we know that
\begin{equation}
	\mathcal{F}_{B_T}(R') = 
	2^{R'} \left[\sigma\bigg(\frac{R'}{2}\bigg)\right]^2 \, \frac{e^{-\gamma_E R'}}{\Gamma(1+R')} \ \ .
\end{equation}
Therefore, using eqs.~\eqref{eq:sigmaR'}, \eqref{eq:chiR'} and \eqref{eq:rhoR'} and setting $R'_1 = R'_2 = R'/2$ we find
\begin{equation} \label{eq:chiT}
	\chi_T(R') = 
	-2 + \chi\bigg(\frac{R'}{2}\bigg) - \rho\bigg(\frac{R'}{2}\bigg) + \psi(1+R') + \gamma_E + \frac{1}{R'} \ \ ,
\end{equation}
and in agreement with \cite{Dokshitzer:1998qp} we obtain
\begin{equation} \label{eq:hBT-analytic}
	\cNP{B_T} = 
	\ln\frac{1}{B_T} + \eta_0^{(B)} - 2 + \chi\bigg(\frac{R'}{2}\bigg) - \rho\bigg(\frac{R'}{2}\bigg) + \psi(1+R') + \gamma_E + \frac{1}{R'} \ \ .
\end{equation}
Using eq.~\eqref{eq:limitsato0} we deduce the leading $R' \to 0$ behaviour of $\cNP{B_T}$
\begin{equation} \label{eq:hBT-leading}
	\cNP{B_T} = 
	\ln\frac{1}{B_T} + \eta_0^{(B)} + \frac{1}{R'} + \mathcal{O}(R')  \ \ .
\end{equation} 

\subsection{Thrust Major}
\label{subsec:NP-major}
As explained in section~\ref{sec:subtraction}, the shift for the thrust major cannot be obtained in closed form. Nevertheless we can analytically determine its behaviour in the limit $R' \to 0$. We start with eqs.~\eqref{eq:cTM-rescaled} and~\eqref{eq:chiM} which we repeat here for clarity
\begin{equation}
	\cNP{T_M} = 
	\frac{4}{\pi}\left(\ln\frac{2}{T_M} + \ln2 - 2\right) + \chi_M(R') \ \ ,
\end{equation}
where
\begin{align}
	\nonumber
	\chi_M(R') = \, 
	&\frac{1}{R' \, \mathcal{F}_{T_M}(R')} \, \frac{2}{\pi} \, \int d\mathcal{Z}[\{R'_{\ell_i},k_i\}] \,  \delta\bigg(1-\frac{T_{M,\rm sc}(\{\tilde{p}\},\{k_i\})}{T_M}\bigg) \, \times \\
	&\times \int_{-\infty}^{\infty} \prod_{\ell} d x_{\ell} \, \delta\Bigg(x_{\ell} + \sum_{i\in \mathcal{H}_{\ell}}\zeta_{i}\sin\phi_i\Bigg) \sum_{\ell} \ln\frac{1}{|x_{\ell}|} \ \ .
\end{align}
In the limit $R' \to 0$, one of the hemispheres will contain a single emission, which we denote $k_1$, with a transverse momentum of order $T_M Q/2$ and which sets the thrust-major axis. All other emissions have transverse momenta less than $\delta \frac{T_M Q}{2}$, where $\epsilon \ll \delta \ll 1$, and thus
\begin{equation} \label{eq:T-1emsn}
	T_{M,\rm sc}(\{\tilde{p}\},\{k_i\}) \simeq T_{M,\rm sc}(\{\tilde{p}\},k_1) \ \ .
\end{equation}
 Using eq.~\eqref{eq:dZ-rescaled}, the $\mathcal{H}_1\leftrightarrow\mathcal{H}_2$ symmetry and the fact that $R' = R'_1 + R'_2$, we write $\chi_M(R')$ in the limit $R' \to 0$ as
\begin{align} \label{eq:chiM-smallR'-start}
	\nonumber
	\chi_M(R') \simeq \, 
	&\frac{2}{\pi} \, \int_{-\infty}^{\infty}dx_2 \, \left(\lim_{\epsilon \to 0} \, \epsilon^{R'_2} \, \sum_{n=0}^{\infty} \frac{\left(R'_2\right)^n}{n!} \, \prod_{i=2}^{n+1} \, \int_{\epsilon}^{\delta}\frac{d\zeta_i}{\zeta_i} \, \int_0^{2\pi}\frac{d\phi_i}{2\pi}\right)\times \\
	&\times\delta\Bigg(x_2 + \sum_{i\in \mathcal{H}_2}\zeta_{i}\sin\phi_i\Bigg) \, \ln\frac{1}{|x_2|}  \ \ .
\end{align}
We rescale $\zeta_i$ and introduce a one-dimensional Fourier transform, noting that we must apply a cut on large values of the transverse momenta of the recoil in $\mathcal{H}_2$ due to the small transverse momenta of emissions in this hemisphere. We achieve this by imposing an upper-bound $|x_2| < x_{\rm max}$ (where $x_{\rm max} \sim \delta$) and a corresponding lower-bound on the conjugate parameter $|b_2| > b_{\rm min}$ (where $b_{\rm min} \sim 1/\delta$). We obtain
\begin{align}
	\nonumber
	\chi_M(R') \simeq \,
	&\frac{8}{\pi} \int_{0} ^{x_{\rm max}} dx_2 \int_{b_{\rm min}}^{\infty} \frac{db_2}{2\pi} \left(\lim_{\epsilon \to 0} \, \delta^{R'_2} \, \epsilon^{R'_2} \, \sum_{n=0}^{\infty}\frac{\left(R'_2\right)^n}{n!} \, \prod_{i=2}^{n+1} \, \int_{\epsilon}^{1}\frac{d\zeta_i}{\zeta_i} \, \int_0^{2\pi}\frac{d\phi_i}{2\pi}\right) \times \\ 
	&\times \exp\left(- i b_2x_2 - i \delta b_2 \sum_{i \in \mathcal{H}_2} \zeta_i\sin\phi_i\right) \, \ln\frac{1}{x_2} \ \ .
\end{align}
This may be written in a simplified exponential form to give
\begin{equation} \label{eq:hM-smallR'-factorised}
	\chi_M(R') \simeq 
	\frac{8}{\pi} \, \delta^{R'_2} \int_0^{x_{\rm max}} dx_2 \int_{b_{\rm min}}^{\infty} \frac{db_2}{2\pi} \, \cos(b_2x_2) \, \exp\left(R'_2 \int_0^1\frac{d\zeta}{\zeta} \, \left[J_0(\delta b_2\zeta) - 1\right]\right) \, \ln\frac{1}{x_2} \ \ .
\end{equation}
As emissions in $\mathcal{H}_2$ do not contribute to the observable we notice that there is no damping factor in the $\zeta$-integral. From eq.~\eqref{eq:radiator-large-b} we obtain
\begin{equation}
	\chi_M(R') \simeq 
	\frac{8}{\pi} \int_0^{x_{\rm max}} dx_2 \int_{b_{\rm min}}^{\infty} \frac{db_2}{2 \pi} \, \cos(b_2x_2) \, \ln\frac{1}{x_2} \, b_2^{- R'_2} \ \ .
\end{equation}
Performing the $x_2$- and $b_2$-integrations and extracting the singular $R' \to 0$ behaviour, setting $R'_1 = R'_2 = R'/2$, we obtain
\begin{align} \label{eq:chiM-leading}
	\chi_M(R') \simeq \frac{4}{\pi R'} \ \ .
\end{align}
We note that we do not control terms of $\mathcal{O}(1)$ in this calculation as they may arise from the interplay of unknown terms of order $R'$ multiplied by the leading $1/R'$.

\section{Analytical determination of the counterterms}
\label{sec:NP-counterterms}
This appendix presents analytical computations of the counterterms for the total broadening and thrust major.

\subsection{Counterterm for the Total Broadening}
\label{subsec:ct-BT}
In section~\ref{sec:subtraction} we introduced the counterterm for the total broadening which we wish to determine analytically. We start with eq.~\eqref{eq:chibarT} which we repeat here 
\begin{align}
	\nonumber
	\chi^{\textrm{c.t.}}_T(R') = \,
	&\frac{1}{R' \mathcal{F}_{B_T}(R')} \int d^2\vec{x}_{\bar{\ell}} \int d\mathcal{Z}[\{R'_{\ell_i},k_i\} ] \, \delta\Big(1-\max_{i}\{\zeta_i\}\Big) \, \frac 12 \ln\frac{1}{x_{\bar{\ell}}} \, \times \\ 
	&\times \delta^{(2)}\Bigg(\vec{x}_{\bar{\ell}}+\sum_{i\in \mathcal{H}_{\bar{\ell}}} \vec{\zeta}_{i}\Bigg) \, \Theta\Bigg(1 - \frac12 x_{\bar{\ell}} - \frac12\sum_{i\in \mathcal{H}_{\bar{\ell}}} \zeta_{i}\Bigg) \ \ ,
\end{align}
where $\mathcal{H}_{\bar{\ell}}$ denotes the hemisphere that does not contain the emission with the largest transverse momentum and $x_{\bar{\ell}} \equiv |\vec x_{\bar{\ell}}|$.
We use eq.~\eqref{eq:dZ-rescaled}, the $\mathcal{H}_1\leftrightarrow\mathcal{H}_2$ symmetry and the fact that $R' = R'_1 + R'_2$. Introducing the Fourier and Mellin transforms of the kinematic and observable constraints respectively, we may write $\chi^{\textrm{c.t.}}_T(R')$ in a simplified exponential form to give 
\begin{align}\label{eq:chibarTbase}
	\nonumber 
	\chi^{\textrm{c.t.}}_T(R') = \,
	&\frac{R'_1}{R' \, \mathcal{F}_{B_T}(R')} \, \int_{\mathcal{C}} \frac{d\nu}{2\pi i\nu} \, e^{2\nu} \, \int\frac{d^2\vec{x}_2 \, d^2\vec{b}_2}{(2\pi)^2} \, e^{-\nu x_2} \, e^{-i\vec{b}_2\cdot\vec{x}_2} \, \frac12\ln\frac{1}{x_2} \times \\
	&\times \exp\left(\int_0^{\infty}\frac{d\zeta}{\zeta} \, R'_2(\zeta v) \, \int_0^{2\pi}\frac{d\phi}{2\pi} \, \left[e^{-i\vec{b}_2\cdot\vec{\zeta}} \, e^{-\nu\zeta} - \Theta(1-\zeta)\right]\right) + 1 \leftrightarrow 2\ \ .
\end{align}
By construction, as discussed in section~\ref{sec:subtraction}, $\chi^{\textrm{c.t.}}_T(R')$ must reproduce the same $1/R'$ behaviour of $\cNP{B_T}$ such that it can act as a suitable local counterterm in the numerical procedure. To see this, we take the NLL approximation of the Sudakov radiator, i.e.\ $R'_2(\zeta v) \simeq R'_2(v)$, in eq.~\eqref{eq:chibarTbase} and using the $\mathcal{H}_1\leftrightarrow\mathcal{H}_2$ symmetry we find
\begin{equation}
	\chi^{\textrm{c.t.}}_T(R') = 
	\frac{e^{-\gamma_E R'_2}}{2 \, \mathcal{F}_{B_T}(R')} \, \int_{\mathcal{C}} \frac{d\nu}{2\pi i\nu} \, e^{2\nu} \, \nu^{-R'_2} \, \int\frac{d^2\vec{x}_2 \, d^2\vec{b}_2}{(2\pi)^2} \, e^{-\nu x_2} \, e^{-i\vec{b}_2\cdot\vec{x}_2} \, \ln\frac{1}{x_2} \, \left(\frac{1+\sqrt{1+\frac{b_2^2}{\nu^2}}}{2}\right)^{-R'_2} \ \ .
\end{equation}
We rescale $b_2 \to \nu b_2$ and $x_2 \to x_2/\nu$, after which we perform the $x_2$-integrals directly and find 
\begin{align} 
	\chi^{\textrm{c.t.}}_T(R') = \,
	\frac{e^{-\gamma_E R'_2}}{2 \, \mathcal{F}_{B_T}(R')} \, \int_{\mathcal{C}} \frac{d\nu}{2\pi i \nu} \, e^{2\nu} \, \nu^{-R'_2} \, \int_1^{\infty} \frac{dy}{y^2} \left(\frac{1+y}{2}\right)^{-R'_2} \left[-2 + \gamma_E + \ln\frac{y^2}{1+y} + \ln2\nu + y\right] \ \ ,
\end{align}
where $y \equiv \sqrt{1+b_2^2}$. Using eqs.~\eqref{eq:sigmaR'}, \eqref{eq:chiR'} and \eqref{eq:rhoR'} and setting $R'_1 = R'_2 = R'/2$ we find
\begin{equation} \label{eq:chibarTNLL}
	\chi^{\textrm{c.t.}}_T(R') = 
	\frac12 f_T(R') \left[-2 + \chi\bigg(\frac{R'}{2}\bigg) - \rho\bigg(\frac{R'}{2}\bigg) + \psi\bigg(1+\frac{R'}{2}\bigg) + \gamma_E + \frac{2}{R'}\right] \ \ ,
\end{equation}
where $f_T(R')$ is defined in eq.~\eqref{eq:f_T(R')}. Using eq.~\eqref{eq:limitsato0} we deduce the leading $R' \to 0$ behaviour of $\chi^{\textrm{c.t.}}_T(R')$
\begin{equation} \label{eq:chiTct-leading}
	\chi^{\textrm{c.t.}}_T(R') = \frac{1}{R'} + \mathcal{O}(R') \ \ .
\end{equation}
This is the same $1/R'$ divergence as for the leading behaviour of $\chi_T(R')$ in the limit $R' \to 0$ (in eq.~\eqref{eq:hBT-leading}) and therefore $\chi^{\textrm{c.t.}}_T(R')-\chi_T(R')$ is regular (and in fact tends to 0) in the limit $R' \to 0$. 

As discussed in section~\ref{sec:subtraction}, in the limit $R' \to 0$ one can no longer neglect the higher derivatives of the radiator in eq.~\eqref{eq:chibarTbase} which, using the $\mathcal{H}_1\leftrightarrow\mathcal{H}_2$ symmetry, now reads
\begin{align} \label{eq:chibarTbase2}
	\nonumber
	\chi^{\textrm{c.t.$\,$imp}}_T (R^{(n)} )  = \, 
	&\frac{1}{2 \, \mathcal{F}_{B_T}(R')} \, \int_{\mathcal{C}} \frac{d\nu}{2\pi i\nu} \, e^{2\nu} \int\frac{d^2\vec{x}_2 \, d^2\vec{b}_2}{(2\pi)^2} \, e^{-\nu x_2} \, e^{-i\vec{b}_2\cdot\vec{x}_2} \, \ln\frac{1}{x_2} \times \\
	&\times \exp\left(\int_0^{\infty}\frac{d\zeta}{\zeta} \, R'_2(\zeta v) \, \left(J_0(b_2\zeta) \, e^{-\nu\zeta} - \Theta(1-\zeta)\right)\right) \ \ .
\end{align}
This can be evaluated exactly to give
\begin{align}\label{eq:chiTbarbase3}
	\nonumber
	\chi^{\textrm{c.t.$\,$imp}}_T(R^{(n)}) = \, 
	&\frac{1}{2 \, \mathcal{F}_{B_T}(R')} \, \int_{\mathcal{C}} \frac{d\nu}{2\pi i\nu} \, e^{2\nu} \int\frac{d^2\vec{x}_2 \, d^2\vec{b}_2}{(2\pi)^2} \, e^{-\nu x_2} \, e^{-i\vec{b}_2\cdot\vec{x}_2} \, \ln\frac{1}{x_2} \times \\
	&\times \exp\left(\lim_{\epsilon \to 0} \sum_{n=0}^\infty \frac{1}{n!} \, R^{(n+1)}   \frac{d^n}{d \epsilon^n} \left[\nu^\epsilon \, \Gamma(-\epsilon) \, {}_2 F_1\bigg(\frac{1-\epsilon}{2},-\frac{\epsilon}{2};1;-\frac{b_2^2}{\nu^2}\bigg) + \frac{1}{\epsilon}  \right]\right) \ \ ,
\end{align}
 where $R^{(n)} \equiv R^{(n)}(v)$ denotes the $n$-th logarithmic derivative of the radiator. Rescaling $b_2 \to \nu b_2$ and $x_2 \to x_2/\nu$ and introducing a change of variables $y = \sqrt{1+b_2^2}$, the $x_2$-integrals can be performed exactly to give 
\begin{align}\label{eq:chiTbarbase4}
	\nonumber
	\chi^{\textrm{c.t.$\,$imp}}_T(R^{(n)}) = \, 
	&\frac{1}{2 \, \mathcal{F}_{B_T}(R')} \, \int_{\mathcal{C}} \frac{d\nu}{2\pi i\nu} \, e^{2\nu} \int_1^\infty 
	\frac{dy}{y^2} \left(-2 + \gamma_E + \ln\frac{y^2}{1+y} + \ln 2\nu + y\right) \times \\ 
	&\times \exp\left(\lim_{\epsilon \to 0} \sum_{n=0}^\infty \frac{1}{n!} \, R_2^{(n+1)}   \frac{d^n}{d \epsilon^n} \left[\nu^\epsilon \, \Gamma(-\epsilon) \, {}_2 F_1\bigg(\frac{1-\epsilon}{2},-\frac{\epsilon}{2};1;1-y^2\bigg) + \frac{1}{\epsilon}\right]\right) \ \ .
\end{align}
In eq.~\eqref{eq:chiTbarbase4}, the accuracy sought is to have full
control over all constant terms in the limit $R' \to 0$. Therefore, in
all the terms on the first line of eq.~\eqref{eq:chiTbarbase4} that
converge as $y \to \infty$ we can neglect contributions that involve
two or higher derivatives of the radiator. This leaves the $dy/y$
term, which originally triggered the $R' \to 0$ divergence, where
higher derivatives of the radiator are now required to be taken into
account. As will become apparent below, this requires the exponent
in~\eqref{eq:chiTbarbase4} to be expanded to $\mathcal{O}(R^{(3)})$
as follows
\begin{align} \label{eq:chiTbarbase5}
	\nonumber 
	\chi^{\textrm{c.t.$\,$imp}}_T(R^{(n)}) &\simeq \, 
	\frac{e^{-\gamma_E R_2'}}{2 \, \mathcal{F}_{B_T}(R')}  \, \int_{\mathcal{C}} \frac{d\nu}{2\pi i\nu} \, e^{2\nu} \nu^{-R_2'}\, \Bigg\{ \int_1^\infty \frac{dy}{y^2} \left(-2 + \gamma_E + \ln\frac{y^2}{1+y} + \ln 2\nu \right) \left(\frac{1+y}{2}\right)^{-R_2'} \\ \nonumber
	&\qquad\qquad\quad + \, \int_1^\infty \frac{dy}{y} \left(\frac{1+y}{2}\right)^{-R_2'} \exp\left(-\frac12 R_2^{\prime\prime} \ln^2\frac{2}{(1+y)}\right) \times \\ \nonumber 
	&\qquad\qquad\quad \times \left[1 + R_2^{\prime\prime} (\gamma_E + \ln\nu) \ln\frac{1}{1+y} + \frac16 R_2^{(3)} \ln^3\frac{1}{(1+y)}\right] \Bigg\} \\ \nonumber
	&\simeq \frac{e^{-\gamma_E R_2'}}{2 \, \mathcal{F}_{B_T}(R')} \, \int_{\mathcal{C}} \frac{d\nu}{2\pi i\nu} \, e^{2\nu} \nu^{-R_2'} \, \Bigg\{ \int_1^\infty \frac{dy}{y^2} \left(-2 + \gamma_E + \ln\frac{y^2}{1+y} + \ln 2\nu + \frac{y}{1+y}\right) \left(\frac{1+y}{2}\right)^{-R_2'} \\
	&\qquad\qquad\quad + \int_0^\infty dz \, \exp\left(-R'_2 z - \frac12 R_2^{\prime\prime} z^2\right) \left[1 - R_2^{\prime\prime} (\gamma_E + \ln \nu) \, z - \frac16 R_2^{(3)} z^3\right]\Bigg\} \ \ ,
\end{align}
where we have introduced the change of variable, $z = \ln[(1+y)/2]$. We notice that it is enough to retain $R_2''$ in the exponent to guarantee the convergence of the $z$-integration as $z\to \infty$. Also, in the last line we have neglected any terms beyond our accuracy.
Setting $R^{(n)}_1 = R^{(n)}_2 = R^{(n)}/2$, this evaluates to
\begin{align} \label{eq:chiTctimp-exc-hc}
	\nonumber
	\chi^{\textrm{c.t.$\,$imp}}_T(R^{(n)}) = \, 
	\frac12 f_T(R') \Bigg[&-2 - \rho\bigg(\frac{R'}{2}\bigg) + \psi\bigg(1+\frac{R'}{2}\bigg) + \gamma_E \, + \\ 
	&+ \frac{1}{\sigma\big(\frac{R'}{2}\big)} \left\{\frac{1}{1+\frac{R'}{2}} \, {}_2 F_1\bigg(1,1;2+\frac{R'}{2};-1\bigg) + \tilde{H}_T\Big(R',R^{\prime\prime},R^{(3)}\Big)\right\}\Bigg] \ \ ,
\end{align}
where the functional form of $\tilde{H}_T$ is given by
\begin{align} \label{eq:HTtilde}
	\nonumber
	\tilde{H}_T\Big(R^\prime,R^{\prime\prime},R^{(3)}\Big) = \, 
	&\frac{2\sqrt{\pi} \, s}{R'} \, e^{s^2} \, {\rm Erfc}(s)  + \frac{2}{3} \, \frac{R^{(3)}}{\left(R^{\prime\prime}\right)^2} \, \frac{\sqrt{\pi}}{8} \, \frac{d^3}{ds^3} \, e^{s^2} \, {\rm Erfc}(s)  \\
	&+ \left(\gamma_E -\ln2 + \psi\bigg(1+\frac{R'}{2}\bigg)\right) \left(\sqrt{\pi} \, s \, e^{s^2} \, {\rm Erfc}(s) - 1\right) \ \ , \qquad s \equiv \frac{R'}{2\sqrt{R^{\prime\prime}}} \ \ .
\end{align}
The above discussion highlights the fact that terms which would be subleading from the point of view of logarithmic resummations, such as those containing $R^{\prime\prime}$ and $R^{(3)}$, do contribute to the shift at an accuracy that is within the control we claim. This suggests the need to include in $\chi^{\textrm{c.t.$\,$imp}}_T(R^{(n)})$ further sub-leading effects such as those induced by hard-collinear radiation. This is what we do in the next section.

\subsection{Improved Sudakov and Hard Collinear Element}
\label{subsec:improved-sudakov}
In the limit of $R' \to 0$, our goal is to have control over terms up to and including constants. To do so we need to consider a perturbative configuration with one hard-collinear emission on top of an ensemble of soft-collinear emissions. On the perturbative level, a single hard-collinear emission gives rise to an NNLL contribution which is of the same order as $R^{\prime\prime}$.

To include the effect of a hard-collinear emission, say to leg $\ell$, we start with the emission probability
\begin{align}
	[dk] M^2_{\textrm{h.c.}}(k) = C_F \frac{d\kappa^2}{\kappa^2} \frac{\alpha_s(\kappa)}{2\pi} \frac{d\phi}{2\pi} dz \, p_{gq}(z) \ \ ,
\end{align}
where the vector $\vec{\kappa} \equiv \vec{\kappa}^{(\ell)}$ is defined in eq.~\eqref{eq:kappadef}. In the collinear limit, we also have that $z = z^{(\ell)} $ where $z = 2E/Q$ is the energy fraction of the emission. The leading-order splitting function reads
\begin{align}
	p_{gq}(z) = \frac{2}{z} - (2-z) \ \ .
\end{align}  
In the presence of a hard-collinear emission, i.e.\
$z \sim \mathcal{O}(1)$, the transverse momentum of the ultra-soft
gluon with respect to final state partons, which we denote by
$\vec{\kappa}_{\rm us}^{(\ell)}$, is given by eq.~\eqref{eq:kappadef}
but now $z^{(\ell)}_p \neq 1$. With the hard-collinear emission in the
perturbative ensemble we have
\begin{align}
	\vec{\kappa}_{\rm us}^{(\ell)} = \left(1+\frac{z_{\rm us}^{(\ell)}}{z_p^{(\ell)}} \right)\vec{k}_{t,\rm us} - \frac{z_{\rm us}^{(\ell)}}{z_p^{(\ell)}} \vec{p}_{t,\ell} \simeq \vec{k}_{t,\rm us} - \frac{z_{\rm us}^{(\ell)}}{z_p^{(\ell)}} \vec{p}_{t,\ell} \ , \quad \vec{p}_{t,\ell}  = - \sum_{i \in \ell} \vec{k}_{ti} - \vec{k}_t \ \ ,
\end{align}
where $k_t$ is the transverse momentum of the hard-collinear emission and we used that $z_{\rm us}^{(\ell)} / z_p^{(\ell)} \ll 1$. Therefore our observable function reads
\begin{align}
		\label{eq:h-BT-hc}
		h_{B_T}(\eta^{(\ell)},\phi, \{\tilde{p}\},k, k_1,\dots,k_n)=\frac{1}{2}\left[\sqrt{1+2 e^{\eta^{(\ell)}}\frac{p_{t,\ell}}{z_p ^{(\ell)}Q}\cos\phi+e^{2\eta^{(\ell)}} \left(\frac{ p_{t,\ell}}{z_p^{(\ell)}Q}\right)^2}-e^{\eta^{(\ell)}}\frac{p_{t,\ell}}{z_p^{(\ell) }Q}\right] \ \ .
\end{align}
 It is then more convenient to work in terms of a rescaled vector $\vec{p}_\ell \equiv \vec{p}_{t,\ell}/z_p^{(\ell)}$ in terms of which the broadening of a single hemisphere due to perturbative emissions becomes
\begin{align}
	2 Q B_\ell = p_\ell + \sum_{i\in \mathcal{H}_\ell} k_{ti} + (k_t - z p_\ell) \ \ ,
\end{align}
where we remind that $k_{ti}$ refer to soft-collinear emissions and thus $z_p^{(\ell)} \simeq 1-z$. In terms of the rescaled vector, $\vec{p}_\ell$, the conservation of transverse momentum in a hemisphere $\ell$ reads
\begin{align}
	\vec{p}_\ell + \vec{\kappa} + \sum_{i\in \mathcal{H}_\ell} k_{t,i} = 0 \ \ .
\end{align} 
The hard-collinear effect will add a contribution to the function $\chi^{\textrm{c.t.$\,$imp}}_T(R^{(n)})$. We will show how this arises in this appendix. Firstly, we construct the equivalent counterterm to eq.~\eqref{eq:chibarT}
\begin{align}
  \label{eq:hchbt}
  \nonumber
  \Delta \chi^{\textrm{c.t.}}_{T}(R') = \,
  &\frac{1}{R' \mathcal{F}_{B_T}(R')} \int  d^2\vec{x}_{\bar{\ell}} \int d\mathcal{Z}[\{R'_{\ell_i},k_i\} ] \int \frac{C_F}{2\pi} \frac{d\zeta}{\zeta} \alpha_s(Q) \frac{d\phi}{2\pi} dz \, p_{qg}(z) \, \, \delta\Big(1-\max_{i}\{\zeta_{i}\}\Big) \times \\
  &\times \frac 12 \ln\frac{1}{x_{\bar{\ell}}}  \, \delta^{(2)}\Bigg(\vec{x}_{\bar{\ell}}+\sum_{i\in \mathcal{H}_{\bar{\ell}}} \vec{\zeta}_{i} + \vec{\zeta}\Bigg) \, \Theta\Bigg(1 - \frac12 \left(x_{\bar{\ell}} - z x_{\bar{\ell}}  \right)- \frac12\sum_{i\in \mathcal{H}_{\bar{\ell}}} \zeta_{i} -\frac12 | \zeta + z x_{\bar{\ell}} |\Bigg) \ \ ,
\end{align}
where $\zeta = \kappa/QB$ and we freeze the strong coupling at $Q$ as this is sufficient for our accuracy. The above expression takes care of both the hard-collinear modification to the squared matrix element in addition to the change in recoil due to the hard-collinear emission.
As written, eq.~\eqref{eq:hchbt} contains double-counting with the soft region which is already accounted for in the improved counterterm in eq.~\eqref{eq:chibarTbase2}. Therefore, we have two contributions in the above equation which we find convenient to separate out based on their physical origin
\begin{align}\label{eq:chihcsep}
	\Delta \chi^{\textrm{c.t.}}_{T}(R')  = \Delta \chi^{\textrm{c.t.}}_{T, \rm h.c.}(R') + \Delta \chi^{\textrm{c.t.}}_{T, \rm rec.}(R') \ \ .
\end{align}
The change in the observable due to the hard-collinear emission leads to the following contribution
\begin{multline}
  \label{eq:chi-rec}
  \Delta \chi^{\textrm{c.t.}}_{T, \rm rec.}(R') = \, \frac{1}{R'
    \mathcal{F}_{B_T}(R')} \int d^2\vec{x}_{\bar{\ell}} \int
  d\mathcal{Z}[\{R'_{\ell_i},k_i\} ] \times \\ \times \int \frac{C_F}{\pi}
  \frac{d\zeta}{\zeta} \alpha_s(Q) \frac{d\phi}{2\pi}\, dz p_{qg}(z)
  \, \, \delta\Big(1-\max_{i}\{\zeta_{i}\}\Big)  \frac 12 \ln\frac{1}{x_{\bar{\ell}}} \,
\delta^{(2)}\Bigg(\vec{x}_{\bar{\ell}}+\sum_{i\in
\mathcal{H}_{\bar{\ell}}} \vec{\zeta}_{i} + \vec{\zeta}\Bigg) \times \\ \times \left(\Theta\Bigg(1 - \frac12 \left(x_{\bar{\ell}} - z
        x_{\bar{\ell}} \right)- \frac12\sum_{i\in
        \mathcal{H}_{\bar{\ell}}} \zeta_{i} -\frac12 | \zeta + z
      x_{\bar{\ell}} |\Bigg) \right. \\ \left. - \, \Theta\Bigg(1 -
      \frac12 x_{\bar{\ell}} - \frac12\sum_{i\in
        \mathcal{H}_{\bar{\ell}}} \zeta_{i} -\frac12 \zeta \Bigg)
    \right) \ \ .
\end{multline}
The above expression is manifestly finite in the singular limits $\zeta \to 0$ and $z \to 0$. To understand the contribution of the above term in the limit $R' \to 0$ we merely need to Taylor expand the first step function around small recoil, $x_{\bar
\ell} \to 0$. Thanks to the fact that the difference between the step functions vanishes at $x_{\bar
\ell} = 0$, the Taylor expansion at leading order is linear in $x_{\bar\ell}$. This leads us to conclude that eq.~\eqref{eq:chi-rec}, in the limit $x_{\bar
\ell} \to 0$, yields an $\mathcal{O}(\alpha_s)$ contribution to the shift which is beyond our accuracy.

Notice that in eq.~\eqref{eq:chi-rec} the double-counting is properly subtracted, albeit that the subtraction is only needed for the $2/z$ part of the splitting function. Therefore, for the regular portion of the splitting function we need to consider the following contribution
\begin{multline}\label{eq:chi-hc}
	\Delta \chi^{\textrm{c.t.}}_{T, \rm h.c.}(R',R^{\prime\prime}) = \,
	\frac{1}{R' \mathcal{F}_{B_T}(R')} \int  d^2\vec{x}_{\bar{\ell}} \int d\mathcal{Z}[\{R'_{\ell_i},k_i\} ] \int \frac{C_F}{\pi} \frac{d\zeta}{\zeta} \alpha_s(Q) \frac{d\phi}{2\pi} dz\, (z-2) \, \, \delta\Big(1-\max_{i}\{\zeta_{i}\}\Big) \times \\
	\times \frac 12 \ln\frac{1}{x_{\bar{\ell}}} \,\bigg[\delta^{(2)}\Bigg(\vec{x}_{\bar{\ell}}+\sum_{i\in \mathcal{H}_{\bar{\ell}}} \vec{\zeta}_{i} + \vec{\zeta}\Bigg) \Theta\Bigg(1 - \frac12 x_{\bar{\ell}} - \frac12\sum_{i\in \mathcal{H}_{\bar{\ell}}} \zeta_{i} -\frac12  \zeta \Bigg) \\
	- \delta^{(2)}\Bigg(\vec{x}_{\bar{\ell}}+\sum_{i\in \mathcal{H}_{\bar{\ell}}} \vec{\zeta}_{i} \Bigg) \Theta\Bigg(1 - \frac12 x_{\bar{\ell}} - \frac12\sum_{i\in \mathcal{H}_{\bar{\ell}}} \zeta_{i}  \Bigg) \Theta(1-\zeta) \bigg] \ \ ,
\end{multline}
where the term entering with negative weight is nothing but the hard-collinear virtual corrections. It should be understood that the soft-collinear phase space measure is upgraded by including higher derivatives of the radiator, as was done in the previous subsection. Once again, the above expression is manifestly finite in the singular limit $\zeta \to 0$. As we mentioned before, our goal is to have full control over any constants in the limit $R' \to 0$. Therefore, we end up with the following expression (setting $R_1^{(n)} = R_2^{(n)} = R^{(n)}/2$)
\begin{align}
	\Delta \chi^{\textrm{c.t.}}_{T, \rm h.c.}(R',R^{\prime\prime}) = \frac{2^{\frac{R'}{2}}}{2 \mathcal{F}_{B_T}(R')} \frac{e^{-\gamma_E \frac{R'}{2}} }{\Gamma(1+\frac{R'}{2})} \frac{3C_F \alpha_s}{2\pi}  \int_0^\infty 
	dz \, z  \,\exp\left(- \frac{R'}{2} z  - \frac{R^{\prime\prime}}{4} z^2\right) \ \ ,
\end{align} 
which easily evaluates to 
\begin{align} \label{eq:hc-contribution}
	\Delta \chi^{\textrm{c.t.}}_{T, \rm h.c.}(R',R^{\prime\prime}) = \frac12 f_T(R') \, \frac{1}{\sigma\big(\frac{R'}{2}\big)} \, \frac{3C_F \alpha_s}{\pi R''} \left(1 - \sqrt{\pi} \, s \, e^{s^2} {\rm Erfc}(s)\right) \ \ .
\end{align}
We add this contribution to eq.~\eqref{eq:chiTctimp-exc-hc} and obtain our final result
\begin{align} \label{eq:chiTctimp}
	\nonumber
	\chi^{\textrm{c.t.$\,$imp}}_T(R^{(n)}) = \, 
	\frac12 f_T(R') \Bigg[&-2 - \rho\bigg(\frac{R'}{2}\bigg) + \psi\bigg(1+\frac{R'}{2}\bigg) + \gamma_E \, + \\ 
	&+ \frac{1}{\sigma\big(\frac{R'}{2}\big)} \left\{\frac{1}{1+\frac{R'}{2}} \, {}_2 F_1\bigg(1,1;2+\frac{R'}{2};-1\bigg) + H_T\Big(R',R^{\prime\prime},R^{(3)}\Big)\right\}\Bigg] \ \ ,
\end{align}
where $H_T$ is given by
\begin{equation} \label{eq:HT}
	H_T\Big(R^\prime,R^{\prime\prime},R^{(3)}\Big) = \tilde{H}_T\Big(R^\prime,R^{\prime\prime},R^{(3)}\Big) - \frac{3C_F \alpha_s}{\pi R''} \left(\sqrt{\pi} \, s \, e^{s^2} {\rm Erfc}(s) - 1\right) \ \ .
\end{equation}

\subsection{Counterterm for the Thrust Major}
\label{subsec:ct-TM}
In section~\ref{sec:subtraction} we introduced the counterterm for the thrust major which we wish to determine analytically. We start with eq.~\eqref{eq:chibarM}
\begin{align}
	\nonumber
	\chi^{\textrm{c.t.}}_M(R') = \,
	&\frac{1}{R' \mathcal{F}_{T_M}(R')} \int dx_{\bar{\ell}} \int d\mathcal{Z}[\{R'_{\ell_i},k_i\} ] \, \delta\Big(1-\max_{i}\{\zeta_{i}\}\Big) \, \frac{2}{\pi} \ln\frac{1}{|x_{\bar{\ell}}|} \, \times \\ 
	&\times \delta\Bigg(x_{\bar{\ell}}+\sum_{i\in \mathcal{H}_{\bar{\ell}}} \zeta_{i}\sin\phi_{i,\max}\Bigg) \, \Theta\Bigg(1 - |x_{\bar{\ell}}| - \sum_{i\in \mathcal{H}_{\bar{\ell}}} \zeta_{i}\Bigg) \ \ , 
\end{align}
where $\phi_{i,\max}$ denotes the angle between $\vec{k}_{t,i}$ and $\vec{k}_{t,\max}$.
We use eq.~\eqref{eq:dZ-rescaled}, the $\mathcal{H}_1\leftrightarrow\mathcal{H}_2$ symmetry and the fact that $R' = R'_1 + R'_2$. Introducing the Fourier and Mellin transforms of the kinematic and observable constraints, respectively, we may write $\chi^{\textrm{c.t.}}_M(R')$ in a simplified exponential form to give 
\begin{align} \label{eq:chibarMbase}
	\nonumber
	\chi^{\textrm{c.t.}}_M(R') = 
	&\frac{2}{\pi} \frac{R'_1}{R' \, \mathcal{F}_{T_M}(R')} \, \int_{\mathcal{C}} \frac{d\nu}{2\pi i\nu} \, e^{\nu} \int_{-\infty}^{\infty}\frac{d x_2 \, db_2}{2\pi} \, e^{-\nu |x_2|} \, e^{-ib_2x_2} \, \ln\frac{1}{|x_2|} \times \\
	&\times \exp\left(\int_0^{\infty}\frac{d\zeta}{\zeta} \, R'_2(\zeta v) \int_0^{2\pi}\frac{d\phi}{2\pi} \, \left[e^{-ib_2\zeta\sin\phi} \, e^{-\nu\zeta} - \Theta(1-\zeta)\right]\right) + 1 \leftrightarrow 2 \ \ .
\end{align}
As was the case for $\chi^{\textrm{c.t.}}_T(R')$, by construction $\chi^{\textrm{c.t.}}_M(R')$ must exactly reproduce the leading $1/R'$ behaviour of $\cNP{T_M}$ such that it can act as a local counterterm in the numerical procedure. To demonstrate this, we take the NLL approximation of the Sudakov radiator, i.e.\ $R'_2(\zeta v) \simeq R'_2(v)$, in eq.~\eqref{eq:chibarMbase} and using the $\mathcal{H}_1\leftrightarrow\mathcal{H}_2$ symmetry we find
\begin{equation}
	\chi^{\textrm{c.t.}}_M(R') = 
	\frac{2}{\pi} \frac{e^{-\gamma_E R'_2}}{\mathcal{F}_{T_M}(R')} \, \int_{\mathcal{C}} \frac{d\nu}{2\pi i\nu} \, e^{\nu} \, \nu^{-R'_2} \, \int_{-\infty}^{\infty}\frac{d x_2 \, db_2}{2\pi} \, e^{-\nu |x_2|} \, e^{-ib_2x_2} \, \ln\frac{1}{|x_2|} \,  \left(\frac{1+\sqrt{1+\frac{b_2^2}{\nu^2}}}{2}\right)^{-R'_2} \ .
\end{equation}
We rescale $b_2 \to \nu b_2$ and $x_2 \to x_2/\nu$, after which we perform the $x_2$-integral directly and find 
\begin{align} 
	\nonumber
	\chi^{\textrm{c.t.}}_M(R') = \,
	&\frac{2}{\pi}\frac{e^{-\gamma_E R'_2}}{\mathcal{F}_{T_M}(R')} \, \int_{\mathcal{C}} \frac{d\nu}{2\pi i \nu} \, e^{\nu} \, \nu^{-R'_2} \, \times \\
	&\times \frac{2}{\pi} \int_1^\infty \frac{dy}{y} \, \bigg(\frac{1+y}{2}\bigg)^{-R'_2} \, \left[\frac{\gamma_E + \ln\nu}{\sqrt{y^2-1}} + \frac{\pi}{2} + \left(\frac{\ln y}{\sqrt{y^2-1}} - \tan^{-1}\frac{1}{\sqrt{y^2-1}}\right)\right] \ \ ,
\end{align}
where $y \equiv \sqrt{1+b_2^2}$. To evaluate the $y$-integrals we make use of the following
\begin{align} 
	\label{eq:b2integral1}
	\rho_1(a) &\equiv 
	\frac{2}{\pi} \int_1^{\infty} \frac{dy}{y \sqrt{y^2 - 1}} \, \left(\frac{1+y}{2}\right)^{-a} \ \ , \\
	\label{eq:b2integral2}
	\rho_2(a) &\equiv 
	\int_1^{\infty}\frac{dy}{y} \, \left(\frac{1+y}{2}\right)^{-a} = \sigma(a)\bigg(\chi(a) + \frac{1}{a}\bigg) = \frac{1}{a} \left[2 - \sigma(a)\right]  \ \ , \\
	\label{eq:b2integral3}
	\rho_3(a) &\equiv
	\frac{2}{\pi} \int_1^{\infty}\frac{dy}{y} \, \left(\frac{1+y}{2}\right)^{-a} \left(\frac{\ln y}{\sqrt{y^2 - 1}} - \tan^{-1}\frac{1}{\sqrt{y^2 - 1}}\right) \ \ .
\end{align}
Setting $R'_1 = R'_2 = R'/2$ we therefore find
\begin{equation} \label{eq:chibarMNLL}
	\chi^{\textrm{c.t.}}_M(R') = 
	\frac{2}{\pi \, \mathcal{F}_{T_M}(R')} \, \frac{e^{-\gamma_E \frac{R'}{2}}}{\Gamma\big(1+\frac{R'}{2}\big)} \left[\left\{\gamma_E + \psi\bigg(1+\frac{R'}{2}\bigg)\right\}\rho_1\bigg(\frac{R'}{2}\bigg) + \rho_2\bigg(\frac{R'}{2}\bigg) + \rho_3\bigg(\frac{R'}{2}\bigg)\right] \ \ .
\end{equation}
To determine the leading $R' \to 0$ behaviour of $\chi^{\textrm{c.t.}}_M(R')$ we observe the following limits
\begin{align}
	\lim_{a \to 0} \rho_1(a) = 1, \quad 
	\lim_{a \to 0} \rho_2(a) = \frac{1}{a} + \ln 2, \quad
	\lim_{a \to 0} \rho_3(a)  = 0 \ \  .
\end{align}
Therefore we find
\begin{equation} \label{eq:chiMbar-leading}
	\chi^{\textrm{c.t.}}_M(R') = 
	\frac{4}{\pi} \left(\frac{1}{R'} + \frac{\ln 2}{2}\right) + \mathcal{O}(R') \ \ .
\end{equation}
We observe that this is the same $1/R'$ divergence as for the leading behaviour of $\chi_M(R')$ in the limit $R' \to 0$ in eq.~\eqref{eq:chiM-leading}. We note from section~\ref{subsec:TMsubtraction} that in the limit $R' \to 0$, $(\chi_M(R') - \chi^{\textrm{c.t.}}_M(R')) \to 0$ and we may therefore deduce the following
\begin{equation} \label{eq:chiM-leading-full}
	\chi_M(R') = 
	\frac{4}{\pi} \left(\frac{1}{R'} + \frac{\ln 2}{2}+\mathcal{O}(R')\right) \ \ .
\end{equation} 

As discussed in section~\ref{sec:subtraction}, in the limit $R' \to 0$ one can no longer neglect the higher derivatives of the radiator in eq.~\eqref{eq:chibarMbase} which, using the $\mathcal{H}_1\leftrightarrow\mathcal{H}_2$ symmetry, now reads
\begin{align} \label{eq:chiMbarbase2}
	\nonumber
	\chi^{\textrm{c.t.$\,$imp}}_M(R^{(n)}) = \, 
	&\frac{2}{\pi \, \mathcal{F}_{T_M}(R')} \, \int_{\mathcal{C}} \frac{d\nu}{2\pi i\nu} \, e^{\nu} \, \int_{-\infty}^{\infty}\frac{d x_2 \, db_2}{2\pi} \, e^{-\nu |x_2|} \, e^{-ib_2x_2} \, \ln\frac{1}{|x_2|} \, \times \\
	&\times \exp\left( \int_0^{\infty}\frac{d\zeta}{\zeta} \, R'_2(\zeta v) \, \left[J_0(b_2\zeta) \,  e^{-\nu\zeta} - \Theta(1-\zeta)\right]\right) \ \ .
\end{align}
This may be evaluated exactly to give
\begin{align} \label{eq:chiMbarbase3}
	\nonumber 
	\chi^{\textrm{c.t.$\,$imp}}_M(R^{(n)}) = \, 
	&\frac{2}{\pi \, \mathcal{F}_{T_M}(R')} \, \int_{\mathcal{C}} \frac{d\nu}{2\pi i\nu} \, e^{\nu} \int_{-\infty}^{\infty}\frac{d x_2 \, db_2}{2\pi} \, e^{-\nu |x_2|} \, e^{-ib_2x_2} \, \ln\frac{1}{|x_2|} \, \times \\
	&\times \exp\left(\lim_{\epsilon \to 0} \sum_{n=0}^\infty \frac{1}{n!} \, R^{(n+1)}   \frac{d^n}{d \epsilon^n} \left[\nu^\epsilon \, \Gamma(-\epsilon) \, {}_2 F_1\bigg(\frac{1-\epsilon}{2},-\frac{\epsilon}{2};1;-\frac{b_2^2}{\nu^2}\bigg) + \frac{1}{\epsilon}  \right]\right) \ \ ,
\end{align}
where $R^{(n)}$ denotes the $n$-th logarithmic derivative of the radiator. Rescaling $b_2 \to \nu b_2$ and $x_2 \to x_2/\nu$ and introducing a change of variables $y = \sqrt{1+b_2^2}$, the $x_2$-integral can be performed exactly to give 
\begin{align} \label{eq:chiMbarbase4}
	\nonumber 
	\chi^{\textrm{c.t.$\,$imp}}_M(R^{(n)}) = \, 
	&\frac{2}{\pi \, \mathcal{F}_{T_M}(R')} \, \int_{\mathcal{C}} \frac{d\nu}{2\pi i\nu} \, e^{\nu} \int_1^\infty \frac{dy}{y} \, \frac{2}{\pi}\left[\frac{\gamma_E + \ln\nu}{\sqrt{y^2-1}} + \frac{\pi}{2} + \left(\frac{\ln y}{\sqrt{y^2-1}} - \tan^{-1}\frac{1}{\sqrt{y^2-1}}\right)\right] \times \\ 
	&\times \exp\left(\lim_{\epsilon \to 0} \sum_{n=0}^\infty \frac{1}{n!} \, R_2^{(n+1)}   \frac{d^n}{d \epsilon^n} \left[\nu^\epsilon \, \Gamma(-\epsilon) \, {}_2 F_1\bigg(\frac{1-\epsilon}{2},-\frac{\epsilon}{2};1;1-y^2\bigg) + \frac{1}{\epsilon} \right]\right) \ \ .
\end{align}
As was the case for $\chi^{\textrm{c.t.$\,$imp}}_T(R^{(n)})$, the
accuracy sought in eq.~\eqref{eq:chiMbarbase4} is to have full control
over all constant terms in the limit $R' \to 0$. Therefore, in all the
terms on the first line of eq.~\eqref{eq:chiMbarbase4} that converge
as $y \to \infty$ we can neglect contributions containing two or
higher derivatives of the radiator. This leaves the $dy/y$ term, which
originally triggered the $R' \to 0$ divergence, where higher
derivatives of the radiator are now required to be taken into
account. As will become apparent below, this requires the exponent
in~\eqref{eq:chiMbarbase4} to be expanded to $\mathcal{O}(R^{(3)})$,
thus
\begin{align} 
	\nonumber 
	\chi^{\textrm{c.t.$\,$imp}}_M(R^{(n)}) &\simeq \, 
	\frac{2 \, e^{-\gamma_E R_2'}}{\pi \, \mathcal{F}_{T_M}(R')} \, \int_{\mathcal{C}} \frac{d\nu}{2\pi i\nu} \, e^{\nu} \, \nu^{-R_2'} \, \times \\ \nonumber 
	&\quad\,\, \times \Bigg\{ \int_1^\infty \frac{dy}{y} \, \frac{2}{\pi} \left(\frac{\gamma_E + \ln\nu}{\sqrt{y^2-1}} + \frac{\ln y}{\sqrt{y^2-1}} - \tan^{-1}\frac{1}{\sqrt{y^2-1}} \right) \left(\frac{1+y}{2}\right)^{-R_2'} \\ \nonumber
	&\qquad\quad + \, \int_1^\infty \frac{dy}{y} \left(\frac{1+y}{2}\right)^{-R_2'} \exp\left(-\frac12 R_2^{\prime\prime} \ln^2\frac{2}{(1+y)}\right) \times \\ 
	&\qquad\qquad \times \left[1 + R_2^{\prime\prime} (\gamma_E + \ln\nu) \ln\frac{1}{1+y} + \frac16 R_2^{(3)} \ln^3\frac{1}{(1+y)}\right] \Bigg\} \ \ ,
\end{align}
which we write as
\begin{align} \label{eq:chiMbarbase5}
	\nonumber 
	\chi^{\textrm{c.t.$\,$imp}}_M(R^{(n)}) &\simeq \,
	\frac{2 \, e^{-\gamma_E R_2'}}{\pi \, \mathcal{F}_{T_M}(R')} \, \int_{\mathcal{C}} \frac{d\nu}{2\pi i\nu} \, e^{\nu} \, \nu^{-R_2'} \, \times \\ \nonumber 
	&\quad\,\, \times \Bigg\{ \int_1^\infty \frac{dy}{y} \, \frac{2}{\pi} \left(\frac{\gamma_E + \ln\nu}{\sqrt{y^2-1}} + \frac{\ln y}{\sqrt{y^2-1}} - \tan^{-1}\frac{1}{\sqrt{y^2-1}} + \frac{\pi}{2(1+y)} \right) \left(\frac{1+y}{2}\right)^{-R_2'} \\
	&\qquad\quad + \int_0^\infty dz \, \exp\left(-R'_2 z - \frac12 R_2^{\prime\prime} z^2\right) \left[1 - R_2^{\prime\prime} (\gamma_E + \ln \nu) \, z - \frac16 R_2^{(3)} z^3\right]\Bigg\} \ \ ,
\end{align}	
where we have introduced the change of variable, $z = \ln[(1+y)/2]$. We notice that it is enough to retain $R^{\prime\prime}_2$ in the exponent to guarantee convergence of the $z$-integration as $z \to \infty$. In addition, in the final line we have neglected any terms beyond our accuracy. Setting $R^{(n)}_1 = R^{(n)}_2 = R^{(n)}/2$, this evaluates to 
\begin{align} \label{eq:chiMctimp-exc-hc}
	\nonumber 
	\chi^{\textrm{c.t.$\,$imp}}_M(R^{(n)}) = \,
	&\frac{2}{\pi \, \mathcal{F}_{T_M}(R')} \, \frac{e^{-\gamma_E}\frac{R'}{2}}{\Gamma\big(1+\frac{R'}{2}\big)} \Bigg[\left\{\gamma_E + \psi\left(1+\frac{R'}{2}\right)\right\}\rho_1\bigg(\frac{R'}{2}\bigg) + \rho_3\bigg(\frac{R'}{2}\bigg) \\
	&+ \frac{1}{1+\frac{R'}{2}} \, {}_2 F_1\bigg(1,1;2+\frac{R'}{2};-1\bigg) + \tilde{H}_M\Big(R',R^{\prime\prime},R^{(3)}\Big)\Bigg] \ \ ,
\end{align}
where the functional form of $\tilde{H}_M$ is given by 
\begin{align} \label{eq:HMtilde}
	\nonumber
	\tilde{H}_M\Big(R^\prime,R^{\prime\prime},R^{(3)}\Big) = \, 
	&\frac{2\sqrt{\pi} \, s}{R'} \, e^{s^2} \, {\rm Erfc}(s)  + \frac{2}{3} \, \frac{R^{(3)}}{\left(R^{\prime\prime}\right)^2} \, \frac{\sqrt{\pi}}{8} \, \frac{d^3}{ds^3} \, e^{s^2} \, {\rm Erfc}(s)  \\
	&+ \left(\gamma_E + \psi\bigg(1+\frac{R'}{2}\bigg)\right) \left(\sqrt{\pi} \, s \, e^{s^2} \, {\rm Erfc}(s) - 1\right) \ \ , \qquad s \equiv \frac{R'}{2\sqrt{R^{\prime\prime}}} \ \ .
\end{align}
As for $\tilde{H}_T(R^\prime,R^{\prime\prime},R^{(3)})$, we must also consider the contribution of one hard-collinear emission on top of an ensemble of soft-collinear emissions. The calculation of this contribution follows the same steps as is set-out in appendix~\ref{subsec:improved-sudakov} and we find analogously that
\begin{align} \label{eq:hc-contribution-TM}
	\Delta \chi^{\textrm{c.t.}}_{M, \rm h.c.}(R') = \frac{2}{\pi \, \mathcal{F}_{T_M}(R')} \, \frac{e^{-\gamma_E}\frac{R'}{2}}{\Gamma\big(1+\frac{R'}{2}\big)} \, \frac{3C_F \alpha_s}{\pi R''} \left(1 - \sqrt{\pi} \, s \, e^{s^2} {\rm Erfc}(s)\right) \ \ .
\end{align}
We add this contribution to eq.~\eqref{eq:chiMctimp-exc-hc} and obtain our final result
	\begin{align} \label{eq:chiMctimp}
		\nonumber 
		\chi^{\textrm{c.t.$\,$imp}}_M(R^{(n)}) = \,
		&\frac{2}{\pi \, \mathcal{F}_{T_M}(R')} \, \frac{e^{-\gamma_E}\frac{R'}{2}}{\Gamma\big(1+\frac{R'}{2}\big)} \Bigg[\left\{\gamma_E + \psi\left(1+\frac{R'}{2}\right)\right\}\rho_1\bigg(\frac{R'}{2}\bigg) + \rho_3\bigg(\frac{R'}{2}\bigg) \\
		&+ \frac{1}{1+\frac{R'}{2}} \, {}_2 F_1\bigg(1,1;2+\frac{R'}{2};-1\bigg) + H_M\Big(R',R^{\prime\prime},R^{(3)}\Big)\Bigg] \ \ ,
	\end{align}
	where $H_M$ is given by
	\begin{equation} \label{eq:HM}
		H_M\Big(R^\prime,R^{\prime\prime},R^{(3)}\Big) = \tilde{H}_M\Big(R^\prime,R^{\prime\prime},R^{(3)}\Big) - \frac{3C_F \alpha_s}{\pi R''} \left(\sqrt{\pi} \, s \, e^{s^2} {\rm Erfc}(s) - 1\right) \ \ .
	\end{equation}

\section{Analytical determination of NP corrections to the means}
\label{sec:NP-means}
This appendix presents analytical computations of the non-perturbative corrections to the mean values for the jet broadenings and the thrust major.

\subsection{Single-Jet Broadening}
\label{subsec:mean-B1}
Without loss of generality, for the single-jet broadening we shall consider $B_1$. We start with eq.~\eqref{eq:mean-NP-simplified}
\begin{equation}
	\langle B_1 \rangle_{\rm NP} \simeq 
	\mathcal{M}\frac{\langle \kappa \rangle_{\rm NP}}{Q} \int\!d\eta^{(1)} \, \frac{d\phi}{2\pi} \int dZ_{\rm sc}[\{k_i\}] \, h_{B_1}(\eta^{(1)},\phi,\{\tilde{p}\},\{k_i\}) \, \Theta\Big(B_{1,\max} - B_{1,\rm sc}(\{\tilde{p}\},\{k_i\})\Big) \ \ .
\end{equation}
Following the approach of section~\ref{sec:observables}, from eq.~\eqref{eq:B-eta-phi-integral}
\begin{equation}
	\int\!d\eta^{(1)} \, \int\!\frac{d\phi}{2\pi} \, h_{B_1}(\eta^{(1)},\phi,\{\tilde{p}\},\{k_i\}) = 
	\frac12 \left[\ln\frac{Q}{p_{t, 1}} + \eta_0^{(B)}\right] + \mathcal{O}\bigg(\frac{p_{t, 1}}{Q}\bigg) \ \ .
\end{equation}
Therefore we obtain
\begin{equation}
	\langle B_1 \rangle_{\rm NP} \simeq 
	\mathcal{M}\frac{\langle \kappa \rangle_{\rm NP}}{Q} \int dZ_{\rm sc}[\{k_i\}] \, \frac{1}{2}\left[\ln\frac{Q}{p_{t, 1}} + \eta_0^{(B)}\right] \, \Theta\bigg(1-\frac{B_{1,\rm sc}(\{\tilde{p}\},\{k_i\})}{B_{1,\max}}\bigg) \ \ .
\end{equation}
We note the presence of $B_{1,\max}$ in the denominator of the observable constraint. As $B_{1,\max}$ corresponds to $R' = 0$ we may write our above equation as
\begin{equation} \label{eq:B1-mean-base}
	\langle B_1 \rangle_{\rm NP} \simeq 
	\lim_{R'\to 0}\left[\mathcal{M}\frac{\langle \kappa \rangle_{\rm NP}}{Q} \int dZ_{\rm sc}[\{k_i\}]\,\frac{1}{2} \left[\ln\frac{Q}{p_{t, 1}} + \eta_0^{(B)}\right] \, \Theta\bigg(1-\frac{B_{1,\rm sc}(\{\tilde{p}\},\{k_i\})}{B_1}\bigg)\right] \ \ .
\end{equation}
It is possible to solve eq.~\eqref{eq:B1-mean-base} by evaluating the square brackets first, for general $R'$, before taking the limit $R' \to 0$. However, this approach will only be amenable for observables that allow an analytic computation. Instead, we may consider the limit $R' \to 0$ at first. 

In the limit $R' \to 0$, our hemisphere of interest will contain a single emission, which we denote $k_1$, with a transverse momentum of order $B_1 Q$. We consider the observable when all other emissions, $k_i$, have transverse momenta less than $\delta B_1 Q$ where $\epsilon \ll \delta \ll 1$ thus 
\begin{equation}
	B_{1,{\rm sc}}(\{\tilde{p}\},\{k_i\}) \simeq
	B_{1,{\rm sc}}(\{ \tilde{p}\},k_1) \ \ .
\end{equation}
We use eqs.~\eqref{eq:dZsc-final} and~\eqref{eq:dZ-rescaled} and, in a similar way as for $\cNP{B_T}$, introduce the rescaled variables $\vec{x}_1 = \vec{p}_{t,1}/(B_1 Q)$ and the two-dimensional vectors $\vec{\zeta}_i \equiv \zeta_i(\cos\phi_i,\sin\phi_i)$ and find
\begin{align} \label{eq:B1-mean-NLL}
	\langle B_1 \rangle_{\rm NP} \simeq 
	\lim_{R' \to 0} \Bigg[&\mathcal{M}\frac{\langle \kappa \rangle_{\rm NP}}{Q} \, \int_0^{\infty}\frac{d\zeta_1}{\zeta_1} \, \zeta_1^{R'_1(\zeta_1 v)} \, R'_1(\zeta_1 v) \, \frac12\left[\ln\frac{1}{\zeta_1} + \eta_0^{(B)} \right] \, \Theta(1-\zeta_1)\Bigg] \ \ .
\end{align}
We note that the factor of $e^{-R(v)}$ in eq.~\eqref{eq:dZsc-final} has been dropped as we are in the limit $R' \to 0$, where $B_1 \to B_{1,\max}$ and thus $e^{-R\left(v_{\max}\right)} \to 1$. We also drop the $\ln( 1/B_1)$ term as this will vanish in the limit $R' \to 0$ as $B_{1,\max} \simeq 1$ (with any resultant discrepancies of order $\sqrt{\alpha_s}$ and thus outside of our desired accuracy).

To identify the precise form of any $1/R'$ behaviour we take the NLL approximation of the Sudakov radiator, i.e. $R'_1(\zeta_1 v) \simeq R'_1(v)$, in eq.~\eqref{eq:B1-mean-NLL}. Performing the $\zeta_1$-integration gives
\begin{equation}
	\langle B_1 \rangle_{\rm NP} \simeq 
	\mathcal{M}\frac{\langle \kappa \rangle_{\rm NP}}{Q} \, \frac12 \left(\frac{1}{R'_1} + \eta_0^{(B)}\right) \ \ .
\end{equation}
As discussed in section~\ref{sec:subtraction}, in the limit $R' \to 0$ one can no longer neglect the higher derivatives of the radiator in eq.~\eqref{eq:B1-mean-NLL}. The accuracy that we require is to have full control over all constant terms in the limit $R' \to 0$. We therefore return to eq.~\eqref{eq:B1-mean-NLL} and, where appropriate, perform the logarithmic expansion of $R'_1(\zeta_1 v)$, expanding to $\mathcal{O}(R^{(3)})$ to achieve the required accuracy. We find
\begin{align}
	\nonumber
	\langle B_1 \rangle_{\rm NP} \simeq 
	\mathcal{M}\frac{\langle \kappa \rangle_{\rm NP}}{Q} \Bigg\{&\frac12\eta_0^{(B)} + \lim_{R' \to 0} \Bigg[\int_0^{\infty}\frac{d\zeta_1}{\zeta_1} \, \exp\left(-R'_1 \ln\frac{1}{\zeta_1} - \frac{R_1^{\prime\prime}}{2}\ln^2\frac{1}{\zeta_1} - \frac{R_1^{(3)}}{6}\ln^3\frac{1}{\zeta_1}\right) \times \\
	&\times \, \left(R'_1 + R_1^{\prime\prime}\ln\frac{1}{\zeta_1} + \frac{R_1^{(3)}}{2}\ln^2\frac{1}{\zeta_1}\right) \, \frac{1}{2}\ln\frac{1}{\zeta_1} \, \Theta(1-\zeta_1)\Bigg]\Bigg\} \ \ .
\end{align}
Performing the $\zeta_1$-integration by parts and setting $R_1^{(n)} = R_2^{(n)} = R^{(n)}/2$ we obtain  
\begin{align}
	\langle B_1 \rangle_{\rm NP} \simeq \,\, 
	&\mathcal{M}\frac{\langle \kappa \rangle_{\rm NP}}{Q} \, \frac{1}{2}\Bigg(\eta_0^{(B)} + \lim_{R' \to 0} \Bigg[\frac{2\sqrt{\pi} \, s}{R'} \,e^{s^2} \, {\rm Erfc}(s)  + \frac{2}{3}\frac{R^{(3)}}{\left(R^{\prime\prime}\right)^2} \, \frac{\sqrt{\pi}}{8}\frac{d^3}{ds^3} e^{s^2} \, {\rm Erfc}(s) \Bigg]\Bigg) \ \ ,
\end{align}
where $s = R'/2\sqrt{R^{\prime\prime}}$. We must also include the contribution of one hard-collinear emission on top of an ensemble of soft-collinear emissions, the calculation of which is set-out in appendix~\ref{subsec:improved-sudakov}. We therefore obtain
\begin{align}
	\nonumber
	\langle B_1 \rangle_{\rm NP} \simeq \,\,
	&\mathcal{M}\frac{\langle \kappa \rangle_{\rm NP}}{Q} \, \frac{1}{2}\Bigg(\eta_0^{(B)} + \lim_{R' \to 0} \Bigg[\frac{2\sqrt{\pi} \, s}{R'} \,e^{s^2} \, {\rm Erfc}(s)  + \frac{2}{3}\frac{R^{(3)}}{\left(R^{\prime\prime}\right)^2} \, \frac{\sqrt{\pi}}{8}\frac{d^3}{ds^3} e^{s^2} \, {\rm Erfc}(s) \\
	&- \frac{3C_F\alpha_s}{\pi\, R^{\prime\prime}} \left(\sqrt{\pi} \, s \, e^{s^2}  {\rm Erfc}(s) - 1 \right) \Bigg]\Bigg) \ \ .
\end{align} In the limit $R' \to 0$, $R' \ll \sqrt{R^{\prime\prime}}$ and from eq.~\eqref{eq:erfc-two-regimes} we find, in agreement with \cite{Dokshitzer:1998qp} up to terms of order $\sqrt{\alpha_s}$, that
\begin{equation} \label{eq:mean-single}
	\langle B_1 \rangle_{\rm NP} \simeq
	\mathcal{M}\frac{\langle \kappa \rangle_{\rm NP}}{Q} \, \frac12 \left[\eta_0^{(B)} + \frac{\pi}{2\sqrt{C_F\,\alpha_s(Q)}} + \frac34 - \frac{2\pi\beta_0}{3 C_F}\right] + \mathcal{O}(\sqrt{\alpha_s}) \ \ .
\end{equation}

\subsection{Wide-Jet Broadening}
\label{subsec:mean-BW}
The analytic determination of the non-perturbative correction to the mean value for the wide-jet broadening follows a similar approach to that for the single-jet broadening in appendix~\ref{subsec:mean-B1}, but with one important difference that we shall see below. As before we start with eq.~\eqref{eq:mean-NP-simplified}
\begin{equation}
	\langle B_W \rangle_{\rm NP} \simeq 
	\mathcal{M}\frac{\langle \kappa \rangle_{\rm NP}}{Q} \sum_{\ell} \int\!d\eta^{(\ell)} \, \frac{d\phi}{2\pi} \int dZ_{\rm sc}[\{k_i\}] \, h_{B_W}(\eta^{(\ell)},\phi,\{\tilde{p}\},\{k_i\}) \, \Theta\Big(B_{W,\max} - B_{W,\rm sc}(\{\tilde{p}\},\{k_i\})\Big) \ \ .
\end{equation}
Following the approach of section~\ref{sec:observables}, from eq.~\eqref{eq:B-eta-phi-integral}
\begin{equation}
	\sum_{\ell} \int\!d\eta^{(\ell)} \, \frac{d\phi}{2\pi} \, h_{B_W}(\eta^{(\ell)},\phi,\{\tilde{p}\},\{k_i\}) = 
	\frac12 \left[\ln\frac{Q}{p_{t, 1}} + \eta_0^{(B)} + \mathcal{O}\bigg(\frac{p_{t, 1}}{Q}\bigg)\right] \Theta(B_1-B_2) + 1 \leftrightarrow 2 \ \ .
\end{equation}
We find
\begin{align}
	\nonumber
	\langle B_W \rangle_{\rm NP} \simeq \, 
	&\lim_{R'\to 0}\left[\mathcal{M}\frac{\langle \kappa \rangle_{\rm NP}}{Q} \int dZ_{\rm sc}[\{k_i\}] \, \frac12 \left[\ln\frac{Q}{p_{t, 1}} + \eta_0^{(B)}\right] \Theta(B_1-B_2) \, \Theta\bigg(1-\frac{B_{1,\rm sc}(\{\tilde{p}\},\{k_i\})}{B_{1}}\bigg)\right] \\
	&+ 1 \leftrightarrow 2 \ \ .
\end{align}
In the limit $R' \to 0$, our wide-hemisphere will contain a single emission, which we denote $k_1$, with a transverse momentum of order $B_W Q$. We consider the observable when all other emissions, $k_i$, have transverse momenta less than $\delta B_W Q$ where $\epsilon \ll \delta \ll 1$ thus 
\begin{equation}
	B_{W,{\rm sc}}(\{\tilde{p}\},\{k_i\})\simeq 
	B_{1,{\rm sc}}(\{ \tilde{p}\},k_1) \ \ .
\end{equation}
We use eqs.~\eqref{eq:dZsc-final} and~\eqref{eq:dZ-rescaled} and introduce the rescaled variables $\vec{x}_{\ell} = \vec{p}_{t,\ell}/(B_W Q)$ and the two-dimensional vectors $\vec{\zeta}_i \equiv \zeta_i(\cos\phi_i,\sin\phi_i)$. We perform the rescaling $\zeta_i \to \zeta_1 \zeta_i$ and, as $R' = R'_1 + R'_2$, we find
\begin{align} \label{eq:BW-mean-NLL}
	\nonumber
	\langle B_W \rangle_{\rm NP} \simeq 
	&\lim_{R' \to 0} \Bigg(\mathcal{M}\frac{\langle \kappa \rangle_{\rm NP}}{Q} \, \Bigg[\int_0^{\infty} \frac{d\zeta_1}{\zeta_1} \, \zeta_1^{R'(\zeta_1 v)} \, R'_1(\zeta_1 v) \, \frac12 \left[\ln\frac{1}{\zeta_1}+\eta_0^{(B)}\right] \, \Theta(1-\zeta_1)\Bigg] \times \\ \nonumber
	&\times \Bigg[\int_{-\infty}^{\infty}d^2\vec{x}_2 \, \left(\lim_{\epsilon \to 0} \, \epsilon^{R'_2} \, \sum_{n=0}^{\infty}\frac{\left(R'_2\right)^n}{n!} \, \prod_{i=2}^{n+1} \, \int_{\epsilon}^{\delta}\frac{d\zeta_i}{\zeta_i} \, \int_0^{2\pi}\frac{d\phi_i}{2\pi}\right) \, \delta^{(2)}\bigg(\vec{x}_2+\sum_{i\in \mathcal{H}_2}\vec{\zeta}_{i}\bigg) \times \\
	&\qquad\times \Theta\Bigg(1-\frac12|\vec{x}_2|-\frac12\sum_{i \in \mathcal{H}_2}{\zeta}_i\Bigg)\Bigg]\Bigg) + 1 \leftrightarrow 2 \ \ .
\end{align}
We note that the factor of $e^{-R(v)}$ in eq.~\eqref{eq:dZsc-final} has been dropped as we are in the limit $R' \to 0$ where $B_W \to B_{W,\max}$ and thus $e^{-R\left(v_{\max}\right)} \to 1$. We also drop the $\ln (1/B_W)$ term as this will vanish in the limit $R' \to 0$ as $B_{W,\max} \simeq 1$ as before.

We note a factorisation of the integrals over emissions in $\mathcal{H}_1$ and $\mathcal{H}_2$. The second set of square brackets, containing the integrals over $\mathcal{H}_2$, evaluates to $1 + \mathcal{O}\big((R'_2)^2\big)$ in the limit $R' \to 0$ and may therefore be dropped. For the first set of square brackets, containing the integrals over $\mathcal{H}_1$, we note that the rescaling of the emissions in the non-wide hemisphere produces an exponent of $R'(\zeta_1 v)$ rather than $R'_1(\zeta_1 v)$ as was the case for $\langle B_1 \rangle_{\rm NP}$. 

To identify the precise form of any $1/R'$ behaviour we take the NLL approximation of the Sudakov radiator, i.e. $R'_{\ell}(\zeta_1 v) \simeq R'_{\ell}(v)$, in eq.~\eqref{eq:BW-mean-NLL}. Performing the $\zeta_1$-integration gives 
\begin{equation}
	\langle B_W \rangle_{\rm NP} \simeq 
	\mathcal{M}\frac{\langle \kappa \rangle_{\rm NP}}{Q} \, \frac12 \left(\frac{1}{R'_1 + R'_2} + \eta_0^{(B)}\right) \ \ .
\end{equation}
As discussed in section~\ref{sec:subtraction}, in the limit $R' \to 0$ one can no longer neglect the higher derivatives of the radiator in eq.~\eqref{eq:BW-mean-NLL}. The accuracy that we require is to have full control over all constant terms in the limit $R' \to 0$. We therefore return to eq.~\eqref{eq:BW-mean-NLL} and, where appropriate, perform the logarithmic expansion of $R'_{\ell}(\zeta_1 v)$, expanding to $\mathcal{O}(R^{(3)})$ to achieve the required accuracy. Using the $\mathcal{H}_1 \leftrightarrow \mathcal{H}_2$ symmetry and setting $R_1^{(n)} = R_2^{(n)} = R^{(n)}/2$, we find
\begin{align}
	\nonumber
	\langle B_W \rangle_{\rm NP} \simeq 
	\mathcal{M}\frac{\langle \kappa \rangle_{\rm NP}}{Q} \Bigg\{&\frac12\eta_0^{(B)} + \lim_{R' \to 0} \Bigg[\int_0^{\infty}\frac{d\zeta_1}{\zeta_1} \, \exp\bigg(-R' \ln\frac{1}{\zeta_1} - \frac{R^{\prime\prime}}{2}\ln^2\frac{1}{\zeta_1} - \frac{R^{(3)}}{6}\ln^3\frac{1}{\zeta_1}\bigg) \times \\
	&\times \, \frac12 \left(R' + R^{\prime\prime}\ln\frac{1}{\zeta_1} + \frac{R^{(3)}}{6}\ln^2\frac{1}{\zeta_1}\right) \, \ln\frac{1}{\zeta_1} \, \Theta(1-\zeta_1)\Bigg]\Bigg\}  \ \ .
\end{align}
Performing the $\zeta_1$-integration by parts we obtain 
\begin{align}
	\nonumber
	\langle B_W \rangle_{\rm NP} \simeq \,\,
	&\mathcal{M}\frac{\langle \kappa \rangle_{\rm NP}}{Q} \, \frac{1}{2}\Bigg(\eta_0^{(B)} + \lim_{R' \to 0} \Bigg[\frac{\sqrt{\pi} \, s}{R'} \,e^{s^2} \, {\rm Erfc}(s)  + \frac{1}{3}\frac{R^{(3)}}{\left(R^{\prime\prime}\right)^2} \, \frac{\sqrt{\pi}}{8}\frac{d^3}{ds^3} e^{s^2} \, {\rm Erfc}(s) \Bigg]\Bigg) \ \ ,
\end{align}
where we highlight that crucially here $s = R'/\sqrt{2R^{\prime\prime}}$. We must also include the contribution of one hard-collinear emission on top of an ensemble of soft-collinear emissions, the calculation of which is set-out in appendix~\ref{subsec:improved-sudakov}. We therefore obtain
\begin{align}
	\nonumber
	\langle B_W \rangle_{\rm NP} \simeq \,\, 
	&\mathcal{M}\frac{\langle \kappa \rangle_{\rm NP}}{Q} \, \frac{1}{2}\Bigg(\eta_0^{(B)} + \lim_{R' \to 0} \Bigg[\frac{\sqrt{\pi} \, s}{R'} \,e^{s^2} \, {\rm Erfc}(s)  + \frac{1}{3}\frac{R^{(3)}}{\left(R^{\prime\prime}\right)^2} \, \frac{\sqrt{\pi}}{8}\frac{d^3}{ds^3} e^{s^2} \, {\rm Erfc}(s) \\
	&- \frac{3C_F\alpha_s}{\pi\, R^{\prime\prime}} \left(\sqrt{\pi} \, s \, e^{s^2}  {\rm Erfc}(s) - 1 \right) \Bigg]\Bigg) \ \ .
\end{align} 
In the limit $R' \to 0$, $R' \ll \sqrt{R^{\prime\prime}}$ and from eq.~\eqref{eq:erfc-two-regimes} we find, in agreement with \cite{Dokshitzer:1998qp} up to terms of order $\sqrt{\alpha_s}$, that
\begin{equation} \label{eq:mean-wide}
	\langle B_W \rangle_{\rm NP} \simeq 
	\mathcal{M}\frac{\langle \kappa \rangle_{\rm NP}}{Q} \, \frac12 \left[\eta_0^{(B)} + \frac{\pi}{2\sqrt{2C_F\,\alpha_s(Q)}} + \frac34 - \frac{\pi\beta_0}{3 C_F}\right] + \mathcal{O}(\sqrt{\alpha_s}) \ \ .
\end{equation}
We note that eq.~\eqref{eq:mean-wide} takes a form similar to that in eq.~\eqref{eq:mean-single} but with $C_F \to 2C_F$ due to the rescaling of the emissions in the non-wide hemisphere producing an exponent of $R'(\zeta_1 v)$ rather than $R'_1(\zeta_1 v)$.

\subsection{Total Broadening}
\label{subsec:mean-BT}
The analytic determination of the non-perturbative correction to the mean value for the total broadening follows a similar approach to that for the single-jet broadening in appendix~\ref{subsec:mean-B1}. As before we start with eq.~\eqref{eq:mean-NP-simplified}
\begin{equation}
	\langle B_T \rangle_{\rm NP} \simeq 
	\mathcal{M}\frac{\langle \kappa \rangle_{\rm NP}}{Q} \sum_{\ell} \int\!d\eta^{(\ell)} \, \frac{d\phi}{2\pi} \int dZ_{\rm sc}[\{k_i\}] \, h_{B_T}(\eta^{(\ell)},\phi,\{\tilde{p}\},\{k_i\}) \, \Theta\Big(B_{T,\max} - B_{T,\rm sc}(\{\tilde{p}\},\{k_i\})\Big) \ \ .
\end{equation}
Following the approach of section~\ref{sec:observables}, from eq.~\eqref{eq:B-eta-phi-integral}
\begin{equation}
	\sum_{\ell} \int\!d\eta^{(\ell)}\,\frac{d\phi}{2\pi}\,h_{B_T}(\eta^{(\ell)},\phi,\{\tilde{p}\},\{k_i\}) = 
	\frac12\ln\frac{Q}{p_{t, 1}} + \frac12\ln\frac{Q}{p_{t, 2}} + \eta_0^{(B)} + \mathcal{O}\bigg(\frac{p_{t,\ell}}{Q}\bigg) \ \ .
\end{equation}
In the limit $R' \to 0$, one hemisphere, say $\mathcal{H}_1$, will contain a single emission, which we denote $k_1$, with a transverse momentum of order $B_T Q$. We consider the observable when all other emissions, $k_i$, have transverse momenta less than $\delta B_T Q$ where $\epsilon \ll \delta \ll 1$ thus 
\begin{equation}
	B_{T,{\rm sc}}(\{\tilde{p}\},\{k_i\})\simeq 
	B_{T,{\rm sc}}(\{ \tilde{p}\},k_1) \ \ .
\end{equation}
We use eqs.~\eqref{eq:dZsc-final} and~\eqref{eq:dZ-rescaled} and introduce the rescaled variables $\vec{x}_{\ell} = \vec{p}_{t,\ell}/(B_T Q)$ and the two-dimensional vectors $\vec{\zeta}_i \equiv \zeta_i(\cos\phi_i,\sin\phi_i)$. We perform the rescaling $\zeta_i \to \zeta_1 \zeta_i$ and, as $R' = R'_1 + R'_2$, we find
\begin{align} \label{eq:BT-mean-NLL}
	\nonumber
	\langle B_T \rangle_{\rm NP} \simeq 
	&\lim_{R' \to 0} \Bigg[\mathcal{M}\frac{\langle \kappa \rangle_{\rm NP}}{Q} \, \int_0^{\infty}\frac{d\zeta_1}{\zeta_1} \, \zeta_1^{R'(\zeta_1 v)} \, R'_1(\zeta_1 v) \, \int_{-\infty}^{\infty}d^2\vec{x}_2 \, \delta^{(2)}\bigg(\vec{x}_2+\sum_{i\in \mathcal{H}_2}\vec{\zeta}_{i}\bigg) \, \Theta(1-\zeta_1) \, \times \\
	&\times \left(\lim_{\epsilon \to 0} \, \epsilon^{R'_2} \, \sum_{n=0}^{\infty}\frac{\left(R'_2\right)^n}{n!} \, \prod_{i=2}^{n+1} \, \int_{\epsilon}^{1}\frac{d\zeta_i}{\zeta_i} \, \int_0^{2\pi}\frac{d\phi_i}{2\pi}\right) \, \left(\ln\frac{1}{\zeta_1} + \frac12\ln\frac{1}{|\vec{x}_2|} + \eta_0^{(B)}\right) \, \Bigg] + 1 \leftrightarrow 2 \ \ .
\end{align}
We note that the factor of $e^{-R(v)}$ in eq.~\eqref{eq:dZsc-final} has been dropped as we are in the limit $R' \to 0$ where $B_T \to B_{T,\max}$ and thus $e^{-R\left(v_{\max}\right)} \to 1$. We also drop the $\ln (1/B_T)$ term as this will vanish in the limit $R' \to 0$ as $B_{T, \rm{max}} \simeq 1$ as before.

We consider the three terms in the final pair of brackets:
\begin{itemize}
	\item 
	For the $\ln (1/\zeta_1)$ and $\eta_0^{(B)}$ terms there is no dependence on emissions in $\mathcal{H}_2$ and therefore the integrals over $\mathcal{H}_2$ will trivially evaluate to 1.
	\item 
	For the $\ln (1/|\vec{x}_2|)$ term, we of course have a dependence on emissions in $\mathcal{H}_2$ but we note that the integrals over $\mathcal{H}_1$ and $\mathcal{H}_2$ will still factorise. We set the upper-bound of the $\zeta_i$-integration to $\delta$ (to reflect the imposed restriction on transverse momenta of emissions in $\mathcal{H}_2$) and recognise the integrals over $\mathcal{H}_2$ as precisely that in eq.~\eqref{eq:chiT-smallR'-start}. We recall from eq.~\eqref{eq:chiT-smallR'-result} that this will give $1/(2R'_2) + \mathcal{O}(R')$.
\end{itemize}
For the integrals over $\mathcal{H}_1$ we note that, as was the case for $\langle B_W \rangle_{\rm NP}$, rescaling the emissions in $\mathcal{H}_2$ produces an exponent of $R'(\zeta_1 v)$ rather than $R'_1(\zeta_1 v)$ (as was present for $\langle B_1 \rangle_{\rm NP}$). 

To identify the precise form of any $1/R'$ behaviour we take the NLL approximation of the Sudakov radiator, i.e. $R'_{\ell}(\zeta_1 v) \simeq R'_{\ell}(v)$, in eq.~\eqref{eq:BT-mean-NLL}. Performing the $\zeta_1$-integration we find     
\begin{equation}
	\langle B_T \rangle_{\rm NP} \simeq 
	\mathcal{M}\frac{\langle \kappa \rangle_{\rm NP}}{Q} \, \left[\frac{R'_1}{(R'_1 + R'_2)^2} + \eta_0^{(B)}\frac{R'_1}{R'_1+R'_2} + \frac{R'_1}{2R'_2(R'_1+R'_2)}\right] + 1 \leftrightarrow 2 \ \ ,
\end{equation}
which simplifies to give
\begin{equation}
	\langle B_T \rangle_{\rm NP} \simeq 
	\frac{\langle \kappa \rangle_{\rm NP}}{Q} \, \left[\frac{1}{2R'_1} + \frac{1}{2R'_2} + \eta_0^{(B)}\right] \ \ .
\end{equation}
As discussed in section~\ref{sec:subtraction}, in the limit $R' \to 0$ one can no longer neglect the higher derivatives of the radiator in eq.~\eqref{eq:BT-mean-NLL}. The accuracy that we require is to have full control over all constant terms in the limit $R' \to 0$. We therefore return to eq.~\eqref{eq:BT-mean-NLL} which, using the  $\mathcal{H}_1 \leftrightarrow \mathcal{H}_2$ symmetry, we write in a simplified manner as 
\begin{align} \label{eq:BT-mean-improved-Sudakov}
	\nonumber
	\langle B_T \rangle_{\rm NP} \simeq 
	\mathcal{M}\frac{\langle \kappa \rangle_{\rm NP}}{Q} \Bigg\{&\eta_0^{(B)} + \lim_{R' \to 0}\Bigg[ \int_{-\infty}^{\infty}d^2\vec{x}_2 \, \left(\lim_{\epsilon \to 0} \, \epsilon^{R'_2(\zeta v)} \, \sum_{n=0}^{\infty}\frac{\left(R'_2\big(\zeta v\big)\right)^n}{n!} \, \prod_{i=2}^{n+1} \, \int_{\epsilon}^{\delta}\frac{d\zeta_i}{\zeta_i} \, \int_0^{2\pi}\frac{d\phi_i}{2\pi}\right) \times \\
	&\times \delta^{(2)}\bigg(\vec{x}_2+\sum_{i\in \mathcal{H}_2}\vec{\zeta}_{i}\bigg) \, \ln\frac{1}{|\vec{x}_2|}\Bigg]\Bigg\} \ \ .
\end{align}
To achieve the required accuracy we will perform the logarithmic expansion of $R'_{\ell}(\zeta_1 v)$, expanding to $\mathcal{O}(R^{(3)})$. To enable convenient analytic computation we may introduce the same theta-constraint as we had for $\chi^{\textrm{c.t.}}_T(R')$ and  $\chi^{\textrm{c.t.$\,$imp}}_T(R')$. This constraint is trivially satisfied in the limit $R' \to 0$ and thus will not affect the result to the given accuracy. We therefore write
\begin{align}
	\nonumber
	\langle B_T \rangle_{\rm NP} \simeq 
	\mathcal{M}\frac{\langle \kappa \rangle_{\rm NP}}{Q} \Bigg\{&\eta_0^{(B)} + \lim_{R' \to 0}\Bigg[\int_{-\infty}^{\infty}d^2\vec{x}_2 \, \left(\lim_{\epsilon \to 0} \, \epsilon^{R'_2(\zeta v)} \, \sum_{n=0}^{\infty}\frac{\left(R'_2\big(\zeta v\big)\right)^n}{n!} \, \prod_{i=2}^{n+1} \, \int_{\epsilon}^{\infty}\frac{d\zeta_i}{\zeta_i} \, \int_0^{2\pi}\frac{d\phi_i}{2\pi}\right) \times \\
	&\times \delta^{(2)}\bigg(\vec{x}_2+\sum_{i\in \mathcal{H}_2}\vec{\zeta}_{i}\bigg) \, \ln\frac{1}{|\vec{x}_2|} \, \Theta\Bigg(1 - \frac12 |\vec{x}_2| - \frac12\sum_{i\in \mathcal{H}_2} \zeta_{i}\Bigg)\Bigg]\Bigg\} \ \ ,
\end{align}
noting that the theta-constraint allows us to send the upper-bound of the $\zeta_i$-integration to infinity. Setting $R_1^{(n)} = R_2^{(n)} = R^{(n)}/2$, we deduce from eq.~\eqref{eq:chibarTbase2} that
\begin{equation}
	\langle B_T \rangle_{\rm NP} \simeq 
	\mathcal{M}\frac{\langle \kappa \rangle_{\rm NP}}{Q} \left(\eta_0^{(B)} + \lim_{R' \to 0}\Big[2 \, \mathcal{F}_{B_T}(R') \, \chi^{\textrm{c.t.$\,$imp}}_T(R^{(n)}) \Big]\right) \ \ .
\end{equation}
In the limit $R' \to 0$, $R' \ll \sqrt{R^{\prime\prime}}$ and from eq.~\eqref{eq:erfc-two-regimes} we find, in agreement with \cite{Dokshitzer:1998qp} up to terms of order $\sqrt{\alpha_s}$, that
\begin{equation} \label{eq:mean-total}
	\langle B_T \rangle_{\rm NP} \simeq 
	\mathcal{M}\frac{\langle \kappa \rangle_{\rm NP}}{Q} \, \left[\eta_0^{(B)} + \frac{\pi}{2\sqrt{C_F\,\alpha_s(Q)}} + \frac34 - \frac{2\pi\beta_0}{3 C_F}\right] + \mathcal{O}(\sqrt{\alpha_s}) \ \ .
\end{equation}
We note, as expected, that $\langle B_T \rangle_{\rm NP} = 2\langle B_1 \rangle_{\rm NP}$. 

\subsection{Thrust Major}
\label{subsec:mean-TM}
The analytic determination of the non-perturbative correction to the mean value for the thrust major follows a similar approach to that for the total broadening in appendix~\ref{subsec:mean-BT}. As before we start with eq.~\eqref{eq:mean-NP-simplified}
\begin{equation}
	\langle T_M \rangle_{\rm NP} \simeq 
	\mathcal{M}\frac{\langle \kappa \rangle_{\rm NP}}{Q} \, \sum_{\ell} \int\!d\eta^{(\ell)} \, \frac{d\phi}{2\pi} \int dZ_{\rm sc}[\{k_i\}] \, h_{T_M}(\eta^{(\ell)},\phi,\{\tilde{p}\},\{k_i\}) \, \Theta\Big(T_{M,\max} - T_{M,\rm sc}(\{\tilde{p}\},\{k_i\})\Big) \ \ .
\end{equation}
Following the approach of Section~\ref{sec:observables}, from eq.~\eqref{eq:TM-eta-phi-integrals}
\begin{equation}
	\sum_{\ell}\int\!d\eta^{(\ell)} \, \frac{d\phi}{2\pi} \, h_{T_M}(\eta^{(\ell)},\phi,\{\tilde{p}\},\{k_i\}) = 
	\frac{2}{\pi} \left(\ln\frac{Q}{|p_{y,1}|} + \ln\frac{Q}{|p_{y,2}|} + 2\left(\ln 2 - 2\right)\right) + \mathcal{O}\bigg(\frac{|p_{y,\ell}|}{Q}\bigg) \ \ . 
\end{equation}
In the limit $R' \to 0$, one hemisphere, say $\mathcal{H}_1$, will contain a single emission, which we denote $k_1$, with a transverse momentum of order $T_M Q/2$ that will set the thrust-major axis. We consider the observable when all other emissions, $k_i$, have transverse momenta less than $\delta T_M Q/2$ where $\epsilon \ll \delta \ll 1$ thus 
\begin{equation}
	T_{M,{\rm sc}}(\{\tilde{p}\},\{k_i\})\simeq 
	T_{M,{\rm sc}}(\{ \tilde{p}\},k_1) \ \ .
\end{equation}
We use eqs.~\eqref{eq:dZsc-final} and~\eqref{eq:dZ-rescaled} and introduce the rescaled variables $x_{\ell} = 2p_{y,\ell}/(T_M Q)$ and the two-dimensional vectors $\vec{\zeta}_i \equiv \zeta_i(\cos\phi_i,\sin\phi_i)$. We perform the rescaling $\zeta_i \to \zeta_1 \zeta_i$ and, as $R' = R'_1 + R'_2$, we find
\begin{align} \label{eq:TM-mean-NLL}
	\nonumber
	&\langle T_M \rangle_{\rm NP} \simeq 
	\lim_{R' \to 0}\Bigg[\mathcal{M}\frac{\langle \kappa \rangle_{\rm NP}}{Q} \, \int_0^{\infty}\frac{d\zeta_1}{\zeta_1} \, \zeta_1^{R'(\zeta_1 v)} \, R'_1(\zeta_1 v) \, \int_{-\infty}^{\infty}dx_2 \, \delta\bigg(x_2+\sum_{i\in \mathcal{H}_2}\zeta_{i}\sin\phi_i\bigg) \, \Theta(1-\zeta_1) \, \times \\
	&\times \left(\lim_{\epsilon \to 0} \, \epsilon^{R'_2} \, \sum_{n=0}^{\infty}\frac{\left(R'_2\right)^n}{n!} \, \prod_{i=2}^{n+1} \, \int_{\epsilon}^{1}\frac{d\zeta_i}{\zeta_i} \, \int_0^{2\pi}\frac{d\phi_i}{2\pi}\right) \, \frac{4}{\pi} \left(\ln\frac{1}{\zeta_1} + \frac12\ln\frac{1}{|x_2|} + (2\ln2 - 2)\right) \, \Bigg] + 1 \leftrightarrow 2 \ \ .
\end{align}
We note that the factor of $e^{-R(v)}$ in eq.~\eqref{eq:dZsc-final} has been dropped as we are in the limit $R' \to 0$ where $T_M \to T_{M,\max}$ and thus $e^{-R\left(v_{\max}\right)} \to 1$. We also drop the $\ln (1/T_M)$ term as this will vanish in the limit $R' \to 0$ as $T_{M,\rm{max}} \simeq 1$ as before.

We consider the three terms in the final pair of brackets:
\begin{itemize}
	\item 
	For the $\ln (1/\zeta_1)$ and $(2\ln2 - 2)$ terms there is no dependence on emissions in $\mathcal{H}_2$ and therefore the integrals over $\mathcal{H}_2$ will trivially evaluate to 1.
	\item 
	For the $\ln (1/|x_2|)$ term, we of course have a dependence on emissions in $\mathcal{H}_2$ but we note that the integrals over $\mathcal{H}_1$ and $\mathcal{H}_2$ will still factorise. We set the upper-bound of the $\zeta_i$-integration to $\delta$ (to reflect the imposed restriction on transverse momenta of emissions in $\mathcal{H}_2$) and recognise the integrals over $\mathcal{H}_2$ as precisely that in eq.~\eqref{eq:chiM-smallR'-start}. We recall from eq.~\eqref{eq:chiM-leading-full} that this will give $(2/\pi)(1/R'_2 + \ln2) + \mathcal{O}(R')$.
\end{itemize}
For the integrals over $\mathcal{H}_1$ we note that, as was the case for $\langle B_W \rangle_{\rm NP}$ and $\langle B_T \rangle_{\rm NP}$, rescaling the emissions in $\mathcal{H}_2$ produces an exponent of $R'(\zeta_1 v)$ rather than $R'_1(\zeta_1 v)$ (as was present for $\langle B_1 \rangle_{\rm NP}$). 

To identify the precise form of any $1/R'$ divergence we take the NLL approximation of the Sudakov radiator, i.e. $R'_{\ell}(\zeta_1 v) \simeq R'_{\ell}(v)$, in eq.~\eqref{eq:TM-mean-NLL}. Performing the $\zeta_1$-integration we find
\begin{equation}
	\langle T_M \rangle_{\rm NP} \simeq 
	\mathcal{M}\frac{\langle \kappa \rangle_{\rm NP}}{Q} \, \frac{4}{\pi} \left[\frac{R'_1}{(R'_1 + R'_2)^2} + (2\ln2 - 2)\frac{R'_1}{R'_1+R'_2} + \left(\frac{1}{2R'_2} + \frac{\ln2}{2}\right) \frac{R'_1}{R'_1+R'_2}\right] + 1 \leftrightarrow 2 \ \ ,
\end{equation}
which simplifies to give
\begin{equation}
	\langle T_M \rangle_{\rm NP} \simeq 
	\mathcal{M}\frac{\langle \kappa \rangle_{\rm NP}}{Q} \, \frac{4}{\pi}\left[(2\ln2 - 2) + \frac{1}{2R'_1} + \frac{1}{2R'_2} + \frac{\ln2}{2}\right] \ \ .
\end{equation}
As discussed in section~\ref{sec:subtraction}, in the limit $R' \to 0$ one can no longer neglect the higher derivatives of the radiator in eq.~\eqref{eq:TM-mean-NLL}. The accuracy that we require is to have full control over all constant terms in the limit $R' \to 0$. We therefore return to eq.~\eqref{eq:TM-mean-NLL} which, using the  $\mathcal{H}_1 \leftrightarrow \mathcal{H}_2$ symmetry, we write in a simplified manner as 
\begin{align}
	\nonumber
	\langle T_M \rangle_{\rm NP} \simeq 
	\mathcal{M}\frac{\langle \kappa \rangle_{\rm NP}}{Q} \, \frac{4}{\pi} \Bigg\{&\lim_{R' \to 0}\Bigg[\int_{-\infty}^{\infty}dx_2 \, \left(\lim_{\epsilon \to 0} \, \epsilon^{R'_2(\zeta v)} \, \sum_{n=0}^{\infty}\frac{\left(R'_2\big(\zeta v\big)\right)^n}{n!} \, \prod_{i=2}^{n+1} \, \int_{\epsilon}^{\delta}\frac{d\zeta_i}{\zeta_i} \, \int_0^{2\pi}\frac{d\phi_i}{2\pi}\right) \times \\
	&\times \delta\bigg(x_2+\sum_{i\in \mathcal{H}_2}\zeta_{i}\sin\phi_i\bigg) \, \ln\frac{1}{|x_2|}\Bigg] + (2\ln2 - 2) - \frac{\ln2}{2}\Bigg\} \ \ .
\end{align}
To achieve the required accuracy we will perform the logarithmic expansion of $R'_{\ell}(\zeta_1 v)$, expanding to $\mathcal{O}(R^{(3)})$. To enable convenient analytic computation we may introduce the same theta-constraint as we had for $\chi^{\textrm{c.t.}}_M(R')$ and  $\chi^{\textrm{c.t.$\,$imp}}_M(R')$. This constraint is trivially satisfied in the limit $R' \to 0$ and thus will not affect the result to the given accuracy. We therefore write
\begin{align}
	\nonumber
	\langle T_M \rangle_{\rm NP} \simeq 
	\mathcal{M}\frac{\langle \kappa \rangle_{\rm NP}}{Q} \, \frac{4}{\pi} &\Bigg\{\lim_{R' \to 0}\Bigg[ \int_{-\infty}^{\infty}dx_2 \, \left(\lim_{\epsilon \to 0} \, \epsilon^{R'_2(\zeta v)} \, \sum_{n=0}^{\infty}\frac{\left(R'_2\big(\zeta v\big)\right)^n}{n!} \, \prod_{i=2}^{n+1} \, \int_{\epsilon}^{\infty}\frac{d\zeta_i}{\zeta_i} \, \int_0^{2\pi}\frac{d\phi_i}{2\pi}\right) \times \\
	&\times \delta\bigg(x_2+\sum_{i\in \mathcal{H}_2}\zeta_i\sin\phi_i\bigg) \, \ln\frac{1}{|x_2|} \, \Theta\Bigg(1 - |x_2| - \sum_{i\in \mathcal{H}_2} \zeta_{i}\Bigg)\Bigg] + (2\ln2 - 2) - \frac{\ln2}{2}\Bigg\} \ \ ,
\end{align}
noting that the theta-constraint allows us to send the upper-bound of the $\zeta_i$-integration to infinity. Setting $R_1^{(n)} = R_2^{(n)} = R^{(n)}/2$, we deduce from eq.~\eqref{eq:chiMbarbase2} that
\begin{equation}
	\langle T_M \rangle_{\rm NP} \simeq 
	\mathcal{M}\frac{\langle \kappa \rangle_{\rm NP}}{Q} \left\{\frac{4}{\pi} \left(2\ln2 - 2 - \frac{\ln2}{2}\right) + \lim_{R' \to 0}\Bigg[2 \, \mathcal{F}_{T_M}(R') \, \chi^{\textrm{c.t.$\,$imp}}_M(R^{(n)}) \Bigg]\right\} \ \ .
\end{equation}
In the limit $R' \to 0$, $R' \ll \sqrt{R^{\prime\prime}}$ and from eq.~\eqref{eq:erfc-two-regimes} we find that
\begin{equation} \label{eq:mean-major}
	\langle T_M \rangle_{\rm NP} \simeq 
	\mathcal{M} \frac{\langle \kappa \rangle_{\rm NP}}{Q} \, \frac{4}{\pi} \, \left[(2\ln2 - 2) + \frac{\pi}{2\sqrt{C_F\,\alpha_s(Q)}} + \frac34 - \frac{2\pi\beta_0}{3 C_F} + \frac{\ln2}{2}\right] + \mathcal{O}(\sqrt{\alpha_s}) \ \ .
\end{equation}

\bibliographystyle{JHEP}
\bibliography{hadronisation.bib}

\end{document}